\documentclass[10pt,a4paper]{article}

\usepackage[margin=0.9in]{geometry}
\usepackage{authblk}
\usepackage{lmodern}
\usepackage{amsmath, amssymb}
\usepackage{graphicx}
\usepackage{setspace}
\usepackage{titlesec}
\usepackage{fancyhdr}
\usepackage{array}
\usepackage{booktabs}
\usepackage{latexsym}
\usepackage[authoryear,round]{natbib}
\usepackage{amsfonts}
\usepackage{upgreek}
\usepackage{subfigure}
\usepackage{bm}
\usepackage{bbold}
\usepackage{tcolorbox}
\usepackage{tabularx}
\definecolor{dg}{HTML}{7FA393}
\definecolor{lg}{HTML}{F3F7F5}
\definecolor{urlgreen}{HTML}{166C65}
\usepackage[bookmarks, colorlinks=true, linkcolor=urlgreen, urlcolor=urlgreen, citecolor=urlgreen, backref=page]{hyperref}
\usepackage{comment}
\usepackage{soul}
\usepackage{tikz}
\usepackage{float}

\renewcommand*{\backref}[1]{}
\renewcommand*{\backrefalt}[4]{%
    \ifcase #1 % No citations.
        No citations.%
    \or
        Referenced on page #2.%
    \else
        Referenced on pages #2.%
    \fi%
}

\usepackage{braket}

\newcommand{\ou}{%
  \mathrel{%
    \vcenter{\offinterlineskip
      \ialign{##\cr$<$\cr\noalign{\kern-1.5pt}$>$\cr}%
    }%
  }%
}

\renewcommand{\i}{\mathbf{i}}
\renewcommand{\j}{\mathbf{j}}

\titleformat{\section}{\normalfont\Large\bfseries}{\thesection}{1em}{}
\titleformat{\subsection}{\normalfont\large\bfseries}{\thesubsection}{1em}{}

\title{\Large\textbf{Machine learning applications in cold atom quantum simulators}}
\author[1,2]{Henning Schl\"omer}
\author[1,2]{Annabelle Bohrdt}
\affil[1]{\small Department of Physics and Arnold Sommerfeld Center for Theoretical Physics (ASC), Ludwig-Maximilians-Universit\"at M\"unchen, Theresienstr. 37, M\"unchen D-80333, Germany}
\affil[2]{\small Munich Center for Quantum Science and Technology (MCQST), Schellingstr. 4, D-80799 M\"unchen, Germany}
\date{}

\begin{document}

\maketitle

% Abstract
\begin{abstract}
As ultracold atom experiments become highly controlled and scalable quantum simulators, they require sophisticated control over high-dimensional parameter spaces and generate increasingly complex measurement data that need to be analyzed and interpreted efficiently. Machine learning (ML) techniques have been established as versatile tools for addressing these challenges, offering strategies for data interpretation, experimental control, and theoretical modeling. In this review, we provide a perspective on how machine learning is being applied across various aspects of quantum simulation, with a focus on cold atomic systems. Emphasis is placed on practical use cases---from classifying many-body phases to optimizing experimental protocols and representing quantum states---highlighting the specific contexts in which different ML approaches prove effective. Rather than presenting algorithmic details, we focus on the physical insights enabled by ML and the kinds of problems in quantum simulation where these methods offer tangible benefits.
\end{abstract}

\tableofcontents
%======================================INTRODUCTION====================================================

\section{Introduction}

Quantum simulators based on neutral atoms have emerged as powerful platforms for studying strongly correlated many-body quantum systems in controlled experimental settings~\cite{Bloch2008, Bernien2017}. In particular, these simulators offer control over system parameters, long coherence times, and the capability to resolve individual particles using high-resolution imaging techniques.  As a result, they promise to enhance our understanding of strongly correlated phases of matter---particularly in regimes that are challenging for classical computational methods~\cite{Daley2022}. However, these advances also introduce significant challenges, including the following:

\begin{figure}[t!]
\begin{centering}
\includegraphics[width=0.89\textwidth]{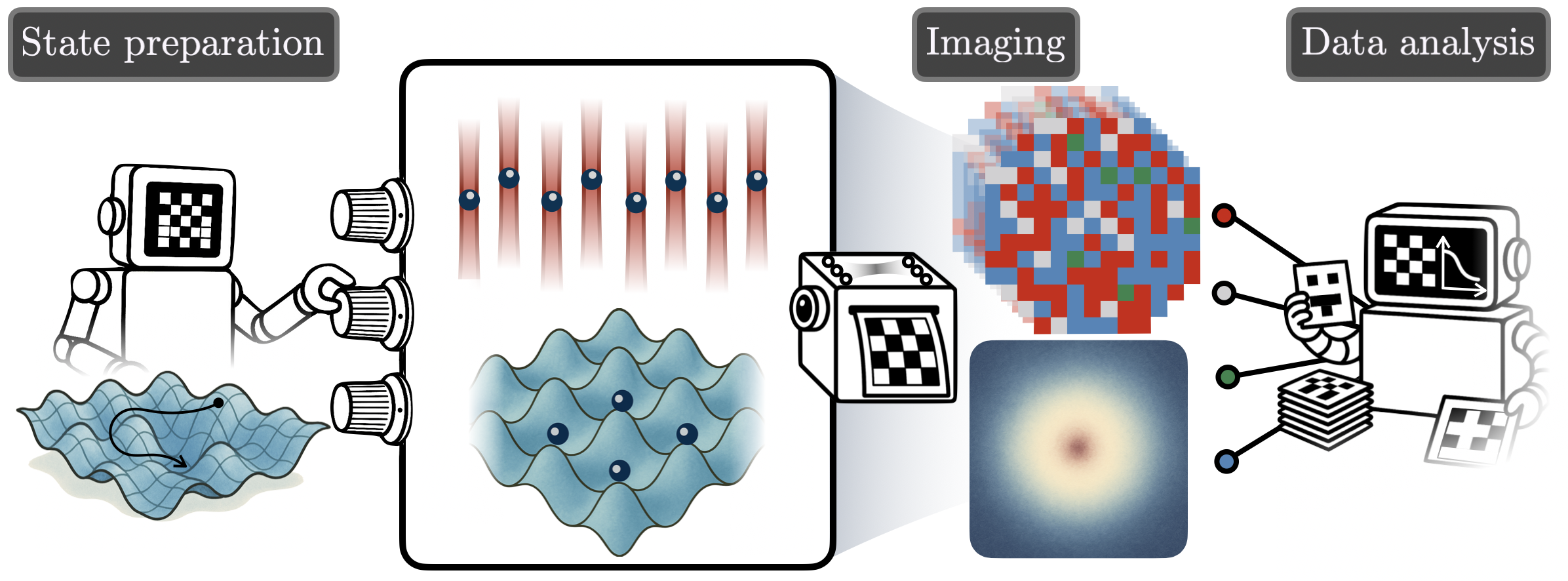}
\caption{\textbf{Applications of machine learning in quantum simulation experiments.} 
Schematic overview of how machine learning can enhance ultracold atom experiments. On the data analysis side, machine learning methods can help to identify phase transitions, uncover physical structures, and interpret experimental measurements---see Sec.~\ref{sec:DA}. On the experimental side, machine learning can assist in the preparation of quantum many-body states by optimizing experimental control sequences, as well as improve detection of individual atom positions during imaging---see Sec.~\ref{sec:ExpAss}. Parts of the figure were generated with OpenAI's 4o model.}
\label{fig:overview}
\end{centering}
\end{figure}

\begin{enumerate}
\item When exploring Hamiltonians that lack complete theoretical understanding, data analysis techniques are needed that are capable of extracting physically meaningful information from experimental results. For instance, many-body phases characterized by unknown, e.g. non-local, order parameters are challenging to interpret using conventional analysis methods, calling for algorithms that can identify phase transitions and extract order parameters directly from the measurement data provided by quantum simulators.
\item Due to the remarkable technical complexity of experimental setups, high-dimensional parameter spaces associated with experimental control (e.g. cooling protocols, trapping methods, state preparation, and imaging) are introduced. Identifying optimal experimental sequences to achieve high-fidelity quantum simulations is a substantial technical challenge.
\end{enumerate}

In recent years, machine learning (ML) techniques have emerged as promising tools to address these challenges, offering new methods for data interpretation and experimental optimization, schematically illustrated in Fig.~\ref{fig:overview}. In general, ML includes a wide range of algorithms that learn patterns from data either through supervised (using labeled data) or unsupervised (without labeled data) methods. In cold atom quantum simulators, ML has been applied across a variety of tasks. In the context of data analysis, examples include classifying many-body quantum phases from experimental snapshots, reconstructing quantum states from noisy measurements, and identifying phase transitions in systems where traditional order parameters fail. 

Beyond data analysis, ML techniques like reinforcement learning and Bayesian optimization have been utilized to directly improve the experimental design and control. For example, finding optimized, non-trivial cooling protocols has been shown to significantly improve experimental efficiency.

\begin{table}[!t]
\caption{Common supervised and unsupervised machine learning techniques used to analyze data in the context of many-body physics and cold atom experiments, along with their respective applications.}
\begin{tabular}{@{}lp{10.2cm}@{}}
\toprule
\textbf{Method} & \textbf{Purpose} \\
\midrule
\multicolumn{2}{@{}l}{\textbf{Supervised Learning}} \\
Neural Networks & Predict physical parameters or phase labels from data. Includes architectures such as feed-forward networks, convolutional networks, and transformers. \textit{Applications:} Classification of phases, extraction of features from quantum gas microscope images. Interpretability through tailored network architectures. \\
Support Vector Machines (SVMs) & Classify labeled data into different phases or regimes using decision boundaries. \textit{Applications:} Detection of phase transitions. Interpretability via decision functions and kernel analysis. \\
Random Forests & Classify data and identify important input features. \textit{Applications:} Robust classification, estimation of parameter importance. Interpretability through decision paths. \\
\midrule\midrule
\multicolumn{2}{@{}l}{\textbf{Unsupervised Learning}} \\
Principal Component Analysis (PCA) & Linear dimensionality reduction along directions of maximal variance in the data. Kernel PCA extends this approach to capture non-linear structures. \textit{Applications:} Visualization, phase separation, identification of simple order parameters. \\
t-distributed Stochastic Neighbor Em-& Nonlinear dimensionality reduction technique that preserves local \\ bedding (t-SNE) & similarities. \textit{Applications:} Visualization and clustering of many-body configurations/experimental snapshots. \\
Diffusion Maps & Nonlinear dimensionality reduction that preserves the intrinsic geometry of the data manifold. \textit{Applications:} Identifying topological phases of matter. \\
Intrinsic Dimension Analysis & Estimate the number of latent variables needed to describe the dataset. \textit{Applications:} Characterization of system complexity and detection of phase transitions. \\
Autoencoders & Neural networks that learn compact, low-dimensional representations of the data. \textit{Applications:} Unsupervised phase discovery, anomaly detection, and denoising of experimental data. \\
Clustering Algorithms (e.g., k-means) & Group similar data points without predefined labels. \textit{Applications:} Phase identification after dimensionality reduction. \\
Gaussian Mixture Models (GMMs) & Model the data as a combination of Gaussian distributions to identify structures. \textit{Applications:} Identification of overlapping or poorly separated phases. \\
Confusion Learning & Identify phase boundaries by training classifiers after manually labeling the data; the maximum accuracy likely corresponds to the true labeling. \textit{Applications:} Automated detection of phase transitions. \\
Discriminative Cooperative Networks & Combine two networks to find the best phase boundary within the learn-by-confusion scheme. \\
Kolmogorov Networks & Estimate algorithmic complexity of data. \textit{Applications:} Characterization of emergent structure, complexity and randomness in many-body quantum states. \\
Tensorial Kernel SVM (TK-SVM) & Extracts interpretable order parameters using tensor kernels. \textit{Applications:} Automated detection of phase transitions with interpretability. \\
Siamese Neural Networks & Learn to compare input pairs by measuring their similarity in a low-dimensional space. \textit{Applications:} Detection of phase similarity, few-shot classification. \\
\bottomrule
\end{tabular}
\label{tab:ml-cold-atoms}
\end{table}

In this review, we highlight these applications: Section~\ref{sec:DA} focuses on data-driven techniques for analyzing quantum systems, from classical Ising models to strongly correlated quantum phases of matter. In this context, we also explore emerging quantum machine learning applications. Subsequently, Section~\ref{sec:ExpAss} discusses how ML methods help and enhance experimental control protocols.

A concise overview of frequently used ML algorithms and their typical applications in the broad context of cold atom quantum simulation experiments is provided in Table~\ref{tab:ml-cold-atoms}. Our primary goal in this review however is not to give detailed technical explanations of machine learning algorithms, but rather to emphasize their practical use in the broad context of quantum simulation. For detailed technical introductions to these ML methods, we refer the reader to review articles such as~\cite{CarleoReview, CarrasquillaReview, NeupertReview, Johnston2022, DawidReview, WetzelReview}.

%=================================DATA ANALYSIS========================================================
\section{Data analysis}
\label{sec:DA}

Unlike traditional condensed matter systems, cold atom experiments offer access to full quantum state statistics on a shot-by-shot basis, rather than only ensemble-averaged quantities. After preparing a many-body state of interest, $\ket{\Psi} = \sum_{n} c_n \ket{n}$, these experiments typically employ projective measurements: laser power is rapidly increased, and fluorescence imaging with simultaneous cooling projects the quantum state onto a Fock basis state $\ket{n}$ of the system’s Hilbert space. The outcome is a large collection—often thousands—of individual snapshots of the many-body state. These snapshots encode far more than local observables; they provide genuine samples of the many-body state and thus enable the extraction of rich, non-trivial information, including non-local and higher-order correlations~\cite{Endres2011, Islam2015, Hilker2017, Rispoli2019}. 

However, leveraging the full potential of this data requires tools for interpretation. Machine learning methods seem particularly well-suited for this task, as they can uncover patterns, correlations, and structures that are often hidden from conventional analysis, especially in systems lacking clear order parameters. Yet, many ML frameworks function as ``black boxes'', offering only limited insight into the physical mechanisms behind their predictions and decisions. This highlights the need for interpretable approaches that are not only accurate but also offer a meaningful physical understanding.

This section explores how machine learning can be (and has been) employed to analyze many-body data across a range of models realized in quantum simulation setups. Beginning with classical spin systems such as the Ising model (Sec.~\ref{sec:cl}), which can be used as a playground for testing various learning strategies, we move on to more complex quantum systems, including topologically nontrivial systems (Sec.~\ref{sec:topo}), Rydberg atom arrays (Sec.~\ref{sec:ryd}), as well as Fermi- and Bose-Hubbard models (Secs.~\ref{sec:FH},~\ref{sec:BH}). Across these cases, we highlight how different ML approaches---supervised and unsupervised learning, dimensionality reduction, anomaly detection, and more, see Table~\ref{tab:ml-cold-atoms}---have enabled the detection and classification of phases of matter from snapshot data, while at the same time giving useful physical insights. In Sec.~\ref{sec:HL} we discuss how Hamiltonian learning techniques can help to verify quantum devices as well as gain physical insights by reconstructing effective Hamiltonians. In the context of Rydberg atom arrays, we further review and discuss how cold atom systems can be used for quantum machine learning applications. Finally, in Sec.~\ref{sec:QST}, we review how machine learning enables quantum state tomography, i.e., reconstructing the underlying quantum state $\ket{\Psi}$ from projective measurements $\ket{n}$ in the Fock basis.

\subsection{Classical systems}
\label{sec:cl}
The analysis of snapshots generated from classical models, such as the Ising model, has proven to be a valuable testing ground for a wide range of data analysis techniques. In this context, thermal equilibrium configurations of the Ising Hamiltonian
\begin{equation}
H = -J \sum_{\langle \mathbf{i},\mathbf{j} \rangle} \sigma_{\mathbf{i}} \sigma_{\mathbf{j}},
\end{equation}
with $\sigma_{\mathbf{i}} = \pm 1$ and $\langle \mathbf{i},\mathbf{j} \rangle$ denoting nearest-neighbor pairs on a lattice (typically the square lattice), are obtained via Monte Carlo sampling. At a critical temperature $T_c/J$, the model exhibits a well-known second-order thermal phase transition from a disordered phase to a long-range ordered phase characterized by spontaneous magnetization. The sampled spin configurations correspond to the type of data produced in quantum gas microscope experiments and can be used to detect the phase transition, for instance by evaluating the magnetization. A broad range of machine learning techniques has been applied to such datasets, demonstrating that even relatively simple models are capable of identifying known phases and locating critical points directly from spin configurations.

\begin{figure}[t!]
\begin{centering}
\includegraphics[width=\textwidth]{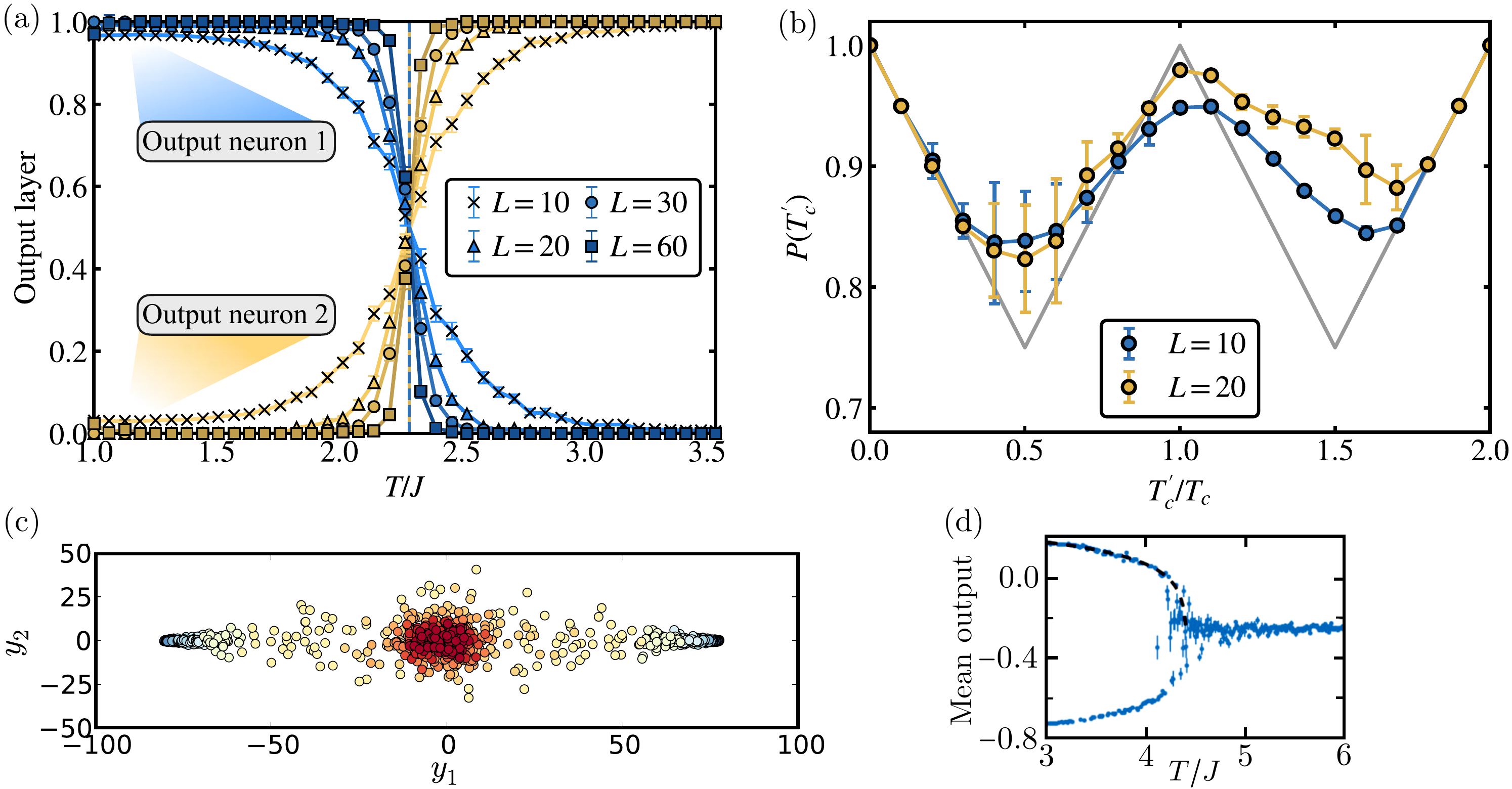}
\caption{\textbf{Applications of machine learning to the classical Ising model.} 
(a) Classification output from supervised training on thermal snapshots using fully connected neural networks. Yellow and blue data correspond to the values of the two output neurons used for classification. Finite-size scaling analysis allows for an estimation of the critical temperature and critical exponents. Data taken from~\cite{Carrasquilla2017}. 
(b) ``Learning by confusion'' scheme applied to the Ising model. The network's classification accuracy shows a characteristic W-shape, with a local maximum when the assumed critical temperature $T_c'$ matches the true $T_c$. Data taken from~\cite{Nieuwenburg2017}. 
(c) Projection of thermal snapshots onto the first two principal components of the dataset. Red and blue data points correspond to high and low temperatures, respectively. The first principal component can capture the total magnetization and therefore shows structure across the phase transition: the central region corresponds to the disordered phase, while the left and right clusters correspond to the two symmetry-broken phases. Figure adapted from~\cite{Wang2016}. 
(d) Latent variable of an autoencoder with a one-dimensional bottleneck applied to the 3D Ising model. The latent representation closely matches the known order parameter across the phase transition (dashed line). Data taken from~\cite{Chng2018}.}
\label{fig:Ising}
\end{centering}
\end{figure}

\textbf{Supervised learning.} In a pioneering work~\cite{Carrasquilla2017}, a supervised feedforward neural network was used to classify configurations of the 2D Ising model into its ordered and disordered phases. The network accurately located the phase transition after finize-size analysis, shown in Fig.~\ref{fig:Ising}~(a).  In a similar spirit, convolutional neural networks (CNN) have been trained to detect the Ising model’s transition~\cite{TanakaTomiya2017}. The CNN was shown to develop an internal order parameter related to the weights of the network, shown to correspond to the magnetization of the system---being an early demonstration of interpretability in terms of physical observables. In~\cite{Wetzel2017Int}, it was shown that further insights into the decision-making process of a neural network can be gained by systematically reducing the filter size, i.e., shrinking the receptive field to increasingly smaller patches. Through this process, it was found that evaluating specific two-point correlations---corresponding to the average energy per spin site---provides more reliable classification than focusing only on the magnetization. Along similar lines, reducing the neural network to minimal sizes can yield interpretability for Ising-type models by analyzing the individual weights~\cite{Suchsland2018, Kim2018, Kashiwa2019}. 

Going beyond the plain-vanilla Ising model, it has been shown that a model trained on the standard Ising model can generalize to related systems that share the same order parameter but exhibit different critical temperatures---a method known as transfer leaning. For example, this includes extensions of the Ising Hamiltonian with an added (uniform or random) longitudinal field of the form $\propto h \sum_{\mathbf{i}} \sigma_{\mathbf{i}}$~\cite{Huembeli2018}, or different lattice geometries such as the triangular lattice~\cite{Carrasquilla2017}. Furthermore, neural networks were shown to identify order in gauge transformed Ising models which, without knowing the gauge transformation, seem disordered; analyzing the network's weights allows for a reconstruction of the underlying gauge from the trained models~\cite{Morishita2022}. 

Support Vector Machines (SVMs) are another popular class of supervised learning algorithms used for classification and regression by identifying the optimal hyperplane that maximally separates data classes. A key strength of SVMs lies in their use of kernel functions, which map input data into higher-dimensional spaces where complex structures become linearly separable. Unlike neural networks, which often lack interpretability, SVMs offer a controlled and transparent framework: kernels can be chosen to correspond to physically meaningful quantities, such as spin-spin correlations. SVMs have been successfully applied to classify and interpret phases in classical systems such as the 2D Ising model~\cite{Ponte2017}. 

The tensorial kernel support vector machine (TK-SVM) extends the standard SVM approach by constructing higher-order correlations from tensorial combinations of local observables---such as higher-rank spin tensors. Applied to classical spin systems, this method has uncovered multipolar orders in frustrated magnets~\cite{Greitemann2019a, Liu2019ML, Greitemann2021} and mapped out phase diagrams with competing spin liquids and nematic phases~\cite{Greitemann2019b, Liu2021, Sadoune2025}. Although based on a supervised learning framework, TK-SVMs work in an effectively unsupervised mode, enabling the discovery and interpretation of phases even in the absence of labeled data or prior knowledge of the underlying Hamiltonian. These methods are discussed in more detail in the following.

\textbf{Unsupervised learning.} To detect phase transitions without requiring prior knowledge of labeled data, several unsupervised and semi-supervised strategies have been developed. One widely used method is the “learning by confusion” strategy~\cite{Nieuwenburg2017}, which uses a neural network classifier trained on (some chosen) mislabeled data to predict the location of phase boundaries. Specifically, snapshots of the system are labeled according to a ``trial critical point'' along some control parameter (such as temperature, Hamiltonian parameters, etc.): configurations with control parameters below the trial point are assigned one label, and those above it are assigned the other. The neural network is then trained to distinguish between these artificially labeled classes. This procedure is repeated for a range of trial critical points. When the trial labeling coincides with the true phase boundary, the classification task becomes easiest (as the two classes now contain different phases with qualitatively different characteristics), and the network achieves a local maximum in test accuracy---corresponding to a minimum in ``confusion''. Thus, the critical point can be identified as the trial value where classification performance peaks, as illustrated in Fig.~\ref{fig:Ising}~(b). This technique was successfully demonstrated on a broad range of models, including the 2D Ising model~\cite{Nieuwenburg2017}. By incorporating a second neural network into the pipeline, the task of finding the optimal data labeling can be automated, called ``Discriminative Cooperative Networks'': In the case of the Ising model, it was demonstrated that starting from an initial guess of the phase transition point, optimizing the second network shifts this guess toward the true critical point~\cite{Liu2018}. When using regression instead of classification, phase transitions can instead be detected via minima in the regression uncertainty~\cite{Guo2023}, which requires only a single trained model. Training a single multi-class classifier (rather than multiple binary classifiers) has further been shown to accelerate the learning-by-confusion scheme in the context of the Ising model~\cite{Arnold2023MTL}.

Another widely used class of unsupervised methods involves dimensionality reduction and clustering. A particularly prominent and simple technique is principal component analysis (PCA)~\cite{AbdiPCA}, which identifies the directions (principal components) along which the variance in the input data is maximized. This allows datasets to be effectively represented in a lower-dimensional space while preserving their most significant structural features.

In the case of the 2D Ising model, PCA shows that the configurations vary predominantly along the first principal component as the temperature is tuned~\cite{Wang2016, Wetzel2017}. As illustrated in Fig.~\ref{fig:Ising}~(c), projections of the data onto the first two principal components show three distinct clusters, corresponding to the disordered phase and the two symmetry-broken ordered phases (with positive and negative magnetization). Indeed, the first principal component was shown to directly correspond to the total magnetization, thereby sufficiently separating the data in the different phases.

PCA can also be applied to scenarios where a simple evaluation of the magnetization does not characterize the different phases. For example, when fixing the total magnetization in the Ising model to be zero, non-trivial domain wall structures that break the $C_4$ symmetry of the underlying lattice form below the Ising transition; the largest four principal components then lead to a successful clustering and characterization of the disordered and ordered phases~\cite{Wang2016}. Other examples of applying PCA to classical spin systems include the Blume-Capel model (and generalizations) and the biquadratic-exchange spin-1 Ising model~\cite{Hu2017}, as well as frustrated~\cite{Hu2017, Wang2017Frustrated}, non-equilibrium~\cite{Casert2019}, and gauge transformed Ising models~\cite{LazGo2022}. In more complex settings where order parameters are nonlinear functions of the input configurations, kernel PCA (which adds a non-linear component to PCA) has been shown to identify phases such as classical $\mathbb{Z}_2$ chiral order~\cite{Wang2018Kernel}.

In the 3D Ising model, conventional PCA analysis was demonstrated to be more challenging compared to the 2D case. Nevertheless, analysis of the PCA entropy $S_{\text{PCA}}$ (defined by the values of all principal components) can give useful insights. In particular, a qualitative similarity of $S_{\text{PCA}}$ with the physical thermodynamic entropy of the Ising model has been established; by analyzing its behavior around the transition point, accurate estimations of the critical temperature could be obtained both in 2D and 3D~\cite{Panda2023}. 

Intrinsic dimension ($I_d$) analysis~\cite{Camastra2016} is an alternative unsupervised method that compliments dimensionality reduction schemes like PCA for studying phase transitions. While PCA projects data onto (linear) subspaces to identify dominant directions of variation, intrinsic dimension methods try to directly quantify the minimal number of variables needed to describe the data manifold (i.e. without projecting or compressing the data). Applied to the 2D Ising model, $I_d$ exhibits a clear, non-monotonic signature near the critical temperature; a finite-size scaling analysis then accurately reproduces both $T_c$ and critical exponents~\cite{MendesSantos2021}. As discussed in Sec.~\ref{sec:topo}, intrinsic dimension analysis can also be used in more subtle scenarios, such as topological phase transitions.

Other unsupervised dimensionality reduction methods that have been applied to classical spin systems include autoencoders and t-distributed stochastic neighbor embedding (t-SNE)~\cite{Carrasquilla2017, Wetzel2017, Chng2018}. For example, an autoencoder (a neural network trained to compress and reconstruct data) applied to Ising spin configurations was shown to learn a latent representation related to temperature: the autoencoder’s reconstruction error and its compressed variables changed behavior near the critical point, analogous to how magnetization or susceptibility do~\cite{Chng2018}, see Fig.~\ref{fig:Ising}~(d). Similarly, t-SNE, a nonlinear dimensionality reduction technique, was used to embed Ising snapshots into two dimensions, revealing distinct clusters associated with the different phases~\cite{Wetzel2017, Chng2018}.

Yet another approach involves training a predictive model---such as a neural network---to predict the underlying system parameters from individual snapshots, for example, $J/T$ in the isotropic Ising model or $(J_x/T, J_y/T)$ in an anisotropic setting. The difference between the predicted and true parameters defines a vector field over the parameter space, whose structure (for instance its divergence) can be used as an indicator for phase transitions~\cite{Schafer2019}. Along similar lines, discriminative classifiers have been replaced by generative classifiers to model the underlying snapshot probability distribution~\cite{Arnold2024}.
   
\subsection{Topological systems}
\label{sec:topo}

Topological phases and transitions---such as those in quantum spin liquids, topological insulators, lattice gauge theories, or systems undergoing Berezinskii–Kosterlitz– Thouless (BKT) transitions---are challenging to capture with conventional methods due to the lack of local order parameters: their identification is often based on more abstract quantities such as vortices, Chern numbers, or other non-local probes. Machine learning methods may help to identify such transitions, including in the context of cold atom experiments where real-space or momentum-space snapshots can be used as input data. A common scheme to realize topological systems in quantum simulators is through Floquet engineering, where periodic driving creates synthetic gauge fields~\cite{Dalibard2011, Cooper2019, Eckardt2017}.

\begin{figure}[t!]
\begin{centering}
\includegraphics[width=\textwidth]{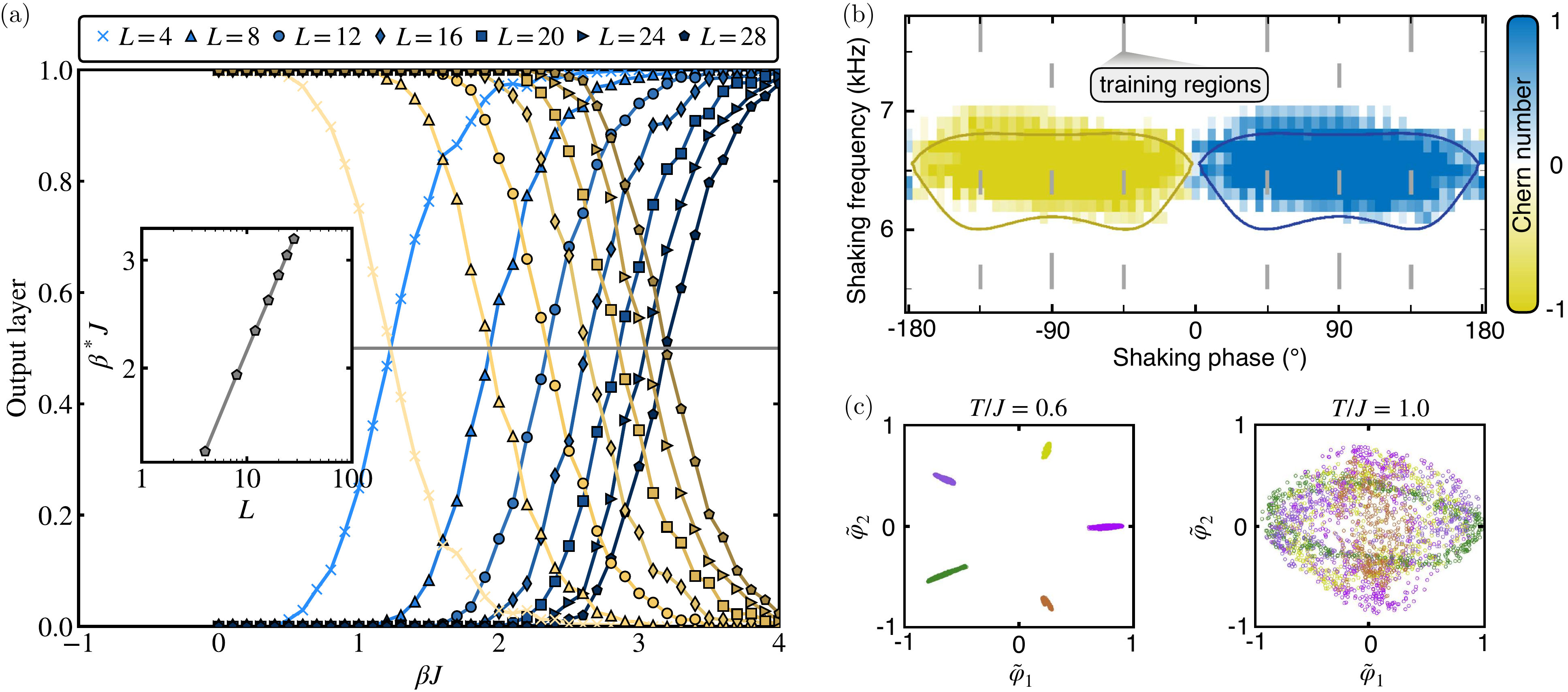}
\caption{\textbf{Applications of machine learning in topological models.} 
(a) Classification probabilities for assigning snapshots of the IGT at various inverse temperatures to either the topological ($T=0$) or trivial ($T=\infty$) phase. Supervised training was performed only at zero and infinite temperature. The model identifies a crossover temperature $\beta^*$ consistent with the expected scaling $\beta^* \propto \ln L$ (inset), where $L$ is the system size, marking the onset of significant thermal excitations. Data taken from from~\cite{Carrasquilla2017}. 
(b) Phase diagram of the experimentally realized Haldane model obtained via supervised learning. The model is trained only on snapshots deep within the distinct phases (gray lines). When applied to the full parameter space, it accurately reproduces the theoretical phase boundaries (solid lines). Data taken from~\cite{Rem2019}. 
(c) Diffusion maps applied to snapshots of the 2D XY model cluster the data into different topological winding numbers. On the left (topological phase), distinct winding sectors are clearly separated. On the right (above the BKT transition, in the trivial phase), no clear clustering is observed. Data taken from~\cite{RodNiev2019}.}
\label{fig:topo}
\end{centering}
\end{figure}

\textbf{Supervised learning.} CNNs trained on labeled lattice configurations of an Ising gauge theory (IGT) in equilibrium have been shown to classify between the two different phases (at $T=0$ and $T=\infty$) without relying of conventional symmetry-breaking order parameters~\cite{Carrasquilla2017}. The IGT Hamiltonian reads
\begin{equation}
    H = -J \sum_{\square} \prod_{\boldsymbol{\ell} \in \square} \sigma_{\boldsymbol{\ell}},
\end{equation}
where the sum is over all NN square lattice plaquettes ($\square$) and the product involves the vertices of each plaquette. The local constraints of minimizing the energy of all plaquette terms is globally fulfilled in the ground state, whereas at high temperatures, spins are disordered.  When training CNNs on snapshot data, the networks effectively learned the local energetic constraints that characterize the gauge theory, allowing them to identify crossover temperatures at which these constraints begin to be globally satisfied, see Fig.~\ref{fig:topo}~(a). Interestingly, fully connected neural networks failed to capture the gauge structure, showing that the spatial structure of convolutional layers can be essential for the classification of certain phases of matter~\cite{Carrasquilla2017}. 

By systematically shrinking a CNN’s filter size, it was identified that the network learns to evaluate certain non-local loop observables when trained to classify phases in an SU(2) gauge theory~\cite{Wetzel2017Int}. Transformer neural networks---known for their ability to capture non-local dependencies---have further been developed in an interpretable framework~\cite{Suresh2025}. Inspired by the correlator convolutional neural network (CCNN)~\cite{Miles2021}, see Sec.~\ref{sec:FH}, this method not only gives accurate classification but also yields physical insight into the network’s decision-making process, e.g. by identifying local Gauss law constraints. Furthermore, SVMs have been applied to classify and interpret phases in the Ising gauge theory~\cite{Ponte2017}, as well as models with emergent gauge structures~\cite{Greitemann2019b}.

Other approaches incorporate specific knowledge into the network: rather than feeding raw configurations into the model, tailored features such as loop observables can be used as input (known as quantum loop topography)~\cite{Zhang2017QLT, Zhang2017QLT2, Zhang2020}. This allowed to identify subtle topological signatures, including those of quantum Hall states and $\mathbb{Z}_2$ spin liquids.

Supervised classification has also been applied to simulated density distribution snapshots obtained after a particle undergoes a quantum walk, which were used as input to a neural network trained to identify the system’s topological phase. The time-evolved density distributions encode information about the Chern number, allowing the network to classify topological phases and detect phase transitions~\cite{Ming2019}.

In the context of analyzing data obtained directly from cold atom experiments, a CNN was trained to classify momentum-space images of ultracold bosons realizing a Floquet-engineered Haldane model~\cite{Rem2019}. By labeling the training data with known Chern numbers of the model, the network was able to reconstruct the entire topological phase diagram of the system from snapshot measurements, see Fig.~\ref{fig:topo}~(b): Even with the presence of experimental imperfections, the model identified subtle differences in momentum distributions corresponding to different phases. 

Along similar lines, CNNs have been trained on detecting topological phase transitions in a one-dimensional symmetry-protected topological (SPT) system realized with spin-orbit-coupled ultracold $^{173}\text{Yb}$ fermions~\cite{Zhao2022}. The network, provided with spin-resolved snapshots taken after time-of-flight expansion, was shown to identify topological phase transitions by effectively calculating the spin imbalance of the input snapshots.

\textbf{Unsupervised learning.} Subsequent efforts to analyze snapshots from the experimentally realized Haldane model in~\cite{Rem2019} shifted toward unsupervised learning approaches~\cite{Kaeming2021}, which used a combination of dimensionality reduction and clustering techniques. While each of these methods alone struggled to reconstruct the full phase diagram, their combination succeeded in revealing the topological transitions in a fully unsupervised manner.

Linear dimensional reduction schemes based on linear distances of individual data points, such as PCA, have been shown to be insufficient to capture the global structure of topological phases. However, other approaches, such as diffusion maps, have been successful in revealing topological structure from Fock configurations in spin models~\cite{RodNiev2019}. These algorithms cluster data according to topological characteristics; in particular, snapshots that are grouped in the same cluster can be continuously deformed into each other, i.e., they are in the same topological phase. In the classical 2D XY model, for example, diffusion maps naturally separated configurations based on the presence or absence of vortices, thereby capturing the BKT transition, see Fig.~\ref{fig:topo}~(c). Similarly, confined and deconfined phases were distinguished in the IGT~\cite{RodNiev2019}. Diffusion maps have further shown to be successful when applying them to the Haldane model~\cite{Lustig2020} as well as experimental data from a lattice of coupled waveguides~\cite{Lustig2020, Noh2017}. Furthermore, it has been shown that certain features of the intrinsic dimension can pinpoint the BKT transition temperature $T_{\text{BKT}}$, even for moderate system sizes~\cite{MendesSantos2021}.

The ability of neural networks to directly recognize topological defects such as vortices in snapshots of many-body systems has also been addressed~\cite{Beach2018}. While standard networks often pick up on spurious features like residual magnetization in finite-size systems, specifically engineered architectures were developed that learn the physically relevant, topological content of the data without having to use feature engineering of the input data~\cite{Beach2018}.

The learn-by-confusion scheme has further been demonstrated on a topological transition: it was shown that it can locate the phase transition in the 1D Kitaev Majorana chain (a prototypical topological superconductor) without any prior input about Majorana modes or winding numbers~\cite{Nieuwenburg2017}. In the case of the XY model, in contrast, it was shown that confusion learning rather identifies the peak structure of the specific heat, calling for more care when analyzing the system's BKT phase transition~\cite{Suchsland2018, Beach2018, Arnold2022PRX}.

Prediction-based schemes have also been explored in topologically nontrivial systems~\cite{Greplova2020}. In particular, they have been applied to both the classical IGT at finite temperature as well as the toric code in the ground state, using prediction errors as indicators of qualitative changes in the system's structure.

TK-SVMs have also been shown to detect topological order from local measurements. Applied to the toric code, it identified vertex and plaquette stabilizers away from the analytically solvable limit; in topologically non-trivial spin models, it reconstructed the phase diagram and uncovers string order parameters characterizing the SPT phase~\cite{Sadoune2023}. Applied directly to experimental data from a trapped-ion quantum simulator, TK-SVMs further distinguished topological from trivial phases from measurement snapshots~\cite{Sadoune2024}.

Alternative strategies focus not on snapshots, but on other representations of quantum states. For instance, neural networks have learned topological invariants directly from momentum-dependent Hamiltonians~\cite{Zhang2018LHam, Che2020}, real-space eigenstates~\cite{Huembeli2018, Holanda2020}, local projections of the density matrix~\cite{Carvalho2018}, or the entanglement spectrum~\cite{Liu2018}.

\subsection{Rydberg atom arrays and transverse field Ising model}
\label{sec:ryd}
Rydberg atom arrays consist of neutral atoms trapped in configurable optical tweezer arrays. With highly tunable interactions and programmable lattice geometries, they have emerged as a versatile platform for simulating and probing strongly correlated quantum many-body phases \cite{Bernien2017, Keesling2019, Scholl2021, Samajdar2020, Bluvstein2021, Ebadi2021, Chen2023} and quantum information processing~\cite{Saffman2010, Endres2016, Barredo2016}. The strong van der Waals interaction between the ground state $\ket{g}$ and highly excited states $\ket{r}$ enables the simulation of long-range interacting spin models. In particular, when coherently driving the $\ket{g} \leftrightarrow \ket{r}$ transition with laser fields with Rabi frequency $\Omega$ and detuning $\delta$, the following Hamiltonian is realized, 
\begin{equation}
    \hat{H}_{\text{Ryd}} = -\frac{\Omega}{2} \sum_{\i} \hat{\sigma}_{\i}^x - \delta \sum_{\i} \hat{n}_{\i} + \sum_{\i,\j} V_{\i\j} \hat{n}_{\i} \hat{n}_{\j}.
\end{equation}
Here, $\hat{\sigma}_{\i}^x = \ket{g}_\i\bra{r} + \ket{r}_{\i}\bra{g}$ acts as a transverse field, and the occupation $\hat{n}_{\i} = \ket{r}_{\i}\bra{r}$ measures Rydberg excitations. Atoms at positions $\i$ and $\j$ interact via a van der Waals potential $V_{\i\j} \propto 1 / |\i - \j|^6$, taking the role of Ising-type interactions. 

The Rydberg Hamiltonian is closely related to the transverse-field Ising model (TFIM), a paradigmatic model for studying quantum phase transitions, and hence provides a natural platform for simulating quantum magnetism in an analog mode. Beyond analyzing phases of matter and phase transitions with Rydberg quantum simulators, the flexibility and scalability of Rydberg tweezer arrays have also generated growing interest in their use for quantum machine learning applications. In the following, we review both directions.

\begin{figure}[t!]
\begin{centering}
\includegraphics[width=0.85\textwidth]{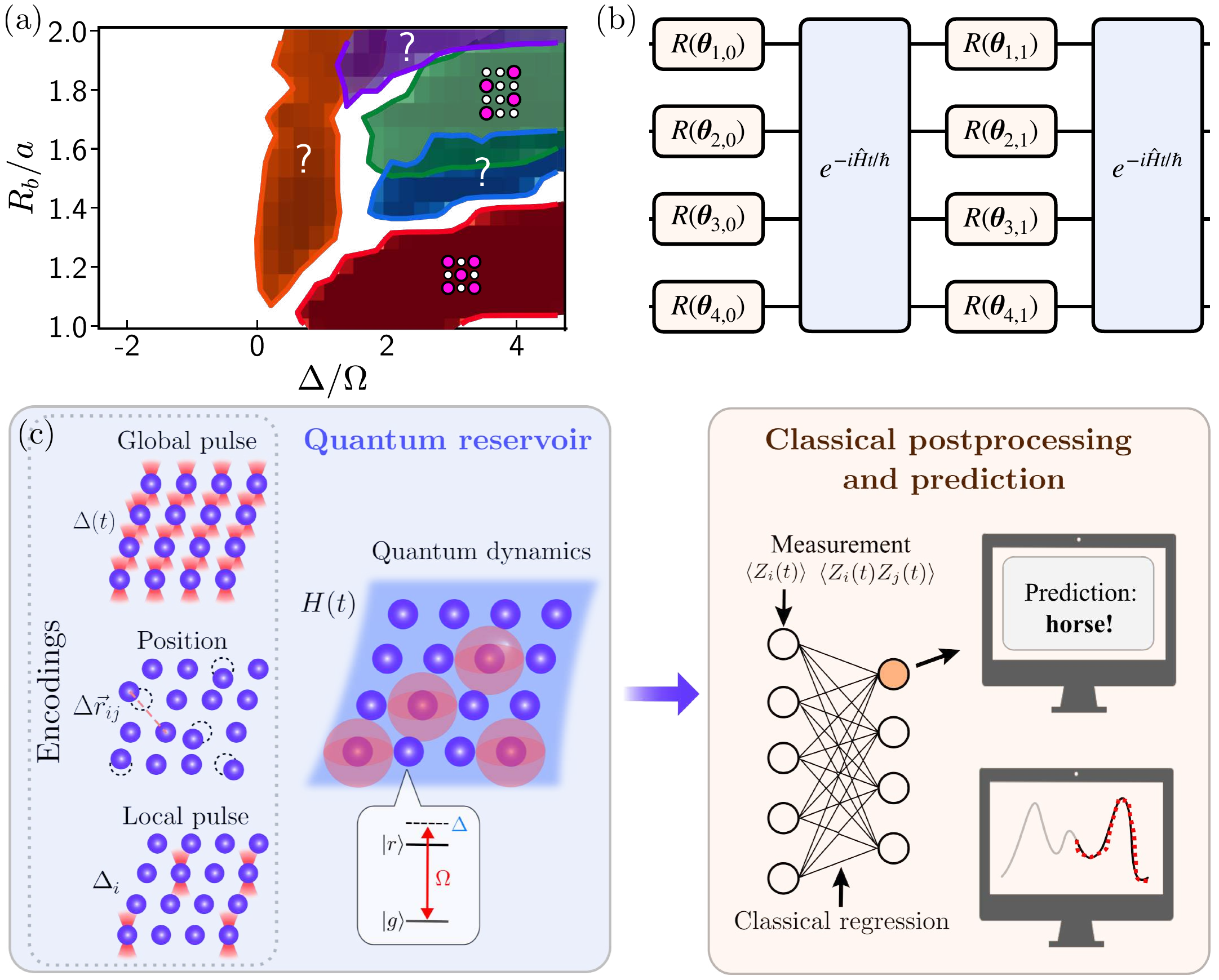}
\caption{\textbf{Applications of machine learning in Rydberg atom arrays.} 
(a) Two-stage phase detection scheme applied to experimental snapshots of a Rydberg atom array. Unsupervised clustering first identifies distinct regions (colored clusters), followed by supervised learning to extract relevant physical correlations. The red and green clusters correspond to the known checkerboard and star phases, while the blue, orange, and purple clusters could be associated with a fluctuating striated phase, a boundary-ordered phase, and a highly entangled nematic phase, respectively. Figure adapted from~\cite{Miles2023}.  
(b) Hybrid digital–analog variational quantum algorithm. Parametrized single-qubit rotations are followed by analog time evolution under a fixed Rydberg Hamiltonian. Data is encoded into the initial state, and the single-qubit gates are trained for optimized classification accuracy~\cite{Lu2024hybrid}.
(c) Quantum reservoir computing (QRC) framework utilizing quantum dynamics as a nonlinear transformation layer. Classical data are encoded into the system (left), and the measurement outputs after unitary evolution serve as features for a classical linear readout layer, enabling tasks such as classification and time-series prediction. Figure adapted from~\cite{Kornjaca2024}.}
\label{fig:ryd}
\end{centering}
\end{figure}

\textbf{Supervised learning.} CNNs have been trained on labeled snapshot data to distinguish between disordered and checkerboard-ordered phases in Rydberg atom arrays using standard classification routines~\cite{Carrasquilla2021Tutorial}. Other methods based on SVMs and random forest classifiers have been employed to identify certain symmetrized Rydberg base states in small atomic clusters~\cite{Chong2021}.

An approach to analyze criticality in the TFIM involves ``neural network scaling''~\cite{Maskara2022}: by systematically increasing the spatial extent of convolutional neural network filters, the algorithm extracts a characteristic classification length scale $\xi_{\text{net}}$, which physically is in analogy with the system's underlying correlation length. When applied to the one-dimensional TFIM, $\xi_{\text{net}}$ is found to diverge at the critical point with a power law, quantitatively reproducing the known critical exponent.

Another approach uses a hybrid quantum-classical supervised learning algorithm: ground states of the TFIM are first prepared and approximated using a variational quantum algorithm (VQA)\footnote{A VQA is an optimization protocol that uses a parameterized quantum circuit and a classical optimizer to minimize a cost function. For instance, when aiming to prepare the ground state of a given Hamiltonian, the variational energy is used as the cost function.}. A learnable unitary is then applied to the ground state, after which the system is measured and classified. This way, the full prepared quantum state is used and manipulated in a learnable way before measurement, instead of directly analyzing projected snapshot data from the ground state~\cite{Uvarov2020}.

From a more theoretical perspective, it has been shown that, for spin systems such as the TFIM, neural network indicators can be interpreted as lower bounds of the quantum Fisher information (QFI), a quantity that can signal phase transitions but is generally difficult to access~\cite{Arnold2023QFI}. 

A further perspective~\cite{Huang2022} shows that classical machine learning models trained on classical shadows (i.e., compact descriptions of quantum states obtained from randomized measurements) can predict ground-state properties and classify quantum phases. In the context of the Rydberg Hamiltonian, supervised models trained on classical shadows for specific parameter regimes were able to generalize and predict local observables in regions of the phase diagram that the model was not trained on~\cite{Huang2022}.

\textbf{Unsupervised learning.} A two-stage hybrid learning pipeline combining unsupervised and supervised methods was introduced to analyze the Rydberg Hamiltonian's phase diagram~\cite{Miles2023}. In this case, dimensionality reduction (PCA) and clustering ($k$-means followed by Gaussian mixture models) to Fourier features of experimental snapshots from a programmable Rydberg simulator was first applied. A subsequent supervised learning stage based on correlator-CNNs (see Sec.~\ref{sec:FH} and~\cite{Miles2021}) refined phase boundaries and provided interpretability of the various phases in terms of their characterizing real-space correlations. This revealed multiple phase regions, including previously unidentified boundary-ordered and rhombic phases, see Fig.~\ref{fig:ryd}~(a). 

Moreover, a combination of CNNs and prediction-based learning methods has been developed into an interpretable unsupervised tool (called TetrisCNN): By using multiple physically motivated kernel shapes in parallel (which map to certain spin-spin correlations), the network selectively activates relevant spin correlators; symbolic regression then provides symbolic formulas of the relevant correlations. Applied to the 1D TFIM (as well as the 2D Ising gauge theory), the framework correctly identifies known order parameters and transition points~\cite{Cybinski2024}.

Siamese neural networks have further been applied to the Rydberg Hamiltonian: These models take as input a pair of measurement outputs, which are projected into a an embedding space using the same neural network for both inputs. After projection, the similarity of the input pair is estimated, from which phase boundaries can be inferred~\cite{Patel2022}.

PCA has further been used to study and classify quantum transport phenomena---such as spin and energy transport---using snapshots sampled after quench dynamics~\cite{Bhakuni2024, Muzzi2024}. Focusing on models like the TFIM and kinetically constrained systems such as the PXP model (see e.g. \cite{Fendley2004}), it was shown that simple quantities derived from PCA grow in time with exponents that match the known dynamical transport exponents of the underlying systems. The dynamics of the studied systems can hence be captured with simple linear dimensionality reduction schemes, from which it was followed that the main driver behind quantum information transfer are conserved quantities. From a different perspective, autoencoders were used to analyze the local complexity of time-evolved quantum states in the TFIM, which has been argued to be useful to probing thermalization properties~\cite{Schmitt2022}. 

The success of nonlinear dimensionality reduction using diffusion maps—first explored in the context of detecting topological phase transitions (see Sec.~\ref{sec:topo})—has also been demonstrated in the study of Rydberg atom arrays~\cite{Lidiak2020}. Applied to a $\mathbb{Z}_3$ TFIM, the method reconstructed the full phase diagram in an automated way, capturing ordered, disordered, and more subtle incommensurate phases. This was further extended to identify valence bond solid phases in Majumdar–Ghosh chains as well as many-body localized phases~\cite{Lidiak2020}.

From a different perspective, quantum many-body spin systems have been analyzed using network theory. Projective measurement snapshots are mapped onto ``wave function networks'', where each configuration is a node, and the links correspond to the similarity between two such configurations. Applying this to experimental data from Rydberg arrays atoms, it was shown that the resulting networks transition from random to scale-free structures as the system crosses a quantum phase transition. This change reflects the buildup of long-range correlations and reduced complexity in the many-body wave function~\cite{MeSa2024}.

\textbf{Applications in quantum machine learning.} Rydberg atom arrays not only offer a tunable platform for quantum simulation but are also promising candidates for implementing quantum machine learning (QML) algorithms. Their inherent analog dynamics, scalability, and control capabilities to implement entangling gates make them particularly suited for testing a variety of algorithms. 

On the one hand, a hybrid digital–analog quantum learning framework tailored to the Rydberg platform has been proposed~\cite{Lu2024hybrid}. The scheme implements a variational quantum algorithm that alternates between digital single-qubit gates and analog time evolution under the native Rydberg Hamiltonian, see Fig.~\ref{fig:ryd}~(b). The method was benchmarked on two tasks: MNIST digit classification (classical image data) and unsupervised phase boundary detection via anomaly detection (quantum many-body snapshot data). In the former case, image data is encoded into the circuit's initial state. After executing the variational circuit, one of the qubits is measured, and its probability distribution used to classify the data. In the case of quantum phase detection, ground state wave functions of some many-body Hamiltonian are used as the direct input, which is then manipulated according to the variational gate sequence and classified. In both cases, the Hamiltonian parameters of the Rydberg atom array (i.p. the lattice parameter governed by the blockade radius) are used as hyperparameters of the algorithm---i.e., they are not changed during training. These digital–analog learning circuits were argued to outperform their purely digital counterparts in terms of both noise robustness and needed circuit depth. 

A related approach is based on quantum reservoir computing (QRC). Here, fully analog quantum dynamics act as a fixed ``reservoir'' that processes and efficiently separates input data~\cite{Fujii2017, Fujii2021}. A large-scale experimental realization of QRC on a neutral-atom analog quantum computer has been presented in~\cite{Kornjaca2024}: Classical input features are encoded into parameters of the Rydberg Hamiltonian---via global pulses, local detunings, or atomic positions---followed by analog quantum evolution and projective measurements. Output observables then correspond to embeddings of the data, which in turn are used as inputs to standard classical classifiers, see Fig.~\ref{fig:ryd}~(c). Applied to a range of tasks including timeseries prediction and image classification~\cite{Kornjaca2024} as well as molecular property prediction~\cite{Beaulieu2025}, the QRC approach showed competitive results to fully classical methods.

A quantum recurrent neural network (qRNN) model implemented with Rydberg atom arrays has also been proposed~\cite{Bravo2022}: the qRNN treats interacting Rydberg atoms as a quantum analog of classical neurons. The dynamics of the system naturally encode memory and decision-making capabilities, enabling the qRNN to perform cognitive tasks such as multitasking, long-term memory, and decision-making.

In parallel to these analog approaches, quantum convolutional neural networks (QCNNs) have been proposed as compact, circuit-based models inspired by classical CNNs~\cite{Cong2019}. There, multi-qubit gates are used in place of classical convolutions, enabling classification of quantum phases (including topological models) and quantum error correction protocols.

\subsection{Fermi-Hubbard systems}
\label{sec:FH}
The Fermi-Hubbard model is a paradigmatic model of strongly correlated fermions and is widely believed to capture the key physics of high-temperature superconductors, whose various exotic phases still lack a complete microscopic understanding~\cite{Lee2006}. It describes fermions hopping on a lattice with on-site interactions, given by the Hamiltonian
\begin{equation}
    \hat{\mathcal{H}} = -t \sum_{\langle \mathbf{i}, \mathbf{j} \rangle, \sigma} \left ( \hat{c}_{\mathbf{i}, \sigma}^{\dagger} \hat{c}_{\mathbf{j}, \sigma} + \text{H.c.} \right) + U \sum_{\mathbf{i}} \hat{n}_{\mathbf{i}, \uparrow} \hat{n}_{\mathbf{i}, \downarrow},
    \label{eq:FH}
\end{equation}
where $\hat{c}_{\mathbf{i}, \sigma}^{(\dagger)}$ annihilates (creates) a fermion with spin $\sigma = \uparrow, \downarrow$ on site $\mathbf{i}$, and $\langle \mathbf{i}, \mathbf{j} \rangle$ denotes nearest-neighbor pairs on a lattice, typically a square lattice. Tunable parameters in the FH model are the particle doping (in most situations, the hole doping $\delta$ away from one particle per site is tuned\footnote{We did not include a chemical potential in Eq.~\eqref{eq:FH}, and assume working in a fixed particle number sector set by $\delta$.}), as well as the interaction strength $U/t$ and temperature $T/t$.

In the context of cold atoms, the FH model can be realized with high precision and level of control by ultracold fermionic atoms in optical lattices~\cite{Bloch2008, Esslinger2010, Bloch2012, Cheuk2015, Gross2017, Bohrdt2020}. Combined with quantum gas microscopes that yield site-resolved images of atoms~\cite{Parsons2015, Haller2015}, these systems can take rich snapshot data for a broad range of doping levels, for system sizes that are typically out of reach with state-of-the-art classical simulation methods. Analyzing snapshots with ML aims to identify the relevant physics in regimes where the microscopic physics is particularly challenging to pin down.

\begin{figure}[t!]
\begin{centering}
\includegraphics[width=\textwidth]{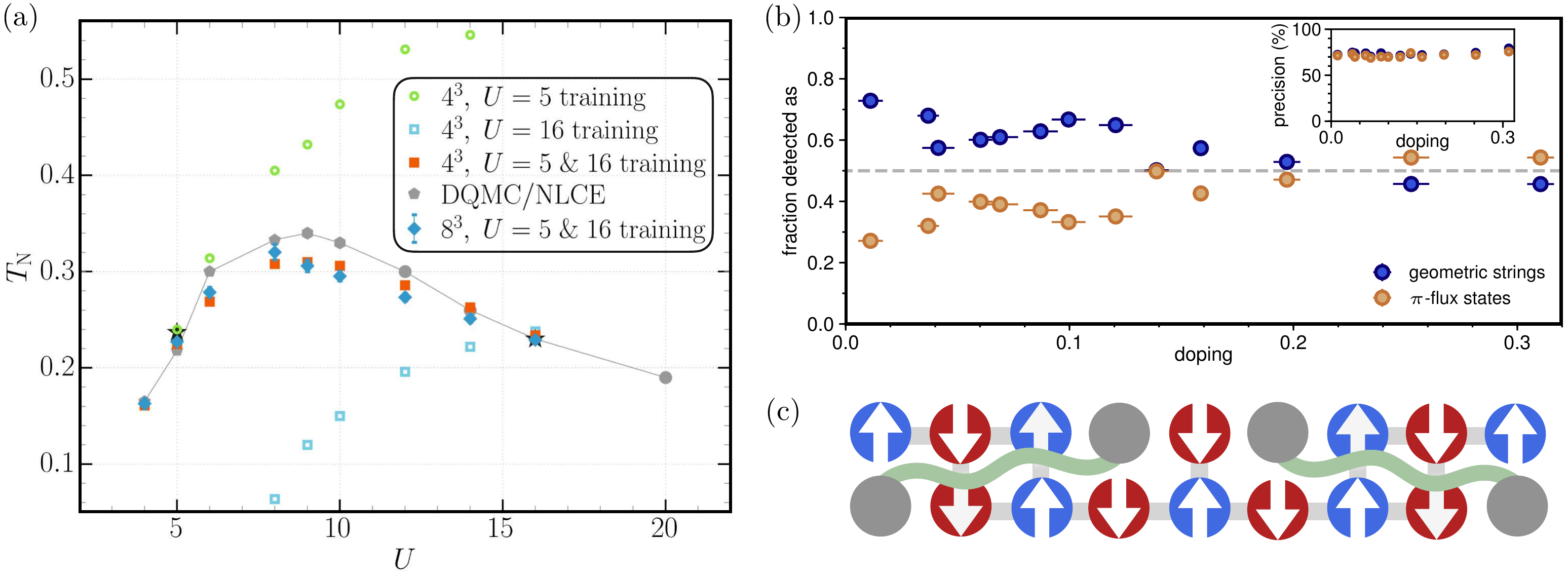}
\caption{\textbf{Applications of machine learning in Fermi-Hubbard models.} 
(a) Learning magnetic phase transitions in the 3D Fermi-Hubbard model at half filling. A 3D CNN is trained on auxiliary spin configurations to predict the N\'eel temperature $T_N$. Different colors and markers indicate training on individual datasets at $U/t = 5$ and $U/t = 16$, as well as joint training on both. In the latter case, the predicted $T_N$ closely matches results from quantum Monte Carlo simulations. Figure adapted from~\cite{Chng2017}. 
(b) Using CNNs to compare experimental snapshots of the doped 2D Fermi-Hubbard model to two candidate theoretical frameworks: the geometric string theory and the $\pi$-flux (doped quantum spin liquid) state. At low to moderate doping, geometric strings provide a better match to the experimental snapshot distribution. Figure adapted from~\cite{Bohrdt2019}. 
(c) Reconstructing effective Hamiltonians using machine learning. Mobile holes in an AFM background displace spins along their path (green lines), introducing frustration. When constrained to move only along one dimension, the resulting effect can be quantitatively captured using gradient descent methods, yielding an effective frustrated $J_1$-$J_2$ model of the background spins. Figure adapted from~\cite{Schloemer2023rec}.}
\label{fig:FH}
\end{centering}
\end{figure}

\textbf{Supervised learning.} In an early demonstration of applying deep learning to FH systems, auxiliary-field configurations generated via determinant quantum Monte Carlo simulations of the 3D Hubbard model were used in combination with CNNs~\cite{Chng2017}. In particular, these configurations (which encode both spatial and imaginary time information) were used to train a neural network to classify snapshots into paramagnetic or antiferromagnetic phases at one particle per site. The model succeeded in reproducing the magnetic phase diagram, in particular identifying the Néel temperature where magnetic ordering sets in, see Fig.~\ref{fig:FH}~(a). Going further, transfer learning enabled the same network---trained only at half-filling---to generalize to doped systems, suggesting that antiferromagnetic order persists up to at least 5\% hole doping~\cite{Chng2017}. In related supervised approaches, CNNs have been employed to detect phase transitions in Hubbard-type models using Green’s functions as input features, identifying boundaries between metallic and charge-ordered phases~\cite{Broecker2017_green}.

CNNs have further been trained to distinguish between high-temperature and the lowest achievable temperature configurations of experimental cold atom snapshots for a range of dopings~\cite{Khatami2020, Striegel2023}. Near one particle per site, where the system forms a Mott insulator with extended AFM spin correlations, CNN filters indeed revealed sensitivity to AFM patterns. However, when going to more challenging regimes (such as non-Fermi-liquid phases expected to emerge when increasing the doping), the network's performance decreases, as in such cases spatial patterns that emerge at low temperatures are much more subtle. The nonlinearity of CNNs (introduced through nonlinear activation functions following the application of learnable filters), while important for classification, further complicates the interpretability of the learned features in these challenging regimes.

From a different perspective, neural networks have been used to identify which theoretical model best describes experimental data from quantum gas microscopes in the doped 2D FH model~\cite{Bohrdt2019}. Specifically, CNNs were trained to distinguish between experimental snapshots and simulated images generated from two competing theories: one based on geometric string theory with hidden spin order, and another describing a doped quantum spin liquid. When applied to experimental snapshots, the trained networks consistently classified them as “string-like” rather than spin-liquid-like up to intermediate doping levels, as shown in Fig.~\ref{fig:FH}~(b). This result supports the interpretation that the small-to-intermediate doped experimental system features hidden AFM correlations consistent with the geometric string picture~\cite{Chiu2019}.

Building on these results, an interpretable neural network architecture, called correlator-CNNs (CCNNs), has been introduced to address the challenge of understanding the neural network's decision making process~\cite{Miles2021}. In contrast to standard black-box networks, CCNNs are designed such that their nonlinear layers explicitly compute $N$-point correlators. When applied to FH model snapshots, this architecture ensures that the network learns and utilizes physically meaningful correlators (such as spin–spin, density–density, or spin–density correlations) when trained for classification tasks. When comparing snapshots generated from two competing theories (geometric string theory and spin-liquid-like theories), the CCNN identified fourth-order correlations as the most significant features for distinguishing between the two data sets, underlining that subtle physical structural differences are expected to play a key role in the intricate phases of the doped FH model.

In another step toward interpretable machine learning in quantum many-body systems, influence functions were used to analyze the internal decision making of a CNN trained to classify phases of the extended 1D spinless FH model~\cite{Dawid2020}. The core idea of the method is to quantify how much each training example influences the prediction for a given test input, which can in turn be used as an indicator whether the network has learned physically meaningful features. A broader set of diagnostic tools based on the Hessian of the loss function was later introduced~\cite{Dawid2022}, generalizing the influence function approach into a more versatile framework for interpretability and reliability assessment when training neural networks.

\textbf{Unsupervised learning.} Unsupervised techniques such as t-SNE and convolutional autoencoders combined with random forest embedding have been applied to Hubbard auxiliary-field configurations~\cite{Chng2018}. Among these, t-SNE was found to perform particularly well in capturing the magnetic phase transition in three dimensions. After applying clustering algorithms to the dimensionally reduced data, certain features of the resulting clusters were shown to closely track physical observables of the underlying model, such as the antiferromagnetic structure factor. Similar patterns emerged in the 2D half-filled FH model, where, despite the absence of a finite-temperature phase transition due the continuous SU(2) symmetry, magnetic correlations begin to develop below characteristic temperature scales. Along similar lines, the structural complexity of FH snapshots (which can be calculated with a series of coarse-graining steps of a given image) was shown to behave similarly as the entropy per site in the system~\cite{Eduardo2024}, possibly facilitating a simple way to estimate the entropy in ultracold atom simulators.

In the strongly interacting half-filled 2D FH model, both the magnetic susceptibility and specific heat show characteristic peaks at temperatures where magnetic correlations become significantly long-range. Using the learning-by-confusion framework combined with interpretable architectures, neural networks were able to detect these subtle thermodynamic signatures~\cite{Schloemer2023fluc}. Notably, these features arise from non-local, long-range properties of the system. Although the convolutional architectures that were used are inherently limited to capturing local correlations, it was shown that analyzing the full counting statistics of many-body snapshots can provide valuable insights into such non-local properties, and hence capture qualitative changes of thermodynamic quantities. 

In the FH model on honeycomb and Lieb lattices, the absence of perfect nesting leads to metal-insulator transitions at a finite critical interaction strength. Here, it was shown that PCA can capture signatures of these transitions~\cite{Costa2017}. In contrast, the 2D attractive Hubbard model exhibits a BKT transition that is less sharply captured~\cite{Costa2017}, as also discussed in Sec.~\ref{sec:topo}. 

By using certain topological information of snapshots, it has further been demonstrated that quantum critical points can be detected in Hubbard-type models from auxiliary-field configurations~\cite{Tirelli2021}. 

Beyond the Fermi-Hubbard model, Falicov-Kimball models~\cite{Falicov1969}, which are relevant in the context of cold atomic Fermi mixtures~\cite{Maska2008}, have been explored using machine learning. In one approach, certain average quantities were identified in prediction-based methods that are essential for accurate classification~\cite{Arnold2021}. Sudden changes in these mean features can then directly serve as reliable indicators of phase boundaries. Next to being computationally much cheaper than full predictive models, this approach further yields interpretable classification. 

\subsection{Bose-Hubbard systems}
\label{sec:BH}
Ultracold bosons in optical lattices, described by the Bose-Hubbard (BH) model and its extensions, provide yet another platform to explore quantum phenomena such as the superfluid–Mott insulator (SF-MI) transition, as well as density wave, supersolid and topologically nontrivial phases~\cite{Chanda2025}. 

The BH model describes interacting bosons hopping on a lattice and is given by the Hamiltonian
\begin{equation}
    \hat{H}_\text{BH} = -t \sum_{\langle \i,\j \rangle} \hat{b}_{\i}^\dagger \hat{b}_{\j} + \frac{U}{2} \sum_{\i} \hat{n}_{\i} (\hat{n}_{\i} - 1) - \mu \sum_{\i} \hat{n}_{\i},
\end{equation}
where $\hat{b}_{\i}^\dagger$ ($\hat{b}_{\i}$) are bosonic creation (annihilation) operators at site $\i$, $\hat{n}_{\i} = \hat{b}_{\i}^\dagger \hat{b}_{\i}$ is the number operator, $t$ is the tunneling amplitude between neighboring sites, $U$ is the on-site interaction strength, and $\mu$ is the chemical potential. The competition between kinetic energy (set by $t$) and interaction energy (set by $U$), together with the filling controlled by $\mu$, gives rise to a rich phase diagram. The two main control parameters are the ratios $U/t$ and $\mu/t$, which drive the transition from the superfluid phase (dominant tunneling, $U/t \ll 1$) to the Mott insulating phase (dominant interactions, $U/t \gg 1$).

\begin{figure}[t!]
\begin{centering}
\includegraphics[width=\textwidth]{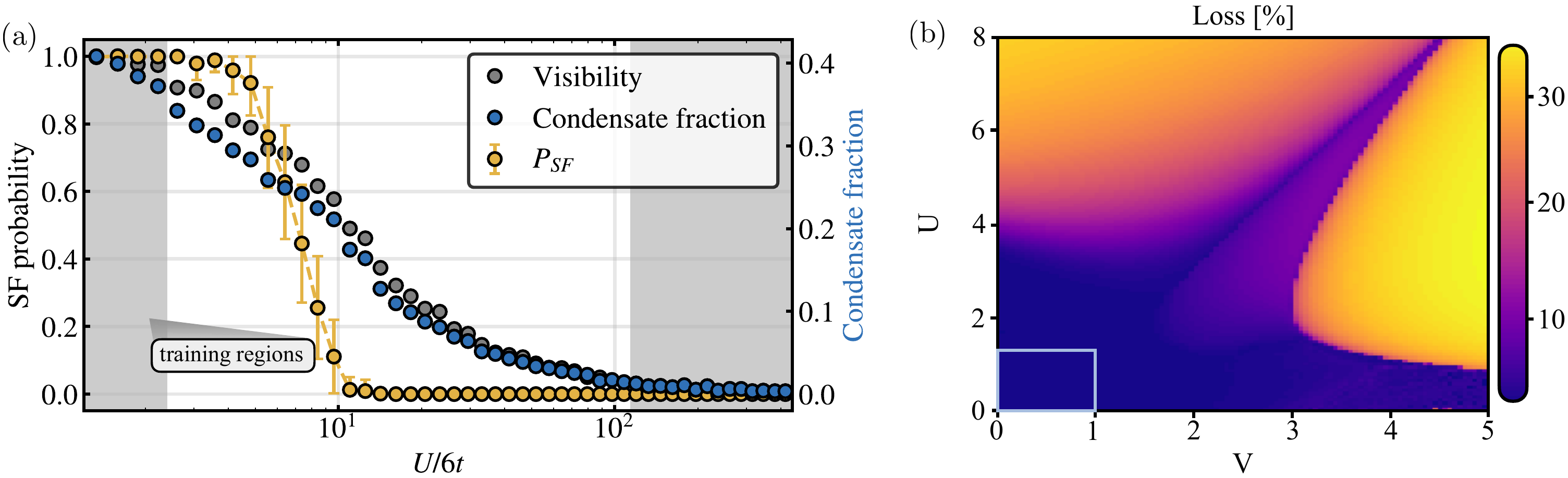}
\caption{\textbf{Applications of machine learning in Bose-Hubbard models.} 
(a) Identifying the Mott insulator–superfluid transition in an ultracold atom experiment on a triangular lattice with harmonic confinement. A neural network is trained on data deep within the superfluid phase (gray area), and its predicted probability $P_{\mathrm{SF}}$ of being in the superfluid phase shows a sharp drop in the transition region (yellow data points). Due to the trap-induced inhomogeneity, multiple local densities coexist; conventional global observables such as the condensate fraction or the visibility of interference peaks provide no clear signature of the transition. Data taken from~\cite{Rem2019}. 
(b) Unsupervised anomaly detection in the extended Bose-Hubbard model with on-site ($U$) and nearest-neighbor ($V$) interactions. An autoencoder trained within the superfluid phase (blue square near the origin) shows increased reconstruction loss in other regions of parameter space, indicating phase transitions.  Data taken from~\cite{Kottmann2020}.}
\label{fig:BH}
\end{centering}
\end{figure}

\textbf{Supervised learning.} Machine learning methods have been applied to identify and classify these phases based on experimental and simulated data, using both real-space and momentum-space measurements. One prominent example is the application of supervised learning to experimental momentum-space images of a bosonic gas realizing the BH model in an optical lattice with harmonic confinement~\cite{Rem2019}. A CNN trained on labeled snapshot images deep in the SF and MI regimes has been shown to predict the quantum phase transition with high accuracy, as shown in Fig.~\ref{fig:BH}~(a). The network was shown to learn subtle features of the momentum distribution, outperforming traditional quantities such as the condensate fraction, especially in the presence of the harmonic trap where global observables wash out the underlying many-body physics due to a locally varying chemical potential.

CNNs have also been used to analyze real-space occupation data~\cite{Huembeli2018}: When trained on a single cut through the phase diagram, the network was shown to reproduce the Mott lobes of the BH model. 

In the context of disorder-induced phenomena, machine learning has also been used to investigate many-body localization (MBL). A CNN trained on BH configurations from two limiting cases---fully thermalized and fully localized states---was able to interpolate its predictions across intermediate disorder strengths~\cite{Bohrdt2021}. Applied to experimental data from a quantum gas microscope, the network produced a sharply defined crossover consistent with previously observed MBL behavior. Notably, whereas conventional indicators vary gradually across the transition, the CNN learned to detect higher-order spatial correlations that can serve as fingerprints of localization. In a related context, it was shown that a feed-forward NN can detect the transition from a metallic to an Anderson localized phase in the (fermionic) Aubry-And\'e model~\cite{Carrasquilla2017}. 

\textbf{Unsupervised learning.} Anomaly detection techniques based on autoencoders have proven to be powerful tools for identifying phase transitions without requiring labeled data. Within this framework, an autoencoder is trained to compress and reconstruct data from a known reference phase. When giving the network data from other regions in the phase diagram, it may either succeed or fail in compressing and reconstructing the data. In the former case, it is inferred that the data belongs to the same phase as the network was trained in. In the latter case, the system is likely in a different phase with qualitatively distinct features~\cite{Kottmann2020}. Applied to an extended 1D BH model with long-range interactions, this method not only recovered known transitions but also uncovered a phase-separated region between the superfluid and supersolid phases, see Fig.~\ref{fig:BH}~(b). 

However, generally speaking, capturing features that are related to phase coherence presents a challenge when using single-site-resolved density snapshots. To name one example, in dipolar BH models, PCA successfully distinguished superfluid and density wave phases by extracting dominant patterns in occupation imbalance~\cite{Rosson2020}. However, the method failed when attempting to identify the supersolid phase---characterized by coexisting density modulation and global phase coherence. To this end, global basis rotation techniques offer a promising route by mapping off-diagonal coherence observables into diagonal ones, making them directly accessible through projective measurements, similar to what has been proposed for fermionic systems~\cite{Schloemer2024}. These transformed observables can then be processed using machine learning-based classification methods together with standard Fock measurements, possibly enabling the detection of phases characterized by a combination of off-diagonal and diagonal correlations.

Another unsupervised strategy applied to data of the BH model is inspired by the learn-by-confusion paradigm, which reconstructed the phase diagram of bosonic lattice systems based on shifts in the network's internal confidence~\cite{Broecker2017}.
Lastly, fully automated generalizations (Discriminative Cooperative Networks) have further successfully labeled the 2D parameter space of the BH model~\cite{Liu2018}, see also Sec.~\ref{sec:cl}.

\subsection{Hamiltonian learning} 
\label{sec:HL}

Another direction in the application of machine learning to quantum simulation is Hamiltonian reconstruction~\cite{DiFranco2009,Zhang2014,Qi2019,Bairey2019,Cao2020, Anshu2021}, where the goal is to learn the microscopic coupling parameters of an underlying Hamiltonian directly from many-body snapshot measurements. In the context of analog quantum simulation, this offers reliable strategies to verify the Hamiltonians implemented by quantum devices~\cite{Carrasco2021}. For instance, in Rydberg atom arrays, it was shown that graph neural networks can accurately reconstruct the underlying Hamiltonian parameters of large-scale systems, whereby the training is only done within small clusters~\cite{Simard2025}. Moreover, Hamiltonian learning has been generalized to dissipative systems, where next to the Hamiltonian content, the Lindblad operators of the Liouvillian can be reconstructed~\cite{Olsacher2025}.

Next to helping in the calibration and certification of quantum simulators, Hamiltonian reconstruction can also provide valuable physical insights. In the context of the FH model, Hamiltonian reconstruction schemes were used to quantify the impact of mobile dopants on an underlying antiferromagnetic background~\cite{Schloemer2023rec}, schematically illustrated in Fig.~\ref{fig:FH}~(c). By directly accessing highly non-local correlation information from many-body snapshots, it was shown that mobile holes drive the spin environment into a strongly frustrated regime---potentially facilitating the emergence of quantum spin liquids in certain regions of the phase diagram. 

Hamiltonian learning schemes have also been developed to explicitly learn entanglement Hamiltonians of strongly correlated systems~\cite{Kokail2021}, which carry information about correlations and quantum entanglement in the system. Here, instead of calculating the entanglement Hamiltonian of a given subsystem classically, the quantum simulator itself is used by locally deforming the Hamiltonian (i.e. locally changing its parameters). Analog quantum dynamics under this modified Hamiltonian combined with classical optimization loops then allow for the reconstruction of the entanglement Hamiltonian for a given subsystem. In a 51-qubit trapped ion quantum simulator, the entanglement Hamiltonian of subsystems was learned experimentally, providing evidence of certain quantum field theoretical predictions and giving insights into the entanglement structure of ground and excited states~\cite{Joshi2023}.

\subsection{Quantum state tomography}
\label{sec:QST}

Thus far, we have mainly focused on extracting relevant physical information using machine learning techniques directly from many-body snapshots in various many-body systems. An alternative approach to analyzing the output of quantum simulators is to reconstruct the full underlying quantum state---either pure or mixed---based on measurement data, which can then be used to extract useful information in a second step.

Neural quantum states (NQS) are a class of variational wavefunction representations that use artificial neural networks to model wave functions. Originally introduced in~\cite{Carleo2017}, NQS aim to represent the exponentially large Hilbert space of many-body quantum systems using a comparatively small number of trainable parameters. This is achieved by leveraging the ability of neural networks to capture patterns, correlations and interdependencies of the corresponding wave function. Since their introduction, NQS have been employed across a broad range of applications, including ground state calculations, time evolution, and excited states---for a dedicated review, we refer to~\cite{Lange2024Rev}.

Reconstructing the full quantum state from projective measurement data is known as ``quantum state tomography''. However, using traditional approaches (such as  linear inversion and maximum likelihood estimation~\cite{Hradil1997}), this quickly becomes infeasible as the number of particles increases, due to an exponential scaling of measurement requirements and the computational cost of storing the quantum state. Using NQS offers an efficient alternative; common architectures are restricted Boltzmann machines (RBMs), recurrent neural networks (RNNs) or other generative models that learn the probability distribution of measurement outcomes. In this section, we briefly review how state tomography through NQS can be used as tools for extracting information from quantum simulation experiments. 

\begin{figure}[t!]
\begin{centering}
\includegraphics[width=0.92\textwidth]{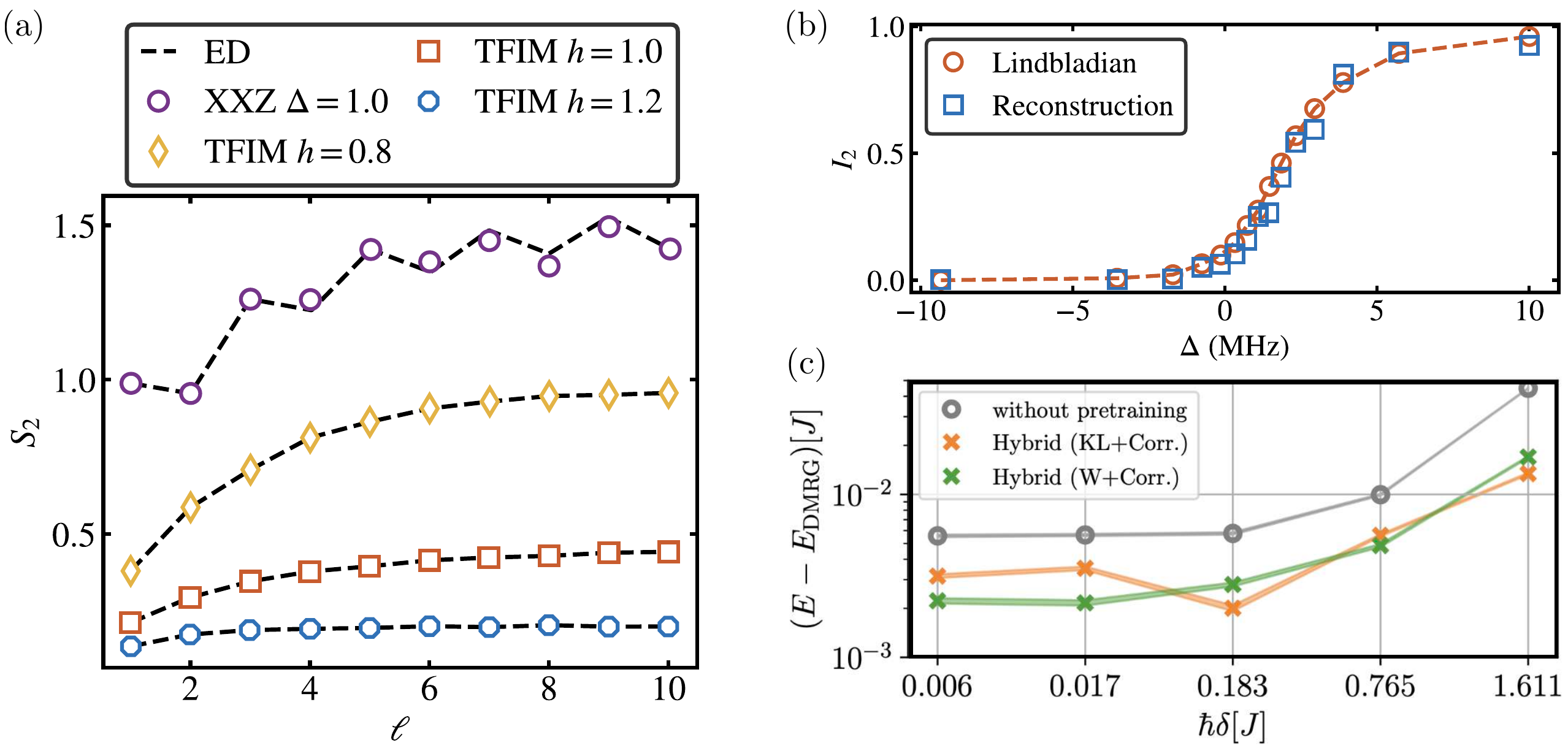}
\caption{\textbf{Quantum State Reconstruction.} 
(a) Reconstruction of entangled spin states in the TFIM and XXZ models. The second Rényi entropy ($S_2$, data points) extracted from the reconstructed states matches exact diagonalization results (dashed lines). Data taken from~\cite{Torlai2018}. 
(b) Reconstruction of the many-body wave function of a Rydberg atom array quantum simulator using RBMs in the presence of noise. The mutual information $I_2$ (defined via the second Rényi entropy) of the reconstruction (blue squares) agrees well with theoretical predictions based on Lindblad master equations (orange circles). Data taken from~\cite{Torlai2019}. 
(c) Ground-state search for the two-dimensional dipolar XY model under a staggered magnetic field of strength $\hbar \delta$ using neural quantum states. Pretraining with experimental snapshots (crosses) allows to reach significantly lower energies compared to standard variational Monte Carlo within a fixed number of training steps. Figure adapted from~\cite{Lange2025hybrid}.}
\label{fig:QST}
\end{centering}
\end{figure}

In a seminal work~\cite{Torlai2016}, the ability of RBMs to capture thermodynamic observables of classical spin models from Monte Carlo samples was demonstrated. By training on sampled spin configurations, the network could reproduce quantities such as energy and magnetization, even around criticality. Building on these insights from classical systems, it was then demonstrated that an RBM can also reconstruct entangled many-body quantum states from projective measurements~\cite{Torlai2018}. To this end, the RBM is trained to model the distribution underlying the snapshot measurements. After training, it was shown that local as well as non-local correlations could be accurately reproduced by the network for a variety of states. One particular advantage of reconstructing the wave function is that, next to correlation functions, quantities such as the entanglement entropy can further be extracted---which otherwise requires involved and tailored measurement protocols~\cite{Brydges2019, Elben2020, Kokail2021Ent}. 

Fig.~\ref{fig:QST}~(a) shows this reconstruction for ground states of the TFIM as well as the XXZ model~\cite{Torlai2018}. Building on this idea, later works studied various architectures and training procedures, such as CNNs~\cite{Schmale2022}, variational autoencoders~\cite{Rocchetto2018} and transformers~\cite{Cha2021}, as well as explicitly incorporating symmetry constraints~\cite{Morawetz2021}. 

Applying state tomography directly to experimental data, an RBM was trained on measurements from a Rydberg atom array, while taking into account experimental noise through a tailored noise layer~\cite{Torlai2019}. The network was able to reconstruct both local observables and quantities like the mutual information from measurements, shown in Fig.~\ref{fig:QST}~(b). Additional applications to experimental data include two-qubit photonic systems~\cite{Neugebauer2020}: There, it was shown that enforcing the positivity of the reconstructed density matrix improves reconstruction accuracy, though it increases computational costs. Simplifying assumptions (e.g., assuming pure states as in~\cite{Torlai2018}) make training more straight-forward, but can lead to biased results and inaccurate modeling. 

In order to optimize reconstruction efficiency from measurement data, adaptive measurement strategies have been proposed~\cite{Lange2023}: There, the current NQS estimate of the state is used to guide the choice of future measurements (such as the measurement basis), selecting those that are expected to lead to the highest information gain. Indeed, it was shown that this active learning approach can reduce the overall number of measurements needed while maintaining high reconstruction accuracy.

Neural networks have also been used to enhance observable estimation without full state reconstruction. For instance, it was shown that a neural network trained on a small number of single-shot images from a quantum gas microscope could accurately predict one- and two-body observables~\cite{Lode2021}. Moreover, the model could infer momentum-space distributions from real-space images, which was argued to replace the need for separate time-of-flight measurements. From a different perspective, transformer-based models have been developed specifically for Rydberg atoms arrays (named RydbergGPT), which take as input interacting Hamiltonians and directly output qubit measurement probabilities associated with the corresponding Rydberg Hamiltonian~\cite{Fitzek2024}. 

Instead of reconstructing many-body states from measurement data, neural network quantum states can also be used as purely variational \textit{ans\"atze} to find variationally optimized ground or thermal states of a target Hamiltonian. To this end, hybrid data- and Hamiltonian-driven approaches have been developed that combine experimental measurements with variational optimization. For instance, a data-enhanced variational Monte Carlo method has been introduced~\cite{Czischek2022}, where an RNN is pretrained on numerically simulated projective measurements from a Rydberg atom array and then fine-tuned via variational optimization, yielding faster convergence in ground state reconstruction compared to mere variational optimization. Building on this idea, the same concept was applied to experimental data from a 16$\times$16 Rydberg array, underlining that explicitly using snapshot information of the targeted systems can guide and improve variational Monte Carlo simulations of strongly correlated phases of matter~\cite{Moss2024}. Along similar lines, a hybrid training scheme for transformer quantum states that incorporates both measurement snapshots and direct information of expectation values from multiple bases (such as spin-spin correlations) has been proposed, leading to robust and efficient ground state learning on large 2D spin systems using data from programmable quantum simulators~\cite{Lange2025hybrid}, Fig.~\ref{fig:QST}~(c). 

Furthermore, a hybrid quantum-neural algorithm for VQA optimization was introduced in~\cite{Zhang2022VQE}. There, the main idea is that after applying the VQA unitary, resulting bitstrings after measurement are post-processed by a neural network, which enables a more efficient way of evaluating the variational energy $\braket{\hat{H}}$ and hence speeds up optimization. Finally, neural error mitigation has been proposed to improve variational algorithms on near-term devices~\cite{Bennewitz2022}. After applying and optimizing the variational gate sequence on a noisy quantum device, the resulting quantum state is reconstructed using NQS. In a second step, this reconstructed state is then used to variationally optimize under the target Hamiltonian to improve the variational result.

%=================================EXPERIMENTAL ASSISTANCE==============================================
\section{Experimental assistance}
\label{sec:ExpAss}

Besides supporting the analysis of experimental data and extraction of relevant physical quantities, ML techniques can further be used to help run and improve quantum gas experiments themselves. To this end, ML algorithms can learn patterns from data and use that knowledge to make better decisions in future experiments. Typically, this involves trying different settings, acquiring data, and gradually improving performance by navigating through a high-dimensional optimization landscape spanned by the experiment's tuning parameters. In this section, we review how ML supports two essential aspects of the quantum gas experiment pipeline: state preparation and imaging.

\subsection{State preparation}
The preparation of specific quantum states---like Bose-Einstein condensate or a strongly correlated lattice state---requires carefully timed sequences of a high dimensional manifold of control parameters, with the goal of efficiently cooling and trapping atoms in optical potentials. Optimizing these sequences manually is a significant challenge, and may not lead to the best possible results in terms of, e.g., state fidelity or particle density. ML methods can help by efficiently exploring the large space of possible settings and identifying effective sequences. 

\textbf{Bayesian optimization.} A widely used machine learning approach for experimental control is Bayesian optimization (BO). BO works by constructing a model of a specified cost function, typically using a Gaussian process or neural network. The model then estimates the expected performance for any given set of control parameters, and aims to optimize the next experimental run. Over successive iterations, the model is updated with new data, progressively improving performance.

\begin{figure}[t!]
\begin{centering}
\includegraphics[width=0.85\textwidth]{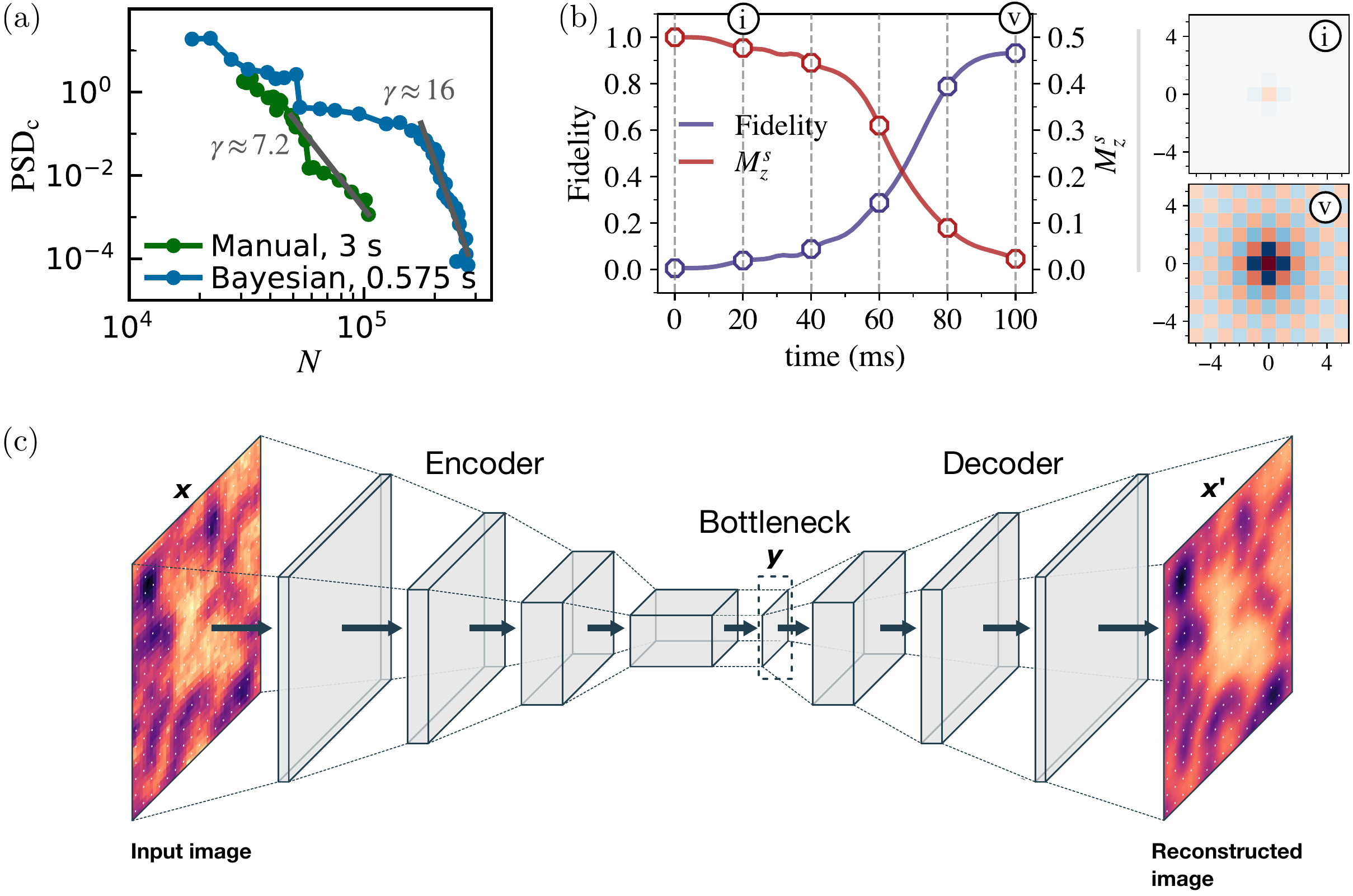}
\caption{\textbf{Machine learning for experimental assistance.} 
(a) Classical phase-space density (PSD$_\text{c}$) versus atom number for a manually tuned 3 s protocol and a 575 ms protocol optimized via Bayesian optimization. The BO-optimized protocol enables an efficient production of BECs with significantly higher PSDs and atom numbers. Figure adapted from~\cite{Vandeiro2022}. 
(b) Optimization of state preparation in the AFM Heisenberg model using BO over a fixed 100 ms window. Both AFM exchange couplings and applied staggered magnetic field are optimized. Left: fidelity of the resulting 2D AFM state as a function of time. Right: buildup of long-range connected AFM correlations, which are absent in the initial product state. Data taken from~\cite{Xie2022}. 
(c) Improved imaging fidelity via neural networks. CCD images are processed by an autoencoder that compresses input to minimize reconstruction error. The bottleneck output corresponds to a high-fidelity binary map of local occupations. Figure adapted from~\cite{Impertro2023}.}
\label{fig:EA}
\end{centering}
\end{figure}

BO has been applied to control up to 55 parameters in a Bose-Einstein condensation (BEC) experiment, which enabled the production of a rubidium condensate in 575\,ms---significantly faster than any previously reported protocols using standard alkali-metal atoms~\cite{Vandeiro2022}, shown in Fig.~\ref{fig:EA}~(a). The optimization algorithm identified sequences that combined Raman cooling with evaporation stages, and achieved significantly higher phase space densities compared to manually optimized sequences. The learned control strategy could be physically interpreted, showing how the optimization process utilizes subtle mechanisms for enhanced cooling.

In a similar spirit, optimization of thulium BEC production revealed a previously unrecognized bottleneck caused by three-body losses. By interpreting this constraint and adjusting the magnetic field to shift the scattering properties, the final atom number could be improved~\cite{Kumpilov2024}. BO has also been applied to optimize purely evaporative cooling~\cite{Wigley2016} as well as combined laser and evaporation sequences~\cite{Barker2020}, where it consistently found parameter sets that increased densities by notable factors compared to manual tuning. In situations with large shot noise, BO has been shown to succeed in situations when only few measurement shots are available~\cite{sauvage2020optimal}.

BO can also help in the preparation of strongly correlated many-body states. In 1D and 2D Heisenberg antiferromagnets, BO has identified high-fidelity control protocols achieving over 96\% fidelity in systems with up to 80 spins. In experiments, AFM states are often initialized from a Néel-ordered product state and built up dynamically by tuning interaction and field parameters. However, nonadiabatic effects during this ramp limit the final state fidelity. Using BO, it was demonstrated that the model efficiently learns optimized control paths that reduce these effects and, consequently, improve AFM ordering~\cite{Xie2022} of the final state, as shown in Fig.~\ref{fig:EA}~(b). 

Further applications include the preparation of fractional quantum Hall (FQH) states. By optimizing the control ramps with a model trained on numerical simulations and explicitly incorporating realistic disorder, a protocol that was notably faster and more robust than manual tuning was designed~\cite{Blatz2024}.

\textbf{Reinforcement learning.} Experimental control can also be framed as a sequential decision-making problem, where the experimental cycle is divided into discrete time steps. At each step, the system's state is observed, and a control decision is made based on that observation. An ``agent'' learns through trial and error, and is trained by giving it rewards for successful outcomes. This training scheme is referred to as reinforcement learning (RL). In RL, the objective is not optimized explicitly. Instead, the agent improves its behavior over time by using feedback to adjust its decisions. In the context of cold atom experiments, this reward usually reflects concrete experimental goals, such as the fidelity of a prepared quantum state or the entropy of a final measurement outcome.

For instance, RL has been used to control the cooling of atoms in a magneto-optical trap in real time. Taking advantage of a live cooling fluorescence image stream, the algorithm adjusted laser detuning and magnetic fields during the preparation process, resulting in more efficient loading of atoms~\cite{Reinschmidt2024}. In a different experiment, by controlling 30 experimental parameters simultaneously, an RL agent learned to produce BECs robustly~\cite{Milson2023}. Similar results have been reported for the cooling of interacting degenerate Fermi gases of $^6$Li, showing more than doubling in atomic density and uncovering non-trivial cooling strategies that balance evaporation with efficient thermalization~\cite{Min2025}. Additional applications of RL include efficient stirring of a superfluid~\cite{Simjanovski2023}, where reinforcement learning was used to optimize stirring protocols in a Bose-Einstein condensate.

RL has further been used to optimize protocols for preparing target quantum states in a gate-based fashion~\cite{Bukov2018}. An agent is trained to find driving protocols that guide the system from a trivial initial state to a desired target state. The space of control protocols was shown to have a physical analogy: for short preparation durations, optimal control is impossible; for long durations, many high-fidelity solutions exist; and in between lies a glassy phase, where finding the best protocol is exponentially hard due to a rugged optimization landscape with many local minima. Nevertheless, even in this latter regime, the RL agent was able to find efficient schemes, indeed outperforming traditional gradient-based methods in some cases.

A more exploratory use of ML was proposed when training an agent to autonomously design new quantum experiments. Using a reinforcement learning agent, the system interacts with a simulated optical table, sequentially placing optical elements to build experiments. Without prior knowledge of quantum optics, the agent was shown to create and optimize experimental setups~\cite{Krenn2016, Melnikov2018}.

\subsection{Imaging}

Precise measurement is essential in quantum gas experiments, particularly in quantum gas microscopy, where the goal is to retrieve the occupation of individual lattice sites after taking fluorescence images. Conventional reconstruction techniques (such as analyzing intensity overlaps between different lattice sites) perform well when atoms are well-separated and the signal-to-noise ratio is high. However, accurate reconstruction of the measured Fock state becomes increasingly harder when the lattice spacing becomes smaller than the imaging resolution or when continuous cooling is not available during imaging.

To overcome these limitations, supervised deep learning approaches have been used to classify the occupation of lattice sites directly from fluorescence images~\cite{Picard2020}. However, though increasing imaging fidelities, this method requires acquiring labeled data for training, which generally is hard to aquire. To address this, an unsupervised method based on convolutional autoencoders was introduced~\cite{Impertro2023}. There, the main idea is the following: obtained fluorescence images are compressed to a low-dimensional latent space, whose size and shape corresponds to the one of the optical lattice. From this latent representation, a decoder then reconstructs the fluorescence image, and the entire network is trained to minimize the reconstruction error, shown in Fig.~\ref{fig:EA}~(c). An additional regularization term in the loss function ensures that the intermediate representation corresponds to a binary occupation map (i.e., non-binary values of the latent image are penalized). After training, unprocessed images can then be fed into the network; the bottleneck layer is then an accurate reconstruction of the atomic occupation matrix. This method was shown to be very effective, resulting in reconstruction fidelities exceeding 96\% at all filling levels of the optical lattice, even when the lattice spacing was more than two times smaller than the imaging resolution. 

Lastly, in a related context, deep neural networks have been shown to reduce interference fringes in dual-special trapped neutral atom setups~\cite{Lee2025dual}.

%=====================Discussion and Outlook=============================
\section{Discussion and Outlook}

Machine learning techniques have had, and will continue to have, a big impact across the entire pipeline of quantum technologies. As highlighted in this review, ML is becoming an essential tool for (i) optimizing experimental procedures---such as state preparation and high-fidelity readout---as well as (ii) for analyzing snapshot data and uncovering physical phenomena in strongly correlated quantum systems.

In the context of many-body physics, ML may play a crucial role in addressing some of the field's most persistent open questions. A prominent example is the unresolved nature of the pseudogap phase and the mechanisms underlying high-temperature superconductivity in cuprate materials and the FH model. Despite decades of theoretical and experimental efforts, a clear picture on whether the pseudogap originates from competing orders, preformed pairs, or fractionalized excitations remains elusive (see~\cite{Chowdhury2015, Schloemer2024GOM} and references therein). By leveraging large-scale quantum simulators, such as ultracold atoms in optical lattices which now reach temperatures in the cryogenic regime~\cite{Xu_cryo2025}, in combination with advanced ML-based analysis, it may soon become possible to test competing theories against experimental data with high precision. In particular, ML tools could help identify hidden order parameters or detect subtle correlations that would otherwise go unnoticed.

Beyond high-$T_c$ physics, ML methods can support the study of topological systems, in particular in identifying topological invariants and phase transitions when conventional order parameters are absent. Similarly, in lattice models such as the Bose-Hubbard model, ML can help explore the rich phase diagram beyond equilibrium conditions, including e.g. thermalization processes. 

Across these domains, machine learning provides a powerful framework not only for optimizing and interpreting experiments, but also for generating physical hypotheses, finding and selecting minimal models that describe systems of interest, and finding effective theories. Its integration with quantum simulation platforms opens a promising frontier for exploring complex quantum matter beyond the reach of conventional methods.

\section*{Acknowledgments} We thank Juan Carrasquilla Alvarez, Han-Ning Dai, Ehsan Khatami, Korbinian Kottmann, Christof Weitenberg, Mathias Scheurer, Evert van Nieuwenburg, Giacomo Torlai, and Lei Wang for correspondence regarding the use of data and figures for this review. We acknowledge funding from the Deutsche Forschungsgemeinschaft (DFG, German Research Foundation) under Germany's Excellence
Strategy—EXC2111—390814868, as well as the European Research Council (ERC) under the European
Union's Horizon 2020 research and innovation programme (Grant Agreement No. 948141-ERC Starting Grant SimUcQuam).

\bibliographystyle{plainnat} 
\bibliography{lit.bib}

\begin{thebibliography}{184}
\providecommand{\natexlab}[1]{#1}
\providecommand{\url}[1]{\texttt{#1}}
\expandafter\ifx\csname urlstyle\endcsname\relax
  \providecommand{\doi}[1]{doi: #1}\else
  \providecommand{\doi}{doi: \begingroup \urlstyle{rm}\Url}\fi

\bibitem[Abdi and Williams(2010)]{AbdiPCA}
Hervé Abdi and Lynne~J. Williams.
\newblock {Principal component analysis}.
\newblock \emph{WIREs Computational Statistics}, 2\penalty0 (4):\penalty0
  433--459, 2010.
\newblock \doi{https://doi.org/10.1002/wics.101}.
\newblock URL
  \url{https://wires.onlinelibrary.wiley.com/doi/abs/10.1002/wics.101}.

\bibitem[Anshu et~al.(2021)Anshu, Arunachalam, Kuwahara, and
  Soleimanifar]{Anshu2021}
Anurag Anshu, Srinivasan Arunachalam, Tomotaka Kuwahara, and Mehdi
  Soleimanifar.
\newblock {Sample-efficient learning of interacting quantum systems}.
\newblock \emph{Nature Physics}, 17\penalty0 (8):\penalty0 931--935, 2021.
\newblock \doi{10.1038/s41567-021-01232-0}.
\newblock URL \url{https://doi.org/10.1038/s41567-021-01232-0}.

\bibitem[Arnold and Sch\"afer(2022)]{Arnold2022PRX}
Julian Arnold and Frank Sch\"afer.
\newblock {Replacing Neural Networks by Optimal Analytical Predictors for the
  Detection of Phase Transitions}.
\newblock \emph{Phys. Rev. X}, 12:\penalty0 031044, Sep 2022.
\newblock \doi{10.1103/PhysRevX.12.031044}.
\newblock URL \url{https://link.aps.org/doi/10.1103/PhysRevX.12.031044}.

\bibitem[Arnold et~al.(2021)Arnold, Sch\"afer, \ifmmode~\check{Z}\else
  \v{Z}\fi{}onda, and Lode]{Arnold2021}
Julian Arnold, Frank Sch\"afer, Martin \ifmmode~\check{Z}\else \v{Z}\fi{}onda,
  and Axel U.~J. Lode.
\newblock {Interpretable and unsupervised phase classification}.
\newblock \emph{Phys. Rev. Res.}, 3:\penalty0 033052, Jul 2021.
\newblock \doi{10.1103/PhysRevResearch.3.033052}.
\newblock URL \url{https://link.aps.org/doi/10.1103/PhysRevResearch.3.033052}.

\bibitem[Arnold et~al.(2023{\natexlab{a}})Arnold, Lörch, Holtorf, and
  Schäfer]{Arnold2023QFI}
Julian Arnold, Niels Lörch, Flemming Holtorf, and Frank Schäfer.
\newblock {Machine learning phase transitions: Connections to the Fisher
  information}, 2023{\natexlab{a}}.
\newblock URL \url{https://arxiv.org/abs/2311.10710}.

\bibitem[Arnold et~al.(2023{\natexlab{b}})Arnold, Schäfer, and
  Lörch]{Arnold2023MTL}
Julian Arnold, Frank Schäfer, and Niels Lörch.
\newblock {Fast Detection of Phase Transitions with Multi-Task
  Learning-by-Confusion}, 2023{\natexlab{b}}.
\newblock URL \url{https://arxiv.org/abs/2311.09128}.

\bibitem[Arnold et~al.(2024)Arnold, Sch\"afer, Edelman, and Bruder]{Arnold2024}
Julian Arnold, Frank Sch\"afer, Alan Edelman, and Christoph Bruder.
\newblock {Mapping Out Phase Diagrams with Generative Classifiers}.
\newblock \emph{Phys. Rev. Lett.}, 132:\penalty0 207301, May 2024.
\newblock \doi{10.1103/PhysRevLett.132.207301}.
\newblock URL \url{https://link.aps.org/doi/10.1103/PhysRevLett.132.207301}.

\bibitem[Bairey et~al.(2019)Bairey, Arad, and Lindner]{Bairey2019}
Eyal Bairey, Itai Arad, and Netanel~H. Lindner.
\newblock {Learning a Local Hamiltonian from Local Measurements}.
\newblock \emph{Phys. Rev. Lett.}, 122:\penalty0 020504, Jan 2019.
\newblock \doi{10.1103/PhysRevLett.122.020504}.
\newblock URL \url{https://link.aps.org/doi/10.1103/PhysRevLett.122.020504}.

\bibitem[Barker et~al.(2020)Barker, Style, Luksch, Sunami, Garrick, Hill, Foot,
  and Bentine]{Barker2020}
A~J Barker, H~Style, K~Luksch, S~Sunami, D~Garrick, F~Hill, C~J Foot, and
  E~Bentine.
\newblock {Applying machine learning optimization methods to the production of
  a quantum gas}.
\newblock \emph{Machine Learning: Science and Technology}, 1\penalty0
  (1):\penalty0 015007, 2020.
\newblock \doi{10.1088/2632-2153/ab6432}.

\bibitem[Barredo et~al.(2016)Barredo, de~Léséleuc, Lienhard, Lahaye, and
  Browaeys]{Barredo2016}
Daniel Barredo, Sylvain de~Léséleuc, Vincent Lienhard, Thierry Lahaye, and
  Antoine Browaeys.
\newblock {An atom-by-atom assembler of defect-free arbitrary two-dimensional
  atomic arrays}.
\newblock \emph{Science}, 354\penalty0 (6315):\penalty0 1021--1023, 2016.
\newblock \doi{10.1126/science.aah3778}.
\newblock URL \url{https://www.science.org/doi/abs/10.1126/science.aah3778}.

\bibitem[Beach et~al.(2018)Beach, Golubeva, and Melko]{Beach2018}
Matthew J.~S. Beach, Anna Golubeva, and Roger~G. Melko.
\newblock {Machine learning vortices at the Kosterlitz-Thouless transition}.
\newblock \emph{Phys. Rev. B}, 97:\penalty0 045207, Jan 2018.
\newblock \doi{10.1103/PhysRevB.97.045207}.
\newblock URL \url{https://link.aps.org/doi/10.1103/PhysRevB.97.045207}.

\bibitem[Beaulieu et~al.(2025)Beaulieu, Kornja{\v c}a, Krunic, Stivaktakis,
  Chen, Ehmer, Wang, and Pham]{Beaulieu2025}
Daniel Beaulieu, Milan Kornja{\v c}a, Zoran Krunic, Michael Stivaktakis, Jing
  Chen, Thomas Ehmer, Sheng-Tao Wang, and Anh Pham.
\newblock Robust quantum reservoir learning for molecular property prediction.
\newblock \emph{Journal of Chemical Information and Modeling}, 08 2025.
\newblock \doi{10.1021/acs.jcim.5c00958}.
\newblock URL \url{https://doi.org/10.1021/acs.jcim.5c00958}.

\bibitem[Bennewitz et~al.(2022)Bennewitz, Hopfmueller, Kulchytskyy,
  Carrasquilla, and Ronagh]{Bennewitz2022}
Elizabeth~R. Bennewitz, Florian Hopfmueller, Bohdan Kulchytskyy, Juan
  Carrasquilla, and Pooya Ronagh.
\newblock {Neural Error Mitigation of Near-Term Quantum Simulations}.
\newblock \emph{Nature Machine Intelligence}, 4\penalty0 (7):\penalty0
  618--624, 2022.
\newblock \doi{10.1038/s42256-022-00509-0}.
\newblock URL \url{https://doi.org/10.1038/s42256-022-00509-0}.

\bibitem[Bernien et~al.(2017)Bernien, Schwartz, Keesling, Levine, Omran,
  Pichler, Choi, Zibrov, Endres, Greiner, Vuleti{\'c}, and Lukin]{Bernien2017}
Hannes Bernien, Sylvain Schwartz, Alexander Keesling, Harry Levine, Ahmed
  Omran, Hannes Pichler, Soonwon Choi, Alexander~S. Zibrov, Manuel Endres,
  Markus Greiner, Vladan Vuleti{\'c}, and Mikhail~D. Lukin.
\newblock {Probing many-body dynamics on a 51-atom quantum simulator}.
\newblock \emph{Nature}, 551\penalty0 (7682):\penalty0 579--584, 2017.
\newblock \doi{10.1038/nature24622}.
\newblock URL \url{https://doi.org/10.1038/nature24622}.

\bibitem[Bhakuni et~al.(2024)Bhakuni, Verdel, Muzzi, Andreoni, Aidelsburger,
  and Dalmonte]{Bhakuni2024}
Devendra~Singh Bhakuni, Roberto Verdel, Cristiano Muzzi, Riccardo Andreoni,
  Monika Aidelsburger, and Marcello Dalmonte.
\newblock {Diagnosing quantum transport from wave function snapshots}.
\newblock \emph{Phys. Rev. B}, 110:\penalty0 144204, Oct 2024.
\newblock \doi{10.1103/PhysRevB.110.144204}.
\newblock URL \url{https://link.aps.org/doi/10.1103/PhysRevB.110.144204}.

\bibitem[Blatz et~al.(2024)Blatz, Kwan, Léonard, and Bohrdt]{Blatz2024}
Tizian Blatz, Joyce Kwan, Julian Léonard, and Annabelle Bohrdt.
\newblock {Bayesian Optimization for Robust State Preparation in Quantum
  Many-Body Systems}.
\newblock \emph{Quantum}, 8:\penalty0 1388, 2024.
\newblock \doi{10.22331/q-2024-06-27-1388}.

\bibitem[Bloch et~al.(2008)Bloch, Dalibard, and Zwerger]{Bloch2008}
Immanuel Bloch, Jean Dalibard, and Wilhelm Zwerger.
\newblock {Many-body physics with ultracold gases}.
\newblock \emph{Rev. Mod. Phys.}, 80:\penalty0 885--964, Jul 2008.
\newblock \doi{10.1103/RevModPhys.80.885}.
\newblock URL \url{https://link.aps.org/doi/10.1103/RevModPhys.80.885}.

\bibitem[Bloch et~al.(2012)Bloch, Dalibard, and Nascimb{\`e}ne]{Bloch2012}
Immanuel Bloch, Jean Dalibard, and Sylvain Nascimb{\`e}ne.
\newblock {Quantum simulations with ultracold quantum gases}.
\newblock \emph{Nature Physics}, 8\penalty0 (4):\penalty0 267--276, 2012.
\newblock \doi{10.1038/nphys2259}.
\newblock URL \url{https://doi.org/10.1038/nphys2259}.

\bibitem[Bluvstein et~al.(2021)Bluvstein, Omran, Levine, Keesling, Semeghini,
  Ebadi, Wang, Michailidis, Maskara, Ho, Choi, Serbyn, Greiner, Vuletić, and
  Lukin]{Bluvstein2021}
D.~Bluvstein, A.~Omran, H.~Levine, A.~Keesling, G.~Semeghini, S.~Ebadi, T.~T.
  Wang, A.~A. Michailidis, N.~Maskara, W.~W. Ho, S.~Choi, M.~Serbyn,
  M.~Greiner, V.~Vuletić, and M.~D. Lukin.
\newblock {Controlling quantum many-body dynamics in driven Rydberg atom
  arrays}.
\newblock \emph{Science}, 371\penalty0 (6536):\penalty0 1355--1359, 2021.
\newblock \doi{10.1126/science.abg2530}.
\newblock URL \url{https://www.science.org/doi/abs/10.1126/science.abg2530}.

\bibitem[Bohrdt et~al.(2021{\natexlab{a}})Bohrdt, Kim, Lukin, Rispoli,
  Schittko, Knap, Greiner, and L\'eonard]{Bohrdt2021}
A.~Bohrdt, S.~Kim, A.~Lukin, M.~Rispoli, R.~Schittko, M.~Knap, M.~Greiner, and
  J.~L\'eonard.
\newblock {Analyzing Nonequilibrium Quantum States through Snapshots with
  Artificial Neural Networks}.
\newblock \emph{Phys. Rev. Lett.}, 127:\penalty0 150504, Oct
  2021{\natexlab{a}}.
\newblock \doi{10.1103/PhysRevLett.127.150504}.
\newblock URL \url{https://link.aps.org/doi/10.1103/PhysRevLett.127.150504}.

\bibitem[Bohrdt et~al.(2019)Bohrdt, Chiu, Ji, Xu, Greif, Greiner, Demler,
  Grusdt, and Knap]{Bohrdt2019}
Annabelle Bohrdt, Christie~S. Chiu, Geoffrey Ji, Muqing Xu, Daniel Greif,
  Markus Greiner, Eugene Demler, Fabian Grusdt, and Michael Knap.
\newblock {Classifying snapshots of the doped Hubbard model with machine
  learning}.
\newblock \emph{Nature Physics}, 15\penalty0 (9):\penalty0 921--924, 2019.
\newblock \doi{10.1038/s41567-019-0565-x}.
\newblock URL \url{https://doi.org/10.1038/s41567-019-0565-x}.

\bibitem[Bohrdt et~al.(2021{\natexlab{b}})Bohrdt, Homeier, Reinmoser, Demler,
  and Grusdt]{Bohrdt2020}
Annabelle Bohrdt, Lukas Homeier, Christian Reinmoser, Eugene Demler, and Fabian
  Grusdt.
\newblock {Exploration of doped quantum magnets with ultracold atoms}.
\newblock \emph{Annals of Physics}, 435:\penalty0 168651, 2021{\natexlab{b}}.
\newblock ISSN 0003-4916.
\newblock \doi{10.1016/j.aop.2021.168651}.
\newblock URL
  \url{https://www.sciencedirect.com/science/article/pii/S0003491621002578}.
\newblock Special issue on Philip W. Anderson.

\bibitem[Bravo et~al.(2022)Bravo, Najafi, Gao, and Yelin]{Bravo2022}
Rodrigo~Araiza Bravo, Khadijeh Najafi, Xun Gao, and Susanne~F. Yelin.
\newblock {Quantum Reservoir Computing Using Arrays of Rydberg Atoms}.
\newblock \emph{PRX Quantum}, 3:\penalty0 030325, Aug 2022.
\newblock \doi{10.1103/PRXQuantum.3.030325}.
\newblock URL \url{https://link.aps.org/doi/10.1103/PRXQuantum.3.030325}.

\bibitem[Broecker et~al.(2017{\natexlab{a}})Broecker, Assaad, and
  Trebst]{Broecker2017}
Peter Broecker, Fakher~F. Assaad, and Simon Trebst.
\newblock {Quantum phase recognition via unsupervised machine learning},
  2017{\natexlab{a}}.
\newblock URL \url{https://arxiv.org/abs/1707.00663}.

\bibitem[Broecker et~al.(2017{\natexlab{b}})Broecker, Carrasquilla, Melko, and
  Trebst]{Broecker2017_green}
Peter Broecker, Juan Carrasquilla, Roger~G. Melko, and Simon Trebst.
\newblock {Machine learning quantum phases of matter beyond the fermion sign
  problem}.
\newblock \emph{Scientific Reports}, 7\penalty0 (1):\penalty0 8823,
  2017{\natexlab{b}}.
\newblock \doi{10.1038/s41598-017-09098-0}.
\newblock URL \url{https://doi.org/10.1038/s41598-017-09098-0}.

\bibitem[Brydges et~al.(2019)Brydges, Elben, Jurcevic, Vermersch, Maier,
  Lanyon, Zoller, Blatt, and Roos]{Brydges2019}
Tiff Brydges, Andreas Elben, Petar Jurcevic, Benoit Vermersch, Christine Maier,
  Ben~P. Lanyon, Peter Zoller, Rainer Blatt, and Christian~F. Roos.
\newblock {Probing Rényi entanglement entropy via randomized measurements}.
\newblock \emph{Science}, 364\penalty0 (6437):\penalty0 260--263, 2019.
\newblock \doi{10.1126/science.aau4963}.
\newblock URL \url{https://www.science.org/doi/abs/10.1126/science.aau4963}.

\bibitem[Bukov et~al.(2018)Bukov, Day, Sels, Weinberg, Polkovnikov, and
  Mehta]{Bukov2018}
Marin Bukov, Alexandre G.~R. Day, Dries Sels, Phillip Weinberg, Anatoli
  Polkovnikov, and Pankaj Mehta.
\newblock {Reinforcement Learning in Different Phases of Quantum Control}.
\newblock \emph{Physical Review X}, 8\penalty0 (3):\penalty0 031086, 2018.
\newblock \doi{10.1103/physrevx.8.031086}.

\bibitem[Camastra and Staiano(2016)]{Camastra2016}
Francesco Camastra and Antonino Staiano.
\newblock {Intrinsic dimension estimation: Advances and open problems}.
\newblock \emph{Information Sciences}, 328:\penalty0 26--41, 2016.
\newblock ISSN 0020-0255.
\newblock \doi{10.1016/j.ins.2015.08.029}.

\bibitem[Cao et~al.(2020)Cao, Hou, Cao, and Zeng]{Cao2020}
Chenfeng Cao, Shi-Yao Hou, Ningping Cao, and Bei Zeng.
\newblock {Supervised learning in Hamiltonian reconstruction from local
  measurements on eigenstates}.
\newblock \emph{Journal of Physics: Condensed Matter}, 33\penalty0
  (6):\penalty0 064002, feb 2020.
\newblock \doi{10.1088/1361-648x/abc4cf}.
\newblock URL \url{https://doi.org/10.1088/1361-648x/abc4cf}.

\bibitem[Carleo and Troyer(2017)]{Carleo2017}
Giuseppe Carleo and Matthias Troyer.
\newblock {Solving the quantum many-body problem with artificial neural
  networks}.
\newblock \emph{Science}, 355\penalty0 (6325):\penalty0 602--606, 2017.
\newblock \doi{10.1126/science.aag2302}.
\newblock URL \url{https://www.science.org/doi/abs/10.1126/science.aag2302}.

\bibitem[Carleo et~al.(2019)Carleo, Cirac, Cranmer, Daudet, Schuld, Tishby,
  Vogt-Maranto, and Zdeborov\'a]{CarleoReview}
Giuseppe Carleo, Ignacio Cirac, Kyle Cranmer, Laurent Daudet, Maria Schuld,
  Naftali Tishby, Leslie Vogt-Maranto, and Lenka Zdeborov\'a.
\newblock {Machine learning and the physical sciences}.
\newblock \emph{Rev. Mod. Phys.}, 91:\penalty0 045002, Dec 2019.
\newblock \doi{10.1103/RevModPhys.91.045002}.
\newblock URL \url{https://link.aps.org/doi/10.1103/RevModPhys.91.045002}.

\bibitem[Carrasco et~al.(2021)Carrasco, Elben, Kokail, Kraus, and
  Zoller]{Carrasco2021}
Jose Carrasco, Andreas Elben, Christian Kokail, Barbara Kraus, and Peter
  Zoller.
\newblock Theoretical and experimental perspectives of quantum verification.
\newblock \emph{PRX Quantum}, 2:\penalty0 010102, Mar 2021.
\newblock \doi{10.1103/PRXQuantum.2.010102}.
\newblock URL \url{https://link.aps.org/doi/10.1103/PRXQuantum.2.010102}.

\bibitem[Carrasquilla(2020)]{CarrasquillaReview}
Juan Carrasquilla.
\newblock {Machine learning for quantum matter}.
\newblock \emph{Advances in Physics: X}, 5\penalty0 (1):\penalty0 1797528, 01
  2020.
\newblock \doi{10.1080/23746149.2020.1797528}.
\newblock URL \url{https://doi.org/10.1080/23746149.2020.1797528}.

\bibitem[Carrasquilla and Melko(2017)]{Carrasquilla2017}
Juan Carrasquilla and Roger~G. Melko.
\newblock {Machine learning phases of matter}.
\newblock \emph{Nature Physics}, 13\penalty0 (5):\penalty0 431--434, 2017.
\newblock \doi{10.1038/nphys4035}.
\newblock URL \url{https://doi.org/10.1038/nphys4035}.

\bibitem[Carrasquilla and Torlai(2021)]{Carrasquilla2021Tutorial}
Juan Carrasquilla and Giacomo Torlai.
\newblock {How To Use Neural Networks To Investigate Quantum Many-Body
  Physics}.
\newblock \emph{PRX Quantum}, 2:\penalty0 040201, Nov 2021.
\newblock \doi{10.1103/PRXQuantum.2.040201}.
\newblock URL \url{https://link.aps.org/doi/10.1103/PRXQuantum.2.040201}.

\bibitem[Carvalho et~al.(2018)Carvalho, Garcia-Martinez, Lado, and
  Fernandez-Rossier]{Carvalho2018}
D.~Carvalho, N.~A. Garcia-Martinez, J.~L. Lado, and J.~Fernandez-Rossier.
\newblock {Real-space mapping of topological invariants using artificial neural
  networks}.
\newblock \emph{Phys. Rev. B}, 97:\penalty0 115453, Mar 2018.
\newblock \doi{10.1103/PhysRevB.97.115453}.
\newblock URL \url{https://link.aps.org/doi/10.1103/PhysRevB.97.115453}.

\bibitem[Casert et~al.(2019)Casert, Vieijra, Nys, and Ryckebusch]{Casert2019}
C.~Casert, T.~Vieijra, J.~Nys, and J.~Ryckebusch.
\newblock {Interpretable machine learning for inferring the phase boundaries in
  a nonequilibrium system}.
\newblock \emph{Phys. Rev. E}, 99:\penalty0 023304, Feb 2019.
\newblock \doi{10.1103/PhysRevE.99.023304}.
\newblock URL \url{https://link.aps.org/doi/10.1103/PhysRevE.99.023304}.

\bibitem[Cha et~al.(2022)Cha, Ginsparg, Wu, Carrasquilla, McMahon, and
  Kim]{Cha2021}
Peter Cha, Paul Ginsparg, Felix Wu, Juan Carrasquilla, Peter~L McMahon, and
  Eun-Ah Kim.
\newblock {Attention-based quantum tomography}.
\newblock \emph{Machine Learning: Science and Technology}, 3\penalty0
  (1):\penalty0 01LT01, 2022.
\newblock \doi{10.1088/2632-2153/ac362b}.
\newblock URL \url{https://dx.doi.org/10.1088/2632-2153/ac362b}.

\bibitem[Chanda et~al.(2025)Chanda, Barbiero, Lewenstein, Mark, and
  Zakrzewski]{Chanda2025}
Titas Chanda, Luca Barbiero, Maciej Lewenstein, Manfred~J Mark, and Jakub
  Zakrzewski.
\newblock {Recent progress on quantum simulations of non-standard Bose--Hubbard
  models}.
\newblock \emph{Reports on Progress in Physics}, 88\penalty0 (4):\penalty0
  044501, 2025.
\newblock \doi{10.1088/1361-6633/adc3a7}.
\newblock URL \url{https://dx.doi.org/10.1088/1361-6633/adc3a7}.

\bibitem[Che et~al.(2020)Che, Gneiting, Liu, and Nori]{Che2020}
Yanming Che, Clemens Gneiting, Tao Liu, and Franco Nori.
\newblock {Topological quantum phase transitions retrieved through unsupervised
  machine learning}.
\newblock \emph{Phys. Rev. B}, 102:\penalty0 134213, Oct 2020.
\newblock \doi{10.1103/PhysRevB.102.134213}.
\newblock URL \url{https://link.aps.org/doi/10.1103/PhysRevB.102.134213}.

\bibitem[Chen et~al.(2023)Chen, Bornet, Bintz, Emperauger, Leclerc, Liu,
  Scholl, Barredo, Hauschild, Chatterjee, Schuler, L{\"a}uchli, Zaletel,
  Lahaye, Yao, and Browaeys]{Chen2023}
Cheng Chen, Guillaume Bornet, Marcus Bintz, Gabriel Emperauger, Lucas Leclerc,
  Vincent~S. Liu, Pascal Scholl, Daniel Barredo, Johannes Hauschild, Shubhayu
  Chatterjee, Michael Schuler, Andreas~M. L{\"a}uchli, Michael~P. Zaletel,
  Thierry Lahaye, Norman~Y. Yao, and Antoine Browaeys.
\newblock {Continuous symmetry breaking in a two-dimensional Rydberg array}.
\newblock \emph{Nature}, 616\penalty0 (7958):\penalty0 691--695, 2023.
\newblock \doi{10.1038/s41586-023-05859-2}.
\newblock URL \url{https://doi.org/10.1038/s41586-023-05859-2}.

\bibitem[Cheuk et~al.(2015)Cheuk, Nichols, Okan, Gersdorf, Ramasesh, Bakr,
  Lompe, and Zwierlein]{Cheuk2015}
Lawrence~W. Cheuk, Matthew~A. Nichols, Melih Okan, Thomas Gersdorf, Vinay~V.
  Ramasesh, Waseem~S. Bakr, Thomas Lompe, and Martin~W. Zwierlein.
\newblock {Quantum-Gas Microscope for Fermionic Atoms}.
\newblock \emph{Phys. Rev. Lett.}, 114:\penalty0 193001, May 2015.
\newblock \doi{10.1103/PhysRevLett.114.193001}.
\newblock URL \url{https://link.aps.org/doi/10.1103/PhysRevLett.114.193001}.

\bibitem[Chiu et~al.(2019)Chiu, Ji, Bohrdt, Xu, Knap, Demler, Grusdt, Greiner,
  and Greif]{Chiu2019}
Christie~S. Chiu, Geoffrey Ji, Annabelle Bohrdt, Muqing Xu, Michael Knap,
  Eugene Demler, Fabian Grusdt, Markus Greiner, and Daniel Greif.
\newblock {String patterns in the doped Hubbard model}.
\newblock \emph{Science}, 365\penalty0 (6450):\penalty0 251--256, 2019.
\newblock \doi{10.1126/science.aav3587}.
\newblock URL \url{https://www.science.org/doi/abs/10.1126/science.aav3587}.

\bibitem[Ch'ng et~al.(2017)Ch'ng, Carrasquilla, Melko, and Khatami]{Chng2017}
Kelvin Ch'ng, Juan Carrasquilla, Roger~G. Melko, and Ehsan Khatami.
\newblock {Machine Learning Phases of Strongly Correlated Fermions}.
\newblock \emph{Phys. Rev. X}, 7:\penalty0 031038, Aug 2017.
\newblock \doi{10.1103/PhysRevX.7.031038}.
\newblock URL \url{https://link.aps.org/doi/10.1103/PhysRevX.7.031038}.

\bibitem[Ch'ng et~al.(2018)Ch'ng, Vazquez, and Khatami]{Chng2018}
Kelvin Ch'ng, Nick Vazquez, and Ehsan Khatami.
\newblock {Unsupervised machine learning account of magnetic transitions in the
  Hubbard model}.
\newblock \emph{Phys. Rev. E}, 97:\penalty0 013306, Jan 2018.
\newblock \doi{10.1103/PhysRevE.97.013306}.
\newblock URL \url{https://link.aps.org/doi/10.1103/PhysRevE.97.013306}.

\bibitem[Chong et~al.(2021)Chong, Kim, Ahn, and Jeong]{Chong2021}
Daryl~Ryan Chong, Minhyuk Kim, Jaewook Ahn, and Heejeong Jeong.
\newblock {Machine learning identification of symmetrized base states of
  Rydberg atoms}.
\newblock \emph{Frontiers of Physics}, 17\penalty0 (1):\penalty0 12504, 2021.
\newblock \doi{10.1007/s11467-021-1099-0}.
\newblock URL \url{https://doi.org/10.1007/s11467-021-1099-0}.

\bibitem[Chowdhury and Sachdev()]{Chowdhury2015}
Debanjan Chowdhury and Subir Sachdev.
\newblock \emph{{The Enigma of the Pseudogap Phase of the Cuprate
  Superconductors}}, pages 1--43.
\newblock \doi{10.1142/9789814704090_0001}.
\newblock URL
  \url{https://www.worldscientific.com/doi/abs/10.1142/9789814704090_0001}.

\bibitem[Cong et~al.(2019)Cong, Choi, and Lukin]{Cong2019}
Iris Cong, Soonwon Choi, and Mikhail~D. Lukin.
\newblock {Quantum convolutional neural networks}.
\newblock \emph{Nature Physics}, 15\penalty0 (12):\penalty0 1273--1278, 2019.
\newblock \doi{10.1038/s41567-019-0648-8}.
\newblock URL \url{https://doi.org/10.1038/s41567-019-0648-8}.

\bibitem[Cooper et~al.(2019)Cooper, Dalibard, and Spielman]{Cooper2019}
N.~R. Cooper, J.~Dalibard, and I.~B. Spielman.
\newblock {Topological bands for ultracold atoms}.
\newblock \emph{Rev. Mod. Phys.}, 91:\penalty0 015005, Mar 2019.
\newblock \doi{10.1103/RevModPhys.91.015005}.
\newblock URL \url{https://link.aps.org/doi/10.1103/RevModPhys.91.015005}.

\bibitem[Costa et~al.(2017)Costa, Hu, Bai, Scalettar, and Singh]{Costa2017}
Natanael~C. Costa, Wenjian Hu, Z.~J. Bai, Richard~T. Scalettar, and Rajiv R.~P.
  Singh.
\newblock {Principal component analysis for fermionic critical points}.
\newblock \emph{Phys. Rev. B}, 96:\penalty0 195138, Nov 2017.
\newblock \doi{10.1103/PhysRevB.96.195138}.
\newblock URL \url{https://link.aps.org/doi/10.1103/PhysRevB.96.195138}.

\bibitem[Cybiński et~al.(2024)Cybiński, Enouen, Georges, and
  Dawid]{Cybinski2024}
Kacper Cybiński, James Enouen, Antoine Georges, and Anna Dawid.
\newblock {Speak so a physicist can understand you! TetrisCNN for detecting
  phase transitions and order parameters}, 2024.
\newblock URL \url{https://arxiv.org/abs/2411.02237}.

\bibitem[Czischek et~al.(2022)Czischek, Moss, Radzihovsky, Merali, and
  Melko]{Czischek2022}
Stefanie Czischek, M.~Schuyler Moss, Matthew Radzihovsky, Ejaaz Merali, and
  Roger~G. Melko.
\newblock {Data-enhanced variational Monte Carlo simulations for Rydberg atom
  arrays}.
\newblock \emph{Phys. Rev. B}, 105:\penalty0 205108, May 2022.
\newblock \doi{10.1103/PhysRevB.105.205108}.
\newblock URL \url{https://link.aps.org/doi/10.1103/PhysRevB.105.205108}.

\bibitem[Daley et~al.(2022)Daley, Bloch, Kokail, Flannigan, Pearson, Troyer,
  and Zoller]{Daley2022}
Andrew~J. Daley, Immanuel Bloch, Christian Kokail, Stuart Flannigan, Natalie
  Pearson, Matthias Troyer, and Peter Zoller.
\newblock Practical quantum advantage in quantum simulation.
\newblock \emph{Nature}, 607\penalty0 (7920):\penalty0 667--676, 2022.
\newblock \doi{10.1038/s41586-022-04940-6}.
\newblock URL \url{https://doi.org/10.1038/s41586-022-04940-6}.

\bibitem[Dalibard et~al.(2011)Dalibard, Gerbier, Juzeli\ifmmode~\bar{u}\else
  \={u}\fi{}nas, and \"Ohberg]{Dalibard2011}
Jean Dalibard, Fabrice Gerbier, Gediminas Juzeli\ifmmode~\bar{u}\else
  \={u}\fi{}nas, and Patrik \"Ohberg.
\newblock {Colloquium: Artificial gauge potentials for neutral atoms}.
\newblock \emph{Rev. Mod. Phys.}, 83:\penalty0 1523--1543, Nov 2011.
\newblock \doi{10.1103/RevModPhys.83.1523}.
\newblock URL \url{https://link.aps.org/doi/10.1103/RevModPhys.83.1523}.

\bibitem[Dawid et~al.(2020)Dawid, Huembeli, Tomza, Lewenstein, and
  Dauphin]{Dawid2020}
Anna Dawid, Patrick Huembeli, Michal Tomza, Maciej Lewenstein, and Alexandre
  Dauphin.
\newblock {Phase detection with neural networks: interpreting the black box}.
\newblock \emph{New Journal of Physics}, 22\penalty0 (11):\penalty0 115001,
  2020.
\newblock \doi{10.1088/1367-2630/abc463}.

\bibitem[Dawid et~al.(2022)Dawid, Huembeli, Tomza, Lewenstein, and
  Dauphin]{Dawid2022}
Anna Dawid, Patrick Huembeli, Michał Tomza, Maciej Lewenstein, and Alexandre
  Dauphin.
\newblock {Hessian-based toolbox for reliable and interpretable machine
  learning in physics}.
\newblock \emph{Machine Learning: Science and Technology}, 3\penalty0
  (1):\penalty0 015002, 2022.
\newblock \doi{10.1088/2632-2153/ac338d}.

\bibitem[Dawid et~al.(2023)Dawid, Arnold, Requena, Gresch, Płodzień,
  Donatella, Nicoli, Stornati, Koch, Büttner, Okuła, Muñoz-Gil,
  Vargas-Hernández, Cervera-Lierta, Carrasquilla, Dunjko, Gabrié, Huembeli,
  van Nieuwenburg, Vicentini, Wang, Wetzel, Carleo, Greplová, Krems,
  Marquardt, Tomza, Lewenstein, and Dauphin]{DawidReview}
Anna Dawid, Julian Arnold, Borja Requena, Alexander Gresch, Marcin Płodzień,
  Kaelan Donatella, Kim~A. Nicoli, Paolo Stornati, Rouven Koch, Miriam
  Büttner, Robert Okuła, Gorka Muñoz-Gil, Rodrigo~A. Vargas-Hernández, Alba
  Cervera-Lierta, Juan Carrasquilla, Vedran Dunjko, Marylou Gabrié, Patrick
  Huembeli, Evert van Nieuwenburg, Filippo Vicentini, Lei Wang, Sebastian~J.
  Wetzel, Giuseppe Carleo, Eliška Greplová, Roman Krems, Florian Marquardt,
  Michał Tomza, Maciej Lewenstein, and Alexandre Dauphin.
\newblock {Modern applications of machine learning in quantum sciences}, 2023.
\newblock URL \url{https://arxiv.org/abs/2204.04198}.

\bibitem[Di~Franco et~al.(2009)Di~Franco, Paternostro, and Kim]{DiFranco2009}
C.~Di~Franco, M.~Paternostro, and M.~S. Kim.
\newblock {Hamiltonian Tomography in an Access-Limited Setting without State
  Initialization}.
\newblock \emph{Phys. Rev. Lett.}, 102:\penalty0 187203, May 2009.
\newblock \doi{10.1103/PhysRevLett.102.187203}.
\newblock URL \url{https://link.aps.org/doi/10.1103/PhysRevLett.102.187203}.

\bibitem[Ebadi et~al.(2021)Ebadi, Wang, Levine, Keesling, Semeghini, Omran,
  Bluvstein, Samajdar, Pichler, Ho, Choi, Sachdev, Greiner, Vuleti{\'c}, and
  Lukin]{Ebadi2021}
Sepehr Ebadi, Tout~T. Wang, Harry Levine, Alexander Keesling, Giulia Semeghini,
  Ahmed Omran, Dolev Bluvstein, Rhine Samajdar, Hannes Pichler, Wen~Wei Ho,
  Soonwon Choi, Subir Sachdev, Markus Greiner, Vladan Vuleti{\'c}, and
  Mikhail~D. Lukin.
\newblock {Quantum phases of matter on a 256-atom programmable quantum
  simulator}.
\newblock \emph{Nature}, 595\penalty0 (7866):\penalty0 227--232, 2021.
\newblock \doi{10.1038/s41586-021-03582-4}.
\newblock URL \url{https://doi.org/10.1038/s41586-021-03582-4}.

\bibitem[Eckardt(2017)]{Eckardt2017}
Andr\'e Eckardt.
\newblock {Colloquium: Atomic quantum gases in periodically driven optical
  lattices}.
\newblock \emph{Rev. Mod. Phys.}, 89:\penalty0 011004, Mar 2017.
\newblock \doi{10.1103/RevModPhys.89.011004}.
\newblock URL \url{https://link.aps.org/doi/10.1103/RevModPhys.89.011004}.

\bibitem[Elben et~al.(2020)Elben, Kueng, Huang, van Bijnen, Kokail, Dalmonte,
  Calabrese, Kraus, Preskill, Zoller, and Vermersch]{Elben2020}
Andreas Elben, Richard Kueng, Hsin-Yuan~(Robert) Huang, Rick van Bijnen,
  Christian Kokail, Marcello Dalmonte, Pasquale Calabrese, Barbara Kraus, John
  Preskill, Peter Zoller, and Benoit Vermersch.
\newblock Mixed-state entanglement from local randomized measurements.
\newblock \emph{Phys. Rev. Lett.}, 125:\penalty0 200501, Nov 2020.
\newblock \doi{10.1103/PhysRevLett.125.200501}.
\newblock URL \url{https://link.aps.org/doi/10.1103/PhysRevLett.125.200501}.

\bibitem[Endres et~al.(2011)Endres, Cheneau, Fukuhara, Weitenberg, Schauß,
  Gross, Mazza, Bañuls, Pollet, Bloch, and Kuhr]{Endres2011}
M.~Endres, M.~Cheneau, T.~Fukuhara, C.~Weitenberg, P.~Schauß, C.~Gross,
  L.~Mazza, M.~C. Bañuls, L.~Pollet, I.~Bloch, and S.~Kuhr.
\newblock {Observation of Correlated Particle-Hole Pairs and String Order in
  Low-Dimensional Mott Insulators}.
\newblock \emph{Science}, 334\penalty0 (6053):\penalty0 200--203, 2011.
\newblock \doi{10.1126/science.1209284}.
\newblock URL \url{https://www.science.org/doi/abs/10.1126/science.1209284}.

\bibitem[Endres et~al.(2016)Endres, Bernien, Keesling, Levine, Anschuetz,
  Krajenbrink, Senko, Vuletic, Greiner, and Lukin]{Endres2016}
Manuel Endres, Hannes Bernien, Alexander Keesling, Harry Levine, Eric~R.
  Anschuetz, Alexandre Krajenbrink, Crystal Senko, Vladan Vuletic, Markus
  Greiner, and Mikhail~D. Lukin.
\newblock {Atom-by-atom assembly of defect-free one-dimensional cold atom
  arrays}.
\newblock \emph{Science}, 354\penalty0 (6315):\penalty0 1024--1027, 2016.
\newblock \doi{10.1126/science.aah3752}.
\newblock URL \url{https://www.science.org/doi/abs/10.1126/science.aah3752}.

\bibitem[Esslinger(2010)]{Esslinger2010}
Tilman Esslinger.
\newblock {Fermi-Hubbard Physics with Atoms in an Optical Lattice}.
\newblock \emph{Annual Review of Condensed Matter Physics}, 1\penalty0
  (1):\penalty0 129--152, 2010.
\newblock \doi{10.1146/annurev-conmatphys-070909-104059}.
\newblock URL \url{https://doi.org/10.1146/annurev-conmatphys-070909-104059}.

\bibitem[Falicov and Kimball(1969)]{Falicov1969}
L.~M. Falicov and J.~C. Kimball.
\newblock {Simple Model for Semiconductor-Metal Transitions:
  Sm${\mathrm{B}}_{6}$ and Transition-Metal Oxides}.
\newblock \emph{Phys. Rev. Lett.}, 22:\penalty0 997--999, May 1969.
\newblock \doi{10.1103/PhysRevLett.22.997}.
\newblock URL \url{https://link.aps.org/doi/10.1103/PhysRevLett.22.997}.

\bibitem[Fendley et~al.(2004)Fendley, Sengupta, and Sachdev]{Fendley2004}
Paul Fendley, K.~Sengupta, and Subir Sachdev.
\newblock {Competing density-wave orders in a one-dimensional hard-boson
  model}.
\newblock \emph{Phys. Rev. B}, 69:\penalty0 075106, Feb 2004.
\newblock \doi{10.1103/PhysRevB.69.075106}.
\newblock URL \url{https://link.aps.org/doi/10.1103/PhysRevB.69.075106}.

\bibitem[Fitzek et~al.(2024)Fitzek, Teoh, Fung, Dagnew, Merali, Moss,
  MacLellan, and Melko]{Fitzek2024}
David Fitzek, Yi~Hong Teoh, Hin~Pok Fung, Gebremedhin~A. Dagnew, Ejaaz Merali,
  M.~Schuyler Moss, Benjamin MacLellan, and Roger~G. Melko.
\newblock Rydberggpt, 2024.
\newblock URL \url{https://arxiv.org/abs/2405.21052}.

\bibitem[Fujii and Nakajima(2017)]{Fujii2017}
Keisuke Fujii and Kohei Nakajima.
\newblock {Harnessing Disordered-Ensemble Quantum Dynamics for Machine
  Learning}.
\newblock \emph{Phys. Rev. Appl.}, 8:\penalty0 024030, Aug 2017.
\newblock \doi{10.1103/PhysRevApplied.8.024030}.
\newblock URL \url{https://link.aps.org/doi/10.1103/PhysRevApplied.8.024030}.

\bibitem[Fujii and Nakajima(2021)]{Fujii2021}
Keisuke Fujii and Kohei Nakajima.
\newblock \emph{{Quantum Reservoir Computing: A Reservoir Approach Toward
  Quantum Machine Learning on Near-Term Quantum Devices}}, pages 423--450.
\newblock Springer Singapore, Singapore, 2021.
\newblock ISBN 978-981-13-1687-6.
\newblock \doi{10.1007/978-981-13-1687-6{\_}18}.
\newblock URL \url{https://doi.org/10.1007/978-981-13-1687-6_18}.

\bibitem[Greitemann et~al.(2019{\natexlab{a}})Greitemann, Liu, Jaubert, Yan,
  Shannon, and Pollet]{Greitemann2019b}
Jonas Greitemann, Ke~Liu, Ludovic D.~C. Jaubert, Han Yan, Nic Shannon, and Lode
  Pollet.
\newblock {Identification of emergent constraints and hidden order in
  frustrated magnets using tensorial kernel methods of machine learning}.
\newblock \emph{Phys. Rev. B}, 100:\penalty0 174408, Nov 2019{\natexlab{a}}.
\newblock \doi{10.1103/PhysRevB.100.174408}.
\newblock URL \url{https://link.aps.org/doi/10.1103/PhysRevB.100.174408}.

\bibitem[Greitemann et~al.(2019{\natexlab{b}})Greitemann, Liu, and
  Pollet]{Greitemann2019a}
Jonas Greitemann, Ke~Liu, and Lode Pollet.
\newblock {Probing hidden spin order with interpretable machine learning}.
\newblock \emph{Phys. Rev. B}, 99:\penalty0 060404, Feb 2019{\natexlab{b}}.
\newblock \doi{10.1103/PhysRevB.99.060404}.
\newblock URL \url{https://link.aps.org/doi/10.1103/PhysRevB.99.060404}.

\bibitem[Greitemann et~al.(2021)Greitemann, Liu, and Pollet]{Greitemann2021}
Jonas Greitemann, Ke~Liu, and Lode Pollet.
\newblock {The view of TK-SVM on the phase hierarchy in the classical kagome
  Heisenberg antiferromagnet}.
\newblock \emph{Journal of Physics: Condensed Matter}, 33\penalty0
  (5):\penalty0 054002, 2021.
\newblock \doi{10.1088/1361-648X/abbe7b}.
\newblock URL \url{https://dx.doi.org/10.1088/1361-648X/abbe7b}.

\bibitem[Greplova et~al.(2020)Greplova, Valenti, Boschung, Sch{\"a}fer,
  L{\"o}rch, and Huber]{Greplova2020}
Eliska Greplova, Agnes Valenti, Gregor Boschung, Frank Sch{\"a}fer, Niels
  L{\"o}rch, and Sebastian~D Huber.
\newblock {Unsupervised identification of topological phase transitions using
  predictive models}.
\newblock \emph{New Journal of Physics}, 22\penalty0 (4):\penalty0 045003,
  2020.
\newblock \doi{10.1088/1367-2630/ab7771}.
\newblock URL \url{https://dx.doi.org/10.1088/1367-2630/ab7771}.

\bibitem[Gross and Bloch(2017)]{Gross2017}
Christian Gross and Immanuel Bloch.
\newblock {Quantum simulations with ultracold atoms in optical lattices}.
\newblock \emph{Science}, 357\penalty0 (6355):\penalty0 995--1001, 2017.
\newblock \doi{10.1126/science.aal3837}.
\newblock URL \url{https://www.science.org/doi/abs/10.1126/science.aal3837}.

\bibitem[Guo and He(2023)]{Guo2023}
Wei-chen Guo and Liang He.
\newblock {Learning phase transitions from regression uncertainty: a new
  regression-based machine learning approach for automated detection of phases
  of matter}.
\newblock \emph{New Journal of Physics}, 25\penalty0 (8):\penalty0 083037,
  2023.
\newblock \doi{10.1088/1367-2630/acef4e}.
\newblock URL \url{https://dx.doi.org/10.1088/1367-2630/acef4e}.

\bibitem[Haller et~al.(2015)Haller, Hudson, Kelly, Cotta, Peaudecerf, Bruce,
  and Kuhr]{Haller2015}
Elmar Haller, James Hudson, Andrew Kelly, Dylan~A. Cotta, Bruno Peaudecerf,
  Graham~D. Bruce, and Stefan Kuhr.
\newblock {Single-atom imaging of fermions in a quantum-gas microscope}.
\newblock \emph{Nature Physics}, 11\penalty0 (9):\penalty0 738--742, 2015.
\newblock \doi{10.1038/nphys3403}.
\newblock URL \url{https://doi.org/10.1038/nphys3403}.

\bibitem[Hilker et~al.(2017)Hilker, Salomon, Grusdt, Omran, Boll, Demler,
  Bloch, and Gross]{Hilker2017}
Timon~A. Hilker, Guillaume Salomon, Fabian Grusdt, Ahmed Omran, Martin Boll,
  Eugene Demler, Immanuel Bloch, and Christian Gross.
\newblock {Revealing hidden antiferromagnetic correlations in doped Hubbard
  chains via string correlators}.
\newblock \emph{Science}, 357\penalty0 (6350):\penalty0 484--487, 2017.
\newblock \doi{10.1126/science.aam8990}.
\newblock URL \url{https://www.science.org/doi/abs/10.1126/science.aam8990}.

\bibitem[Holanda and Griffith(2020)]{Holanda2020}
N.~L. Holanda and M.~A.~R. Griffith.
\newblock {Machine learning topological phases in real space}.
\newblock \emph{Phys. Rev. B}, 102:\penalty0 054107, Aug 2020.
\newblock \doi{10.1103/PhysRevB.102.054107}.
\newblock URL \url{https://link.aps.org/doi/10.1103/PhysRevB.102.054107}.

\bibitem[Hradil(1997)]{Hradil1997}
Z.~Hradil.
\newblock {Quantum-state estimation}.
\newblock \emph{Phys. Rev. A}, 55:\penalty0 R1561--R1564, Mar 1997.
\newblock \doi{10.1103/PhysRevA.55.R1561}.
\newblock URL \url{https://link.aps.org/doi/10.1103/PhysRevA.55.R1561}.

\bibitem[Hu et~al.(2017)Hu, Singh, and Scalettar]{Hu2017}
Wenjian Hu, Rajiv R.~P. Singh, and Richard~T. Scalettar.
\newblock {Discovering phases, phase transitions, and crossovers through
  unsupervised machine learning: A critical examination}.
\newblock \emph{Phys. Rev. E}, 95:\penalty0 062122, Jun 2017.
\newblock \doi{10.1103/PhysRevE.95.062122}.
\newblock URL \url{https://link.aps.org/doi/10.1103/PhysRevE.95.062122}.

\bibitem[Huang et~al.(2022)Huang, Kueng, Torlai, Albert, and
  Preskill]{Huang2022}
Hsin-Yuan Huang, Richard Kueng, Giacomo Torlai, Victor~V. Albert, and John
  Preskill.
\newblock {Provably efficient machine learning for quantum many-body problems}.
\newblock \emph{Science}, 377\penalty0 (6613):\penalty0 eabk3333, 2022.
\newblock \doi{10.1126/science.abk3333}.
\newblock URL \url{https://www.science.org/doi/abs/10.1126/science.abk3333}.

\bibitem[Huembeli et~al.(2018)Huembeli, Dauphin, and Wittek]{Huembeli2018}
Patrick Huembeli, Alexandre Dauphin, and Peter Wittek.
\newblock {Identifying quantum phase transitions with adversarial neural
  networks}.
\newblock \emph{Phys. Rev. B}, 97:\penalty0 134109, Apr 2018.
\newblock \doi{10.1103/PhysRevB.97.134109}.
\newblock URL \url{https://link.aps.org/doi/10.1103/PhysRevB.97.134109}.

\bibitem[Ibarra-Garcia-Padilla et~al.(2024)Ibarra-Garcia-Padilla, Striegel,
  Scalettar, and Khatami]{Eduardo2024}
Eduardo Ibarra-Garcia-Padilla, Stephanie Striegel, Richard~T. Scalettar, and
  Ehsan Khatami.
\newblock Structural complexity of snapshots of two-dimensional fermi-hubbard
  systems.
\newblock \emph{Phys. Rev. A}, 109:\penalty0 053304, May 2024.
\newblock \doi{10.1103/PhysRevA.109.053304}.
\newblock URL \url{https://link.aps.org/doi/10.1103/PhysRevA.109.053304}.

\bibitem[Impertro et~al.(2023)Impertro, Wienand, Häfele, Raven, Hubele,
  Klostermann, Cabrera, Bloch, and Aidelsburger]{Impertro2023}
Alexander Impertro, Julian~F. Wienand, Sophie Häfele, Hendrik~von Raven, Scott
  Hubele, Till Klostermann, Cesar~R. Cabrera, Immanuel Bloch, and Monika
  Aidelsburger.
\newblock {An unsupervised deep learning algorithm for single-site
  reconstruction in quantum gas microscopes}.
\newblock \emph{Communications Physics}, 6\penalty0 (1):\penalty0 166, 2023.
\newblock \doi{10.1038/s42005-023-01287-w}.

\bibitem[Islam et~al.(2015)Islam, Ma, Preiss, Eric~Tai, Lukin, Rispoli, and
  Greiner]{Islam2015}
Rajibul Islam, Ruichao Ma, Philipp~M. Preiss, M.~Eric~Tai, Alexander Lukin,
  Matthew Rispoli, and Markus Greiner.
\newblock {Measuring entanglement entropy in a quantum many-body system}.
\newblock \emph{Nature}, 528\penalty0 (7580):\penalty0 77--83, 2015.
\newblock \doi{10.1038/nature15750}.
\newblock URL \url{https://doi.org/10.1038/nature15750}.

\bibitem[Johnston et~al.(2022)Johnston, Khatami, and Scalettar]{Johnston2022}
Steven Johnston, Ehsan Khatami, and Richard Scalettar.
\newblock A perspective on machine learning and data science for strongly
  correlated electron problems.
\newblock \emph{Carbon Trends}, 9:\penalty0 100231, 2022.
\newblock \doi{https://doi.org/10.1016/j.cartre.2022.100231}.
\newblock URL
  \url{https://www.sciencedirect.com/science/article/pii/S2667056922000876}.

\bibitem[Joshi et~al.(2023)Joshi, Kokail, van Bijnen, Kranzl, Zache, Blatt,
  Roos, and Zoller]{Joshi2023}
Manoj~K. Joshi, Christian Kokail, Rick van Bijnen, Florian Kranzl, Torsten~V.
  Zache, Rainer Blatt, Christian~F. Roos, and Peter Zoller.
\newblock Exploring large-scale entanglement in quantum simulation.
\newblock \emph{Nature}, 624\penalty0 (7992):\penalty0 539--544, 2023.
\newblock \doi{10.1038/s41586-023-06768-0}.
\newblock URL \url{https://doi.org/10.1038/s41586-023-06768-0}.

\bibitem[K{\"a}ming et~al.(2021)K{\"a}ming, Dawid, Kottmann, Lewenstein,
  Sengstock, Dauphin, and Weitenberg]{Kaeming2021}
Niklas K{\"a}ming, Anna Dawid, Korbinian Kottmann, Maciej Lewenstein, Klaus
  Sengstock, Alexandre Dauphin, and Christof Weitenberg.
\newblock {Unsupervised machine learning of topological phase transitions from
  experimental data}.
\newblock \emph{Machine Learning: Science and Technology}, 2\penalty0
  (3):\penalty0 035037, 2021.
\newblock \doi{10.1088/2632-2153/abffe7}.
\newblock URL \url{https://dx.doi.org/10.1088/2632-2153/abffe7}.

\bibitem[Kashiwa et~al.(2019)Kashiwa, Kikuchi, and Tomiya]{Kashiwa2019}
Kouji Kashiwa, Yuta Kikuchi, and Akio Tomiya.
\newblock {Phase transition encoded in neural network}.
\newblock \emph{Progress of Theoretical and Experimental Physics},
  2019\penalty0 (8), 2019.
\newblock \doi{10.1093/ptep/ptz082}.

\bibitem[Keesling et~al.(2019)Keesling, Omran, Levine, Bernien, Pichler, Choi,
  Samajdar, Schwartz, Silvi, Sachdev, Zoller, Endres, Greiner, Vuleti{\'c}, and
  Lukin]{Keesling2019}
Alexander Keesling, Ahmed Omran, Harry Levine, Hannes Bernien, Hannes Pichler,
  Soonwon Choi, Rhine Samajdar, Sylvain Schwartz, Pietro Silvi, Subir Sachdev,
  Peter Zoller, Manuel Endres, Markus Greiner, Vladan Vuleti{\'c}, and
  Mikhail~D. Lukin.
\newblock {Quantum Kibble--Zurek mechanism and critical dynamics on a
  programmable Rydberg simulator}.
\newblock \emph{Nature}, 568\penalty0 (7751):\penalty0 207--211, 2019.
\newblock \doi{10.1038/s41586-019-1070-1}.
\newblock URL \url{https://doi.org/10.1038/s41586-019-1070-1}.

\bibitem[Khatami et~al.(2020)Khatami, Guardado-Sanchez, Spar, Carrasquilla,
  Bakr, and Scalettar]{Khatami2020}
Ehsan Khatami, Elmer Guardado-Sanchez, Benjamin~M. Spar, Juan~Felipe
  Carrasquilla, Waseem~S. Bakr, and Richard~T. Scalettar.
\newblock {Visualizing strange metallic correlations in the two-dimensional
  Fermi-Hubbard model with artificial intelligence}.
\newblock \emph{Phys. Rev. A}, 102:\penalty0 033326, Sep 2020.
\newblock \doi{10.1103/PhysRevA.102.033326}.
\newblock URL \url{https://link.aps.org/doi/10.1103/PhysRevA.102.033326}.

\bibitem[Kim and Kim(2018)]{Kim2018}
Dongkyu Kim and Dong-Hee Kim.
\newblock {Smallest neural network to learn the Ising criticality}.
\newblock \emph{Phys. Rev. E}, 98:\penalty0 022138, Aug 2018.
\newblock \doi{10.1103/PhysRevE.98.022138}.
\newblock URL \url{https://link.aps.org/doi/10.1103/PhysRevE.98.022138}.

\bibitem[Kokail et~al.(2021{\natexlab{a}})Kokail, Sundar, Zache, Elben,
  Vermersch, Dalmonte, van Bijnen, and Zoller]{Kokail2021}
Christian Kokail, Bhuvanesh Sundar, Torsten~V. Zache, Andreas Elben, Benot
  Vermersch, Marcello Dalmonte, Rick van Bijnen, and Peter Zoller.
\newblock Quantum variational learning of the entanglement hamiltonian.
\newblock \emph{Phys. Rev. Lett.}, 127:\penalty0 170501, Oct
  2021{\natexlab{a}}.
\newblock \doi{10.1103/PhysRevLett.127.170501}.
\newblock URL \url{https://link.aps.org/doi/10.1103/PhysRevLett.127.170501}.

\bibitem[Kokail et~al.(2021{\natexlab{b}})Kokail, van Bijnen, Elben, Vermersch,
  and Zoller]{Kokail2021Ent}
Christian Kokail, Rick van Bijnen, Andreas Elben, Benoit Vermersch, and Peter
  Zoller.
\newblock Entanglement hamiltonian tomography in quantum simulation.
\newblock \emph{Nature Physics}, 17\penalty0 (8):\penalty0 936--942,
  2021{\natexlab{b}}.
\newblock \doi{10.1038/s41567-021-01260-w}.
\newblock URL \url{https://doi.org/10.1038/s41567-021-01260-w}.

\bibitem[Kornjaca et~al.(2024)Kornjaca, Hu, Zhao, Wurtz, Weinberg, Hamdan,
  Zhdanov, Cantu, Zhou, Bravo, Bagnall, Basham, Campo, Choukri, DeAngelo,
  Frederick, Haines, Hammett, Hsu, Hu, Huber, Jepsen, Jia, Karolyshyn, Kwon,
  Long, Lopatin, Lukin, Macri, Markovic, Martinez-Martinez, Meng, Ostroumov,
  Paquette, Robinson, Rodriguez, Singh, Sinha, Thoreen, Wan, Waxman-Lenz, Wong,
  Wu, Lopes, Boger, Gemelke, Kitagawa, Keesling, Gao, Bylinskii, Yelin, Liu,
  and Wang]{Kornjaca2024}
Milan Kornjaca, Hong-Ye Hu, Chen Zhao, Jonathan Wurtz, Phillip Weinberg, Majd
  Hamdan, Andrii Zhdanov, Sergio~H. Cantu, Hengyun Zhou, Rodrigo~Araiza Bravo,
  Kevin Bagnall, James~I. Basham, Joseph Campo, Adam Choukri, Robert DeAngelo,
  Paige Frederick, David Haines, Julian Hammett, Ning Hsu, Ming-Guang Hu,
  Florian Huber, Paul~Niklas Jepsen, Ningyuan Jia, Thomas Karolyshyn, Minho
  Kwon, John Long, Jonathan Lopatin, Alexander Lukin, Tommaso Macri, Ognjen
  Markovic, Luis~A. Martinez-Martinez, Xianmei Meng, Evgeny Ostroumov, David
  Paquette, John Robinson, Pedro~Sales Rodriguez, Anshuman Singh, Nandan Sinha,
  Henry Thoreen, Noel Wan, Daniel Waxman-Lenz, Tak Wong, Kai-Hsin Wu, Pedro
  L.~S. Lopes, Yuval Boger, Nathan Gemelke, Takuya Kitagawa, Alexander
  Keesling, Xun Gao, Alexei Bylinskii, Susanne~F. Yelin, Fangli Liu, and
  Sheng-Tao Wang.
\newblock {Large-scale quantum reservoir learning with an analog quantum
  computer}, 2024.
\newblock URL \url{https://arxiv.org/abs/2407.02553}.

\bibitem[Kottmann et~al.(2020)Kottmann, Huembeli, Lewenstein, and
  Acin]{Kottmann2020}
Korbinian Kottmann, Patrick Huembeli, Maciej Lewenstein, and Antonio Acin.
\newblock {Unsupervised Phase Discovery with Deep Anomaly Detection}.
\newblock \emph{Phys. Rev. Lett.}, 125:\penalty0 170603, Oct 2020.
\newblock \doi{10.1103/PhysRevLett.125.170603}.
\newblock URL \url{https://link.aps.org/doi/10.1103/PhysRevLett.125.170603}.

\bibitem[Krenn et~al.(2016)Krenn, Malik, Fickler, Lapkiewicz, and
  Zeilinger]{Krenn2016}
Mario Krenn, Mehul Malik, Robert Fickler, Radek Lapkiewicz, and Anton
  Zeilinger.
\newblock {Automated Search for new Quantum Experiments}.
\newblock \emph{Phys. Rev. Lett.}, 116:\penalty0 090405, Mar 2016.
\newblock \doi{10.1103/PhysRevLett.116.090405}.
\newblock URL \url{https://link.aps.org/doi/10.1103/PhysRevLett.116.090405}.

\bibitem[Kumpilov et~al.(2024)Kumpilov, Pershin, Cojocaru, Khlebnikov, Pyrkh,
  Rudnev, Fedotova, Khoruzhii, Aksentsev, Gaifutdinov, Zykova, Tsyganok, and
  Akimov]{Kumpilov2024}
D.~A. Kumpilov, D.~A. Pershin, I.~S. Cojocaru, V.~A. Khlebnikov, I.~A. Pyrkh,
  A.~E. Rudnev, E.~A. Fedotova, K.~A. Khoruzhii, P.~A. Aksentsev, D.~V.
  Gaifutdinov, A.~K. Zykova, V.~V. Tsyganok, and A.~V. Akimov.
\newblock {Inspiration from machine learning on the example of optimization of
  the Bose-Einstein condensate of thulium atoms in a 1064-nm trap}.
\newblock \emph{Physical Review A}, 109\penalty0 (3):\penalty0 033313, 2024.
\newblock ISSN 2469-9926.
\newblock \doi{10.1103/physreva.109.033313}.

\bibitem[Lange et~al.(2023)Lange, Kebri{\v{c}}, Buser, Schollw{\"{o}}ck,
  Grusdt, and Bohrdt]{Lange2023}
Hannah Lange, Matja{\v{z}} Kebri{\v{c}}, Maximilian Buser, Ulrich
  Schollw{\"{o}}ck, Fabian Grusdt, and Annabelle Bohrdt.
\newblock {Adaptive {Q}uantum {S}tate {T}omography with {A}ctive {L}earning}.
\newblock \emph{{Quantum}}, 7:\penalty0 1129, October 2023.
\newblock ISSN 2521-327X.
\newblock \doi{10.22331/q-2023-10-09-1129}.
\newblock URL \url{https://doi.org/10.22331/q-2023-10-09-1129}.

\bibitem[Lange et~al.(2024)Lange, Van~de Walle, Abedinnia, and
  Bohrdt]{Lange2024Rev}
Hannah Lange, Anka Van~de Walle, Atiye Abedinnia, and Annabelle Bohrdt.
\newblock {From architectures to applications: a review of neural quantum
  states}.
\newblock \emph{Quantum Science and Technology}, 9\penalty0 (4):\penalty0
  040501, 2024.
\newblock \doi{10.1088/2058-9565/ad7168}.
\newblock URL \url{https://dx.doi.org/10.1088/2058-9565/ad7168}.

\bibitem[Lange et~al.(2025)Lange, Bornet, Emperauger, Chen, Lahaye, Kienle,
  Browaeys, and Bohrdt]{Lange2025hybrid}
Hannah Lange, Guillaume Bornet, Gabriel Emperauger, Cheng Chen, Thierry Lahaye,
  Stefan Kienle, Antoine Browaeys, and Annabelle Bohrdt.
\newblock {Transformer neural networks and quantum simulators: a hybrid
  approach for simulating strongly correlated systems}.
\newblock \emph{{Quantum}}, 9:\penalty0 1675, March 2025.
\newblock ISSN 2521-327X.
\newblock \doi{10.22331/q-2025-03-26-1675}.
\newblock URL \url{https://doi.org/10.22331/q-2025-03-26-1675}.

\bibitem[Lee and il~Shin(2025)]{Lee2025dual}
Kyuhwan Lee and Yong il~Shin.
\newblock Dual-species atomic absorption image reconstruction using deep neural
  networks, 2025.
\newblock URL \url{https://arxiv.org/abs/2508.12120}.

\bibitem[Lee et~al.(2006)Lee, Nagaosa, and Wen]{Lee2006}
Patrick~A. Lee, Naoto Nagaosa, and Xiao-Gang Wen.
\newblock {Doping a Mott insulator: Physics of high-temperature
  superconductivity}.
\newblock \emph{Rev. Mod. Phys.}, 78:\penalty0 17--85, Jan 2006.
\newblock \doi{10.1103/RevModPhys.78.17}.
\newblock URL \url{https://link.aps.org/doi/10.1103/RevModPhys.78.17}.

\bibitem[Lidiak and Gong(2020)]{Lidiak2020}
Alexander Lidiak and Zhexuan Gong.
\newblock {Unsupervised Machine Learning of Quantum Phase Transitions Using
  Diffusion Maps}.
\newblock \emph{Phys. Rev. Lett.}, 125:\penalty0 225701, Nov 2020.
\newblock \doi{10.1103/PhysRevLett.125.225701}.
\newblock URL \url{https://link.aps.org/doi/10.1103/PhysRevLett.125.225701}.

\bibitem[Liu et~al.(2019)Liu, Greitemann, and Pollet]{Liu2019ML}
Ke~Liu, Jonas Greitemann, and Lode Pollet.
\newblock {Learning multiple order parameters with interpretable machines}.
\newblock \emph{Phys. Rev. B}, 99:\penalty0 104410, Mar 2019.
\newblock \doi{10.1103/PhysRevB.99.104410}.
\newblock URL \url{https://link.aps.org/doi/10.1103/PhysRevB.99.104410}.

\bibitem[Liu et~al.(2021)Liu, Sadoune, Rao, Greitemann, and Pollet]{Liu2021}
Ke~Liu, Nicolas Sadoune, Nihal Rao, Jonas Greitemann, and Lode Pollet.
\newblock {Revealing the phase diagram of Kitaev materials by machine learning:
  Cooperation and competition between spin liquids}.
\newblock \emph{Phys. Rev. Res.}, 3:\penalty0 023016, Apr 2021.
\newblock \doi{10.1103/PhysRevResearch.3.023016}.
\newblock URL \url{https://link.aps.org/doi/10.1103/PhysRevResearch.3.023016}.

\bibitem[Liu and van Nieuwenburg(2018)]{Liu2018}
Ye-Hua Liu and Evert P.~L. van Nieuwenburg.
\newblock {Discriminative Cooperative Networks for Detecting Phase
  Transitions}.
\newblock \emph{Phys. Rev. Lett.}, 120:\penalty0 176401, Apr 2018.
\newblock \doi{10.1103/PhysRevLett.120.176401}.
\newblock URL \url{https://link.aps.org/doi/10.1103/PhysRevLett.120.176401}.

\bibitem[Lode et~al.(2021)Lode, Lin, B\"uttner, Papariello, L\'ev\^eque,
  Chitra, Tsatsos, Jaksch, and Molignini]{Lode2021}
Axel U.~J. Lode, Rui Lin, Miriam B\"uttner, Luca Papariello, Camille
  L\'ev\^eque, R.~Chitra, Marios~C. Tsatsos, Dieter Jaksch, and Paolo
  Molignini.
\newblock {Optimized observable readout from single-shot images of ultracold
  atoms via machine learning}.
\newblock \emph{Phys. Rev. A}, 104:\penalty0 L041301, Oct 2021.
\newblock \doi{10.1103/PhysRevA.104.L041301}.
\newblock URL \url{https://link.aps.org/doi/10.1103/PhysRevA.104.L041301}.

\bibitem[Lozano-G\'omez et~al.(2022)Lozano-G\'omez, Pereira, and
  Gingras]{LazGo2022}
Daniel Lozano-G\'omez, Darren Pereira, and Michel J.~P. Gingras.
\newblock {Unsupervised machine learning of quenched gauge symmetries: A
  proof-of-concept demonstration}.
\newblock \emph{Phys. Rev. Res.}, 4:\penalty0 043118, Nov 2022.
\newblock \doi{10.1103/PhysRevResearch.4.043118}.
\newblock URL \url{https://link.aps.org/doi/10.1103/PhysRevResearch.4.043118}.

\bibitem[Lu et~al.(2025)Lu, Jiao, Wolinski, Kornja{\v c}a, Hu, Cantu, Liu,
  Yelin, and Wang]{Lu2024hybrid}
Jonathan~Z Lu, Lucy Jiao, Kristina Wolinski, Milan Kornja{\v c}a, Hong-Ye Hu,
  Sergio Cantu, Fangli Liu, Susanne~F Yelin, and Sheng-Tao Wang.
\newblock {Digital--analog quantum learning on Rydberg atom arrays}.
\newblock \emph{Quantum Science and Technology}, 10\penalty0 (1):\penalty0
  015038, 2025.
\newblock \doi{10.1088/2058-9565/ad9177}.
\newblock URL \url{https://dx.doi.org/10.1088/2058-9565/ad9177}.

\bibitem[Lustig et~al.(2020)Lustig, Yair, Talmon, and Segev]{Lustig2020}
Eran Lustig, Or~Yair, Ronen Talmon, and Mordechai Segev.
\newblock Identifying topological phase transitions in experiments using
  manifold learning.
\newblock \emph{Phys. Rev. Lett.}, 125:\penalty0 127401, Sep 2020.
\newblock \doi{10.1103/PhysRevLett.125.127401}.
\newblock URL \url{https://link.aps.org/doi/10.1103/PhysRevLett.125.127401}.

\bibitem[Ma\ifmmode~\acute{s}\else \'{s}\fi{}ka
  et~al.(2008)Ma\ifmmode~\acute{s}\else \'{s}\fi{}ka,
  Lema\ifmmode~\acute{n}\else \'{n}\fi{}ski, Freericks, and
  Williams]{Maska2008}
M.~M. Ma\ifmmode~\acute{s}\else \'{s}\fi{}ka, R.~Lema\ifmmode~\acute{n}\else
  \'{n}\fi{}ski, J.~K. Freericks, and C.~J. Williams.
\newblock {Pattern Formation in Mixtures of Ultracold Atoms in Optical
  Lattices}.
\newblock \emph{Phys. Rev. Lett.}, 101:\penalty0 060404, Aug 2008.
\newblock \doi{10.1103/PhysRevLett.101.060404}.
\newblock URL \url{https://link.aps.org/doi/10.1103/PhysRevLett.101.060404}.

\bibitem[Maskara et~al.(2022)Maskara, Buchhold, Endres, and van
  Nieuwenburg]{Maskara2022}
N.~Maskara, M.~Buchhold, M.~Endres, and E.~van Nieuwenburg.
\newblock {Learning algorithm reflecting universal scaling behavior near phase
  transitions}.
\newblock \emph{Phys. Rev. Res.}, 4:\penalty0 L022032, May 2022.
\newblock \doi{10.1103/PhysRevResearch.4.L022032}.
\newblock URL \url{https://link.aps.org/doi/10.1103/PhysRevResearch.4.L022032}.

\bibitem[Melnikov et~al.(2018)Melnikov, Nautrup, Krenn, Dunjko, Tiersch,
  Zeilinger, and Briegel]{Melnikov2018}
Alexey~A. Melnikov, Hendrik~Poulsen Nautrup, Mario Krenn, Vedran Dunjko, Markus
  Tiersch, Anton Zeilinger, and Hans~J. Briegel.
\newblock {Active learning machine learns to create new quantum experiments}.
\newblock \emph{Proceedings of the National Academy of Sciences}, 115\penalty0
  (6):\penalty0 1221--1226, 2018.
\newblock ISSN 0027-8424.
\newblock \doi{10.1073/pnas.1714936115}.

\bibitem[Mendes-Santos et~al.(2021)Mendes-Santos, Turkeshi, Dalmonte, and
  Rodriguez]{MendesSantos2021}
T.~Mendes-Santos, X.~Turkeshi, M.~Dalmonte, and Alex Rodriguez.
\newblock {Unsupervised Learning Universal Critical Behavior via the Intrinsic
  Dimension}.
\newblock \emph{Phys. Rev. X}, 11:\penalty0 011040, Feb 2021.
\newblock \doi{10.1103/PhysRevX.11.011040}.
\newblock URL \url{https://link.aps.org/doi/10.1103/PhysRevX.11.011040}.

\bibitem[Mendes-Santos et~al.(2024)Mendes-Santos, Schmitt, Angelone, Rodriguez,
  Scholl, Williams, Barredo, Lahaye, Browaeys, Heyl, and Dalmonte]{MeSa2024}
T.~Mendes-Santos, M.~Schmitt, A.~Angelone, A.~Rodriguez, P.~Scholl, H.~J.
  Williams, D.~Barredo, T.~Lahaye, A.~Browaeys, M.~Heyl, and M.~Dalmonte.
\newblock {Wave-Function Network Description and Kolmogorov Complexity of
  Quantum Many-Body Systems}.
\newblock \emph{Phys. Rev. X}, 14:\penalty0 021029, May 2024.
\newblock \doi{10.1103/PhysRevX.14.021029}.
\newblock URL \url{https://link.aps.org/doi/10.1103/PhysRevX.14.021029}.

\bibitem[Miles et~al.(2021)Miles, Bohrdt, Wu, Chiu, Xu, Ji, Greiner,
  Weinberger, Demler, and Kim]{Miles2021}
Cole Miles, Annabelle Bohrdt, Ruihan Wu, Christie Chiu, Muqing Xu, Geoffrey Ji,
  Markus Greiner, Kilian~Q. Weinberger, Eugene Demler, and Eun-Ah Kim.
\newblock {Correlator convolutional neural networks as an interpretable
  architecture for image-like quantum matter data}.
\newblock \emph{Nature Communications}, 12\penalty0 (1):\penalty0 3905, 2021.
\newblock \doi{10.1038/s41467-021-23952-w}.
\newblock URL \url{https://doi.org/10.1038/s41467-021-23952-w}.

\bibitem[Miles et~al.(2023)Miles, Samajdar, Ebadi, Wang, Pichler, Sachdev,
  Lukin, Greiner, Weinberger, and Kim]{Miles2023}
Cole Miles, Rhine Samajdar, Sepehr Ebadi, Tout~T. Wang, Hannes Pichler, Subir
  Sachdev, Mikhail~D. Lukin, Markus Greiner, Kilian~Q. Weinberger, and Eun-Ah
  Kim.
\newblock {Machine learning discovery of new phases in programmable quantum
  simulator snapshots}.
\newblock \emph{Phys. Rev. Res.}, 5:\penalty0 013026, Jan 2023.
\newblock \doi{10.1103/PhysRevResearch.5.013026}.
\newblock URL \url{https://link.aps.org/doi/10.1103/PhysRevResearch.5.013026}.

\bibitem[Milson et~al.(2023)Milson, Tashchilina, Ooi, Czarnecka, Ahmad, and
  LeBlanc]{Milson2023}
N~Milson, A~Tashchilina, T~Ooi, A~Czarnecka, Z~F Ahmad, and L~J LeBlanc.
\newblock {High-dimensional reinforcement learning for optimization and control
  of ultracold quantum gases}.
\newblock \emph{Machine Learning: Science and Technology}, 4\penalty0
  (4):\penalty0 045057, 2023.
\newblock \doi{10.1088/2632-2153/ad1437}.

\bibitem[Min et~al.(2025)Min, Li, Zhong, Xuan, Lin, Feng, and Li]{Min2025}
Yueyang Min, Ziliang Li, Yi~Zhong, Jia-An Xuan, Jian Lin, Lei Feng, and
  Xiaopeng Li.
\newblock {Efficient Preparation of Fermionic Superfluids in an Optical Dipole
  Trap through Reinforcement Learning}, 2025.
\newblock URL \url{https://arxiv.org/abs/2507.12152}.

\bibitem[Ming et~al.(2019)Ming, Lin, Bartlett, and Zhang]{Ming2019}
Yurui Ming, Chin-Teng Lin, Stephen~D. Bartlett, and Wei-Wei Zhang.
\newblock {Quantum topology identification with deep neural networks and
  quantum walks}.
\newblock \emph{npj Computational Materials}, 5\penalty0 (1):\penalty0 88,
  2019.
\newblock \doi{10.1038/s41524-019-0224-x}.
\newblock URL \url{https://doi.org/10.1038/s41524-019-0224-x}.

\bibitem[Morawetz et~al.(2021)Morawetz, De~Vlugt, Carrasquilla, and
  Melko]{Morawetz2021}
Stewart Morawetz, Isaac J.~S. De~Vlugt, Juan Carrasquilla, and Roger~G. Melko.
\newblock {U(1)-symmetric recurrent neural networks for quantum state
  reconstruction}.
\newblock \emph{Phys. Rev. A}, 104:\penalty0 012401, Jul 2021.
\newblock \doi{10.1103/PhysRevA.104.012401}.
\newblock URL \url{https://link.aps.org/doi/10.1103/PhysRevA.104.012401}.

\bibitem[Morishita and Todo(2022)]{Morishita2022}
Tomoyuki Morishita and Synge Todo.
\newblock {Randomized-Gauge Test for Machine Learning of Ising Model Order
  Parameter}.
\newblock \emph{Journal of the Physical Society of Japan}, 91\penalty0
  (4):\penalty0 044001, 2022.
\newblock ISSN 0031-9015.
\newblock \doi{10.7566/jpsj.91.044001}.

\bibitem[Moss et~al.(2024)Moss, Ebadi, Wang, Semeghini, Bohrdt, Lukin, and
  Melko]{Moss2024}
M.~Schuyler Moss, Sepehr Ebadi, Tout~T. Wang, Giulia Semeghini, Annabelle
  Bohrdt, Mikhail~D. Lukin, and Roger~G. Melko.
\newblock {Enhancing variational Monte Carlo simulations using a programmable
  quantum simulator}.
\newblock \emph{Phys. Rev. A}, 109:\penalty0 032410, Mar 2024.
\newblock \doi{10.1103/PhysRevA.109.032410}.
\newblock URL \url{https://link.aps.org/doi/10.1103/PhysRevA.109.032410}.

\bibitem[Muzzi et~al.(2024)Muzzi, Cortes, Bhakuni, Jeli\ifmmode~\acute{c}\else
  \'{c}\fi{}, Gambassi, Dalmonte, and Verdel]{Muzzi2024}
Cristiano Muzzi, Ronald~Santiago Cortes, Devendra~Singh Bhakuni, Asja
  Jeli\ifmmode~\acute{c}\else \'{c}\fi{}, Andrea Gambassi, Marcello Dalmonte,
  and Roberto Verdel.
\newblock {Principal component analysis of absorbing state phase transitions}.
\newblock \emph{Phys. Rev. E}, 110:\penalty0 064121, Dec 2024.
\newblock \doi{10.1103/PhysRevE.110.064121}.
\newblock URL \url{https://link.aps.org/doi/10.1103/PhysRevE.110.064121}.

\bibitem[Neugebauer et~al.(2020)Neugebauer, Fischer, J\"ager, Czischek, Jochim,
  Weidem\"uller, and G\"arttner]{Neugebauer2020}
Marcel Neugebauer, Laurin Fischer, Alexander J\"ager, Stefanie Czischek, Selim
  Jochim, Matthias Weidem\"uller, and Martin G\"arttner.
\newblock {Neural-network quantum state tomography in a two-qubit experiment}.
\newblock \emph{Phys. Rev. A}, 102:\penalty0 042604, Oct 2020.
\newblock \doi{10.1103/PhysRevA.102.042604}.
\newblock URL \url{https://link.aps.org/doi/10.1103/PhysRevA.102.042604}.

\bibitem[Neupert et~al.(2022)Neupert, Fischer, Greplova, Choo, and
  Denner]{NeupertReview}
Titus Neupert, Mark~H Fischer, Eliska Greplova, Kenny Choo, and M.~Michael
  Denner.
\newblock {Introduction to Machine Learning for the Sciences}, 2022.
\newblock URL \url{https://arxiv.org/abs/2102.04883}.

\bibitem[Noh et~al.(2017)Noh, Huang, Leykam, Chong, Chen, and
  Rechtsman]{Noh2017}
Jiho Noh, Sheng Huang, Daniel Leykam, Y.~D. Chong, Kevin~P. Chen, and Mikael~C.
  Rechtsman.
\newblock Experimental observation of optical weyl points and fermi arc-like
  surface states.
\newblock \emph{Nature Physics}, 13\penalty0 (6):\penalty0 611--617, 2017.
\newblock \doi{10.1038/nphys4072}.
\newblock URL \url{https://doi.org/10.1038/nphys4072}.

\bibitem[Olsacher et~al.(2025)Olsacher, Kraft, Kokail, Kraus, and
  Zoller]{Olsacher2025}
Tobias Olsacher, Tristan Kraft, Christian Kokail, Barbara Kraus, and Peter
  Zoller.
\newblock Hamiltonian and liouvillian learning in weakly-dissipative quantum
  many-body systems.
\newblock \emph{Quantum Science and Technology}, 10\penalty0 (1):\penalty0
  015065, 2025.
\newblock \doi{10.1088/2058-9565/ad9ed5}.
\newblock URL \url{https://dx.doi.org/10.1088/2058-9565/ad9ed5}.

\bibitem[Panda et~al.(2023)Panda, Verdel, Rodriguez, Sun, Bianconi, and
  Dalmonte]{Panda2023}
Rajat~K. Panda, Roberto Verdel, Alex Rodriguez, Hanlin Sun, Ginestra Bianconi,
  and Marcello Dalmonte.
\newblock {Non-parametric learning critical behavior in Ising partition
  functions: PCA entropy and intrinsic dimension}.
\newblock \emph{SciPost Phys. Core}, 6:\penalty0 086, 2023.
\newblock \doi{10.21468/SciPostPhysCore.6.4.086}.
\newblock URL \url{https://scipost.org/10.21468/SciPostPhysCore.6.4.086}.

\bibitem[Parsons et~al.(2015)Parsons, Huber, Mazurenko, Chiu, Setiawan,
  Wooley-Brown, Blatt, and Greiner]{Parsons2015}
Maxwell~F. Parsons, Florian Huber, Anton Mazurenko, Christie~S. Chiu, Widagdo
  Setiawan, Katherine Wooley-Brown, Sebastian Blatt, and Markus Greiner.
\newblock {Site-Resolved Imaging of Fermionic $^{6}\mathrm{Li}$ in an Optical
  Lattice}.
\newblock \emph{Phys. Rev. Lett.}, 114:\penalty0 213002, May 2015.
\newblock \doi{10.1103/PhysRevLett.114.213002}.
\newblock URL \url{https://link.aps.org/doi/10.1103/PhysRevLett.114.213002}.

\bibitem[Patel et~al.(2022)Patel, Merali, and Wetzel]{Patel2022}
Zakaria Patel, Ejaaz Merali, and Sebastian~J Wetzel.
\newblock {Unsupervised learning of Rydberg atom array phase diagram with
  Siamese neural networks}.
\newblock \emph{New Journal of Physics}, 24\penalty0 (11):\penalty0 113021,
  2022.
\newblock \doi{10.1088/1367-2630/ac9c7a}.
\newblock URL \url{https://dx.doi.org/10.1088/1367-2630/ac9c7a}.

\bibitem[Picard et~al.(2020)Picard, Mark, Ferlaino, and Bijnen]{Picard2020}
Lewis R~B Picard, Manfred~J Mark, Francesca Ferlaino, and Rick~van Bijnen.
\newblock {Deep learning-assisted classification of site-resolved quantum gas
  microscope images}.
\newblock \emph{Measurement Science and Technology}, 31\penalty0 (2):\penalty0
  025201, 2020.
\newblock ISSN 0957-0233.
\newblock \doi{10.1088/1361-6501/ab44d8}.

\bibitem[Ponte and Melko(2017)]{Ponte2017}
Pedro Ponte and Roger~G. Melko.
\newblock {Kernel methods for interpretable machine learning of order
  parameters}.
\newblock \emph{Phys. Rev. B}, 96:\penalty0 205146, Nov 2017.
\newblock \doi{10.1103/PhysRevB.96.205146}.
\newblock URL \url{https://link.aps.org/doi/10.1103/PhysRevB.96.205146}.

\bibitem[Qi and Ranard(2019)]{Qi2019}
Xiao-Liang Qi and Daniel Ranard.
\newblock {Determining a local {H}amiltonian from a single eigenstate}.
\newblock \emph{{Quantum}}, 3:\penalty0 159, July 2019.
\newblock ISSN 2521-327X.
\newblock \doi{10.22331/q-2019-07-08-159}.
\newblock URL \url{https://doi.org/10.22331/q-2019-07-08-159}.

\bibitem[Reinschmidt et~al.(2024)Reinschmidt, Fortágh, Günther, and
  Volchkov]{Reinschmidt2024}
Malte Reinschmidt, József Fortágh, Andreas Günther, and Valentin~V.
  Volchkov.
\newblock {Reinforcement learning in cold atom experiments}.
\newblock \emph{Nature Communications}, 15\penalty0 (1):\penalty0 8532, 2024.
\newblock \doi{10.1038/s41467-024-52775-8}.

\bibitem[Rem et~al.(2019)Rem, K{\"a}ming, Tarnowski, Asteria, Fl{\"a}schner,
  Becker, Sengstock, and Weitenberg]{Rem2019}
Benno~S. Rem, Niklas K{\"a}ming, Matthias Tarnowski, Luca Asteria, Nick
  Fl{\"a}schner, Christoph Becker, Klaus Sengstock, and Christof Weitenberg.
\newblock {Identifying quantum phase transitions using artificial neural
  networks on experimental data}.
\newblock \emph{Nature Physics}, 15\penalty0 (9):\penalty0 917--920, 2019.
\newblock \doi{10.1038/s41567-019-0554-0}.
\newblock URL \url{https://doi.org/10.1038/s41567-019-0554-0}.

\bibitem[Rispoli et~al.(2019)Rispoli, Lukin, Schittko, Kim, Tai, L{\'e}onard,
  and Greiner]{Rispoli2019}
Matthew Rispoli, Alexander Lukin, Robert Schittko, Sooshin Kim, M.~Eric Tai,
  Julian L{\'e}onard, and Markus Greiner.
\newblock {Quantum critical behaviour at the many-body localization
  transition}.
\newblock \emph{Nature}, 573\penalty0 (7774):\penalty0 385--389, 2019.
\newblock \doi{10.1038/s41586-019-1527-2}.
\newblock URL \url{https://doi.org/10.1038/s41586-019-1527-2}.

\bibitem[Rocchetto et~al.(2018)Rocchetto, Grant, Strelchuk, Carleo, and
  Severini]{Rocchetto2018}
Andrea Rocchetto, Edward Grant, Sergii Strelchuk, Giuseppe Carleo, and Simone
  Severini.
\newblock {Learning hard quantum distributions with variational autoencoders}.
\newblock \emph{npj Quantum Information}, 4\penalty0 (1):\penalty0 28, 2018.
\newblock \doi{10.1038/s41534-018-0077-z}.
\newblock URL \url{https://doi.org/10.1038/s41534-018-0077-z}.

\bibitem[Rodriguez-Nieva and Scheurer(2019)]{RodNiev2019}
Joaquin~F. Rodriguez-Nieva and Mathias~S. Scheurer.
\newblock {Identifying topological order through unsupervised machine
  learning}.
\newblock \emph{Nature Physics}, 15\penalty0 (8):\penalty0 790--795, 2019.
\newblock \doi{10.1038/s41567-019-0512-x}.
\newblock URL \url{https://doi.org/10.1038/s41567-019-0512-x}.

\bibitem[Rosson et~al.(2020)Rosson, Kiffner, Mur-Petit, and Jaksch]{Rosson2020}
Paolo Rosson, Martin Kiffner, Jordi Mur-Petit, and Dieter Jaksch.
\newblock {Characterizing the phase diagram of finite-size dipolar Bose-Hubbard
  systems}.
\newblock \emph{Phys. Rev. A}, 101:\penalty0 013616, Jan 2020.
\newblock \doi{10.1103/PhysRevA.101.013616}.
\newblock URL \url{https://link.aps.org/doi/10.1103/PhysRevA.101.013616}.

\bibitem[Sadoune et~al.(2023)Sadoune, Giudici, Liu, and Pollet]{Sadoune2023}
Nicolas Sadoune, Giuliano Giudici, Ke~Liu, and Lode Pollet.
\newblock {Unsupervised interpretable learning of phases from many-qubit
  systems}.
\newblock \emph{Phys. Rev. Res.}, 5:\penalty0 013082, Feb 2023.
\newblock \doi{10.1103/PhysRevResearch.5.013082}.
\newblock URL \url{https://link.aps.org/doi/10.1103/PhysRevResearch.5.013082}.

\bibitem[Sadoune et~al.(2024)Sadoune, Pogorelov, Edmunds, Giudici, Giudice,
  Marciniak, Ringbauer, Monz, and Pollet]{Sadoune2024}
Nicolas Sadoune, Ivan Pogorelov, Claire~L. Edmunds, Giuliano Giudici, Giacomo
  Giudice, Christian~D. Marciniak, Martin Ringbauer, Thomas Monz, and Lode
  Pollet.
\newblock {Learning symmetry-protected topological order from trapped-ion
  experiments}, 2024.
\newblock URL \url{https://arxiv.org/abs/2408.05017}.

\bibitem[Sadoune et~al.(2025)Sadoune, Liu, Yan, Jaubert, Shannon, and
  Pollet]{Sadoune2025}
Nicolas Sadoune, Ke~Liu, Han Yan, Ludovic D.~C. Jaubert, Nic Shannon, and Lode
  Pollet.
\newblock {Human-machine collaboration: ordering mechanism of rank-2 spin
  liquid on breathing pyrochlore lattice}, 2025.
\newblock URL \url{https://arxiv.org/abs/2402.10658}.

\bibitem[Saffman et~al.(2010)Saffman, Walker, and M\o{}lmer]{Saffman2010}
M.~Saffman, T.~G. Walker, and K.~M\o{}lmer.
\newblock {Quantum information with Rydberg atoms}.
\newblock \emph{Rev. Mod. Phys.}, 82:\penalty0 2313--2363, Aug 2010.
\newblock \doi{10.1103/RevModPhys.82.2313}.
\newblock URL \url{https://link.aps.org/doi/10.1103/RevModPhys.82.2313}.

\bibitem[Samajdar et~al.(2020)Samajdar, Ho, Pichler, Lukin, and
  Sachdev]{Samajdar2020}
Rhine Samajdar, Wen~Wei Ho, Hannes Pichler, Mikhail~D. Lukin, and Subir
  Sachdev.
\newblock {Complex Density Wave Orders and Quantum Phase Transitions in a Model
  of Square-Lattice Rydberg Atom Arrays}.
\newblock \emph{Phys. Rev. Lett.}, 124:\penalty0 103601, Mar 2020.
\newblock \doi{10.1103/PhysRevLett.124.103601}.
\newblock URL \url{https://link.aps.org/doi/10.1103/PhysRevLett.124.103601}.

\bibitem[Sauvage and Mintert(2020)]{sauvage2020optimal}
Fr\'ed\'eric Sauvage and Florian Mintert.
\newblock {Optimal Quantum Control with Poor Statistics}.
\newblock \emph{PRX Quantum}, 1:\penalty0 020322, Dec 2020.
\newblock \doi{10.1103/PRXQuantum.1.020322}.
\newblock URL \url{https://link.aps.org/doi/10.1103/PRXQuantum.1.020322}.

\bibitem[Sch\"afer and L\"orch(2019)]{Schafer2019}
Frank Sch\"afer and Niels L\"orch.
\newblock {Vector field divergence of predictive model output as indication of
  phase transitions}.
\newblock \emph{Phys. Rev. E}, 99:\penalty0 062107, Jun 2019.
\newblock \doi{10.1103/PhysRevE.99.062107}.
\newblock URL \url{https://link.aps.org/doi/10.1103/PhysRevE.99.062107}.

\bibitem[Schl{\"o}mer et~al.(2023)Schl{\"o}mer, Hilker, Bloch, Schollw{\"o}ck,
  Grusdt, and Bohrdt]{Schloemer2023rec}
Henning Schl{\"o}mer, Timon~A. Hilker, Immanuel Bloch, Ulrich Schollw{\"o}ck,
  Fabian Grusdt, and Annabelle Bohrdt.
\newblock {Quantifying hole-motion-induced frustration in doped
  antiferromagnets by Hamiltonian reconstruction}.
\newblock \emph{Communications Materials}, 4\penalty0 (1):\penalty0 64, 2023.
\newblock \doi{10.1038/s43246-023-00382-3}.
\newblock URL \url{https://doi.org/10.1038/s43246-023-00382-3}.

\bibitem[Schl\"omer et~al.(2024)Schl\"omer, Lange, Franz, Chalopin,
  Bojovi\ifmmode~\acute{c}\else \'{c}\fi{}, Wang, Bloch, Hilker, Grusdt, and
  Bohrdt]{Schloemer2024}
Henning Schl\"omer, Hannah Lange, Titus Franz, Thomas Chalopin, Petar
  Bojovi\ifmmode~\acute{c}\else \'{c}\fi{}, Si~Wang, Immanuel Bloch, Timon~A.
  Hilker, Fabian Grusdt, and Annabelle Bohrdt.
\newblock {Local Control and Mixed Dimensions: Exploring High-Temperature
  Superconductivity in Optical Lattices}.
\newblock \emph{PRX Quantum}, 5:\penalty0 040341, Dec 2024.
\newblock \doi{10.1103/PRXQuantum.5.040341}.
\newblock URL \url{https://link.aps.org/doi/10.1103/PRXQuantum.5.040341}.

\bibitem[Schl\"omer et~al.(2025)Schl\"omer, Bohrdt, and
  Grusdt]{Schloemer2024GOM}
Henning Schl\"omer, Annabelle Bohrdt, and Fabian Grusdt.
\newblock Geometric orthogonal metals: Hidden antiferromagnetism and the
  pseudogap from fluctuating stripes.
\newblock \emph{PRX Quantum}, 6:\penalty0 030342, Sep 2025.
\newblock \doi{10.1103/5sq4-r7dk}.
\newblock URL \url{https://link.aps.org/doi/10.1103/5sq4-r7dk}.

\bibitem[Schlömer and Bohrdt(2023)]{Schloemer2023fluc}
Henning Schlömer and Annabelle Bohrdt.
\newblock {Fluctuation based interpretable analysis scheme for quantum
  many-body snapshots}.
\newblock \emph{SciPost Phys.}, 15:\penalty0 099, 2023.
\newblock \doi{10.21468/SciPostPhys.15.3.099}.
\newblock URL \url{https://scipost.org/10.21468/SciPostPhys.15.3.099}.

\bibitem[Schmale et~al.(2022)Schmale, Reh, and G{\"a}rttner]{Schmale2022}
Tobias Schmale, Moritz Reh, and Martin G{\"a}rttner.
\newblock {Efficient quantum state tomography with convolutional neural
  networks}.
\newblock \emph{npj Quantum Information}, 8\penalty0 (1):\penalty0 115, 2022.
\newblock \doi{10.1038/s41534-022-00621-4}.
\newblock URL \url{https://doi.org/10.1038/s41534-022-00621-4}.

\bibitem[Schmitt and Lenar\ifmmode \check{c}\else
  \v{c}\fi{}i\ifmmode~\check{c}\else \v{c}\fi{}(2022)]{Schmitt2022}
Markus Schmitt and Zala Lenar\ifmmode \check{c}\else
  \v{c}\fi{}i\ifmmode~\check{c}\else \v{c}\fi{}.
\newblock {From observations to complexity of quantum states via unsupervised
  learning}.
\newblock \emph{Phys. Rev. B}, 106:\penalty0 L041110, Jul 2022.
\newblock \doi{10.1103/PhysRevB.106.L041110}.
\newblock URL \url{https://link.aps.org/doi/10.1103/PhysRevB.106.L041110}.

\bibitem[Scholl et~al.(2021)Scholl, Schuler, Williams, Eberharter, Barredo,
  Schymik, Lienhard, Henry, Lang, Lahaye, L{\"a}uchli, and
  Browaeys]{Scholl2021}
Pascal Scholl, Michael Schuler, Hannah~J. Williams, Alexander~A. Eberharter,
  Daniel Barredo, Kai-Niklas Schymik, Vincent Lienhard, Louis-Paul Henry,
  Thomas~C. Lang, Thierry Lahaye, Andreas~M. L{\"a}uchli, and Antoine Browaeys.
\newblock {Quantum simulation of 2D antiferromagnets with hundreds of Rydberg
  atoms}.
\newblock \emph{Nature}, 595\penalty0 (7866):\penalty0 233--238, 2021.
\newblock \doi{10.1038/s41586-021-03585-1}.
\newblock URL \url{https://doi.org/10.1038/s41586-021-03585-1}.

\bibitem[Simard et~al.(2025)Simard, Dawid, Tindall, Ferrero, Sengupta, and
  Georges]{Simard2025}
Olivier Simard, Anna Dawid, Joseph Tindall, Michel Ferrero, Anirvan~M.
  Sengupta, and Antoine Georges.
\newblock Learning interactions between rydberg atoms.
\newblock \emph{PRX Quantum}, 6:\penalty0 030324, Aug 2025.
\newblock \doi{10.1103/f58h-zxs3}.
\newblock URL \url{https://link.aps.org/doi/10.1103/f58h-zxs3}.

\bibitem[Simjanovski et~al.(2023)Simjanovski, Gauthier, Davis,
  Rubinsztein-Dunlop, and Neely]{Simjanovski2023}
Simeon Simjanovski, Guillaume Gauthier, Matthew~J. Davis, Halina
  Rubinsztein-Dunlop, and Tyler~W. Neely.
\newblock {Optimizing persistent currents in a ring-shaped Bose-Einstein
  condensate using machine learning}.
\newblock \emph{Physical Review A}, 108\penalty0 (6):\penalty0 063306, 2023.
\newblock ISSN 2469-9926.
\newblock \doi{10.1103/physreva.108.063306}.

\bibitem[Striegel et~al.(2023)Striegel, Ibarra-Garcia-Padilla, and
  Khatami]{Striegel2023}
Stephanie Striegel, Eduardo Ibarra-Garcia-Padilla, and Ehsan Khatami.
\newblock {Machine Learning Detection of Correlations in Snapshots of Ultracold
  Atoms in Optical Lattices}, 2023.
\newblock URL \url{https://arxiv.org/abs/2310.03267}.

\bibitem[Suchsland and Wessel(2018)]{Suchsland2018}
Philippe Suchsland and Stefan Wessel.
\newblock {Parameter diagnostics of phases and phase transition learning by
  neural networks}.
\newblock \emph{Phys. Rev. B}, 97:\penalty0 174435, May 2018.
\newblock \doi{10.1103/PhysRevB.97.174435}.
\newblock URL \url{https://link.aps.org/doi/10.1103/PhysRevB.97.174435}.

\bibitem[Suresh et~al.(2025)Suresh, Schl{\"o}mer, Hashemi, and
  Bohrdt]{Suresh2025}
Abhinav Suresh, Henning Schl{\"o}mer, Baran Hashemi, and Annabelle Bohrdt.
\newblock {Interpretable correlator Transformer for image-like quantum matter
  data}.
\newblock \emph{Machine Learning: Science and Technology}, 6\penalty0
  (2):\penalty0 025006, 2025.
\newblock \doi{10.1088/2632-2153/adc071}.
\newblock URL \url{https://dx.doi.org/10.1088/2632-2153/adc071}.

\bibitem[Tanaka and Tomiya(2017)]{TanakaTomiya2017}
Akinori Tanaka and Akio Tomiya.
\newblock {Detection of Phase Transition via Convolutional Neural Networks}.
\newblock \emph{Journal of the Physical Society of Japan}, 86\penalty0
  (6):\penalty0 063001, 2025/04/09 2017.
\newblock \doi{10.7566/JPSJ.86.063001}.
\newblock URL \url{https://doi.org/10.7566/JPSJ.86.063001}.

\bibitem[Tirelli and Costa(2021)]{Tirelli2021}
Andrea Tirelli and Natanael~C. Costa.
\newblock Learning quantum phase transitions through topological data analysis.
\newblock \emph{Phys. Rev. B}, 104:\penalty0 235146, Dec 2021.
\newblock \doi{10.1103/PhysRevB.104.235146}.
\newblock URL \url{https://link.aps.org/doi/10.1103/PhysRevB.104.235146}.

\bibitem[Torlai and Melko(2016)]{Torlai2016}
Giacomo Torlai and Roger~G. Melko.
\newblock {Learning thermodynamics with Boltzmann machines}.
\newblock \emph{Phys. Rev. B}, 94:\penalty0 165134, Oct 2016.
\newblock \doi{10.1103/PhysRevB.94.165134}.
\newblock URL \url{https://link.aps.org/doi/10.1103/PhysRevB.94.165134}.

\bibitem[Torlai et~al.(2018)Torlai, Mazzola, Carrasquilla, Troyer, Melko, and
  Carleo]{Torlai2018}
Giacomo Torlai, Guglielmo Mazzola, Juan Carrasquilla, Matthias Troyer, Roger
  Melko, and Giuseppe Carleo.
\newblock {Neural-network quantum state tomography}.
\newblock \emph{Nature Physics}, 14\penalty0 (5):\penalty0 447--450, 2018.
\newblock \doi{10.1038/s41567-018-0048-5}.
\newblock URL \url{https://doi.org/10.1038/s41567-018-0048-5}.

\bibitem[Torlai et~al.(2019)Torlai, Timar, van Nieuwenburg, Levine, Omran,
  Keesling, Bernien, Greiner, Vuleti\ifmmode~\acute{c}\else \'{c}\fi{}, Lukin,
  Melko, and Endres]{Torlai2019}
Giacomo Torlai, Brian Timar, Evert P.~L. van Nieuwenburg, Harry Levine, Ahmed
  Omran, Alexander Keesling, Hannes Bernien, Markus Greiner, Vladan
  Vuleti\ifmmode~\acute{c}\else \'{c}\fi{}, Mikhail~D. Lukin, Roger~G. Melko,
  and Manuel Endres.
\newblock {Integrating Neural Networks with a Quantum Simulator for State
  Reconstruction}.
\newblock \emph{Phys. Rev. Lett.}, 123:\penalty0 230504, Dec 2019.
\newblock \doi{10.1103/PhysRevLett.123.230504}.
\newblock URL \url{https://link.aps.org/doi/10.1103/PhysRevLett.123.230504}.

\bibitem[Uvarov et~al.(2020)Uvarov, Kardashin, and Biamonte]{Uvarov2020}
A.~V. Uvarov, A.~S. Kardashin, and J.~D. Biamonte.
\newblock {Machine learning phase transitions with a quantum processor}.
\newblock \emph{Phys. Rev. A}, 102:\penalty0 012415, Jul 2020.
\newblock \doi{10.1103/PhysRevA.102.012415}.
\newblock URL \url{https://link.aps.org/doi/10.1103/PhysRevA.102.012415}.

\bibitem[van Nieuwenburg et~al.(2017)van Nieuwenburg, Liu, and
  Huber]{Nieuwenburg2017}
Evert P.~L. van Nieuwenburg, Ye-Hua Liu, and Sebastian~D. Huber.
\newblock {Learning phase transitions by confusion}.
\newblock \emph{Nature Physics}, 13\penalty0 (5):\penalty0 435--439, 2017.
\newblock \doi{10.1038/nphys4037}.
\newblock URL \url{https://doi.org/10.1038/nphys4037}.

\bibitem[Vendeiro et~al.(2022)Vendeiro, Ramette, Rudelis, Chong, Sinclair,
  Stewart, Urvoy, and Vuletić]{Vandeiro2022}
Zachary Vendeiro, Joshua Ramette, Alyssa Rudelis, Michelle Chong, Josiah
  Sinclair, Luke Stewart, Alban Urvoy, and Vladan Vuletić.
\newblock {Machine-learning-accelerated Bose-Einstein condensation}.
\newblock \emph{Physical Review Research}, 4\penalty0 (4):\penalty0 043216,
  2022.
\newblock \doi{10.1103/physrevresearch.4.043216}.

\bibitem[Wang and Zhai(2017)]{Wang2017Frustrated}
Ce~Wang and Hui Zhai.
\newblock {Machine learning of frustrated classical spin models. I. Principal
  component analysis}.
\newblock \emph{Phys. Rev. B}, 96:\penalty0 144432, Oct 2017.
\newblock \doi{10.1103/PhysRevB.96.144432}.
\newblock URL \url{https://link.aps.org/doi/10.1103/PhysRevB.96.144432}.

\bibitem[Wang and Zhai(2018)]{Wang2018Kernel}
Ce~Wang and Hui Zhai.
\newblock {Machine learning of frustrated classical spin models (II): Kernel
  principal component analysis}.
\newblock \emph{Frontiers of Physics}, 13\penalty0 (5):\penalty0 130507, 2018.
\newblock \doi{10.1007/s11467-018-0798-7}.
\newblock URL \url{https://doi.org/10.1007/s11467-018-0798-7}.

\bibitem[Wang(2016)]{Wang2016}
Lei Wang.
\newblock {Discovering phase transitions with unsupervised learning}.
\newblock \emph{Phys. Rev. B}, 94:\penalty0 195105, Nov 2016.
\newblock \doi{10.1103/PhysRevB.94.195105}.
\newblock URL \url{https://link.aps.org/doi/10.1103/PhysRevB.94.195105}.

\bibitem[Wetzel(2017)]{Wetzel2017}
Sebastian~J. Wetzel.
\newblock {Unsupervised learning of phase transitions: From principal component
  analysis to variational autoencoders}.
\newblock \emph{Phys. Rev. E}, 96:\penalty0 022140, Aug 2017.
\newblock \doi{10.1103/PhysRevE.96.022140}.
\newblock URL \url{https://link.aps.org/doi/10.1103/PhysRevE.96.022140}.

\bibitem[Wetzel and Scherzer(2017)]{Wetzel2017Int}
Sebastian~J. Wetzel and Manuel Scherzer.
\newblock {Machine learning of explicit order parameters: From the Ising model
  to SU(2) lattice gauge theory}.
\newblock \emph{Phys. Rev. B}, 96:\penalty0 184410, Nov 2017.
\newblock \doi{10.1103/PhysRevB.96.184410}.
\newblock URL \url{https://link.aps.org/doi/10.1103/PhysRevB.96.184410}.

\bibitem[Wetzel et~al.(2025)Wetzel, Ha, Iten, Klopotek, and Liu]{WetzelReview}
Sebastian~Johann Wetzel, Seungwoong Ha, Raban Iten, Miriam Klopotek, and Ziming
  Liu.
\newblock {Interpretable Machine Learning in Physics: A Review}, 2025.
\newblock URL \url{https://arxiv.org/abs/2503.23616}.

\bibitem[Wigley et~al.(2016)Wigley, Everitt, Hengel, Bastian, Sooriyabandara,
  McDonald, Hardman, Quinlivan, Manju, Kuhn, Petersen, Luiten, Hope, Robins,
  and Hush]{Wigley2016}
P.~B. Wigley, P.~J. Everitt, A.~van~den Hengel, J.~W. Bastian, M.~A.
  Sooriyabandara, G.~D. McDonald, K.~S. Hardman, C.~D. Quinlivan, P.~Manju,
  C.~C.~N. Kuhn, I.~R. Petersen, A.~N. Luiten, J.~J. Hope, N.~P. Robins, and
  M.~R. Hush.
\newblock {Fast machine-learning online optimization of ultra-cold-atom
  experiments}.
\newblock \emph{Scientific Reports}, 6\penalty0 (1):\penalty0 25890, 2016.
\newblock \doi{10.1038/srep25890}.

\bibitem[Xie et~al.(2022)Xie, Dai, Yuan, Deng, Li, Chen, and Pan]{Xie2022}
Yan-Jun Xie, Han-Ning Dai, Zhen-Sheng Yuan, Youjin Deng, Xiaopeng Li, Yu-Ao
  Chen, and Jian-Wei Pan.
\newblock {Bayesian learning for optimal control of quantum many-body states in
  optical lattices}.
\newblock \emph{Physical Review A}, 106\penalty0 (1):\penalty0 013316, 2022.
\newblock ISSN 2469-9926.
\newblock \doi{10.1103/physreva.106.013316}.

\bibitem[Xu et~al.(2025)Xu, Kendrick, Kale, Gang, Feng, Zhang, Young, Lebrat,
  and Greiner]{Xu_cryo2025}
Muqing Xu, Lev~Haldar Kendrick, Anant Kale, Youqi Gang, Chunhan Feng, Shiwei
  Zhang, Aaron~W. Young, Martin Lebrat, and Markus Greiner.
\newblock {A neutral-atom Hubbard quantum simulator in the cryogenic regime}.
\newblock \emph{Nature}, 642\penalty0 (8069):\penalty0 909--915, 2025.
\newblock \doi{10.1038/s41586-025-09112-w}.
\newblock URL \url{https://doi.org/10.1038/s41586-025-09112-w}.

\bibitem[Zhang and Sarovar(2014)]{Zhang2014}
Jun Zhang and Mohan Sarovar.
\newblock {Quantum Hamiltonian Identification from Measurement Time Traces}.
\newblock \emph{Phys. Rev. Lett.}, 113:\penalty0 080401, Aug 2014.
\newblock \doi{10.1103/PhysRevLett.113.080401}.
\newblock URL \url{https://link.aps.org/doi/10.1103/PhysRevLett.113.080401}.

\bibitem[Zhang et~al.(2018)Zhang, Shen, and Zhai]{Zhang2018LHam}
Pengfei Zhang, Huitao Shen, and Hui Zhai.
\newblock {Machine Learning Topological Invariants with Neural Networks}.
\newblock \emph{Phys. Rev. Lett.}, 120:\penalty0 066401, Feb 2018.
\newblock \doi{10.1103/PhysRevLett.120.066401}.
\newblock URL \url{https://link.aps.org/doi/10.1103/PhysRevLett.120.066401}.

\bibitem[Zhang et~al.(2022)Zhang, Wan, Lee, Hsieh, Zhang, and
  Yao]{Zhang2022VQE}
Shi-Xin Zhang, Zhou-Quan Wan, Chee-Kong Lee, Chang-Yu Hsieh, Shengyu Zhang, and
  Hong Yao.
\newblock {Variational Quantum-Neural Hybrid Eigensolver}.
\newblock \emph{Phys. Rev. Lett.}, 128:\penalty0 120502, Mar 2022.
\newblock \doi{10.1103/PhysRevLett.128.120502}.
\newblock URL \url{https://link.aps.org/doi/10.1103/PhysRevLett.128.120502}.

\bibitem[Zhang and Kim(2017)]{Zhang2017QLT}
Yi~Zhang and Eun-Ah Kim.
\newblock {Quantum Loop Topography for Machine Learning}.
\newblock \emph{Phys. Rev. Lett.}, 118:\penalty0 216401, May 2017.
\newblock \doi{10.1103/PhysRevLett.118.216401}.
\newblock URL \url{https://link.aps.org/doi/10.1103/PhysRevLett.118.216401}.

\bibitem[Zhang et~al.(2017)Zhang, Melko, and Kim]{Zhang2017QLT2}
Yi~Zhang, Roger~G. Melko, and Eun-Ah Kim.
\newblock {Machine learning ${\mathbb{Z}}_{2}$ quantum spin liquids with
  quasiparticle statistics}.
\newblock \emph{Phys. Rev. B}, 96:\penalty0 245119, Dec 2017.
\newblock \doi{10.1103/PhysRevB.96.245119}.
\newblock URL \url{https://link.aps.org/doi/10.1103/PhysRevB.96.245119}.

\bibitem[Zhang et~al.(2020)Zhang, Ginsparg, and Kim]{Zhang2020}
Yi~Zhang, Paul Ginsparg, and Eun-Ah Kim.
\newblock {Interpreting machine learning of topological quantum phase
  transitions}.
\newblock \emph{Phys. Rev. Res.}, 2:\penalty0 023283, Jun 2020.
\newblock \doi{10.1103/PhysRevResearch.2.023283}.
\newblock URL \url{https://link.aps.org/doi/10.1103/PhysRevResearch.2.023283}.

\bibitem[Zhao et~al.(2022)Zhao, Mak, He, Ren, Pak, Liu, and Jo]{Zhao2022}
Entong Zhao, Ting~Hin Mak, Chengdong He, Zejian Ren, Ka~Kwan Pak, Yu-Jun Liu,
  and Gyu-Boong Jo.
\newblock {Observing a topological phase transition with deep neural networks
  from experimental images of ultracold atoms}.
\newblock \emph{Optics Express}, 30\penalty0 (21):\penalty0 37786--37794, 2022.
\newblock \doi{10.1364/OE.473770}.
\newblock URL \url{https://opg.optica.org/oe/abstract.cfm?URI=oe-30-21-37786}.

\end{thebibliography}

\end{document}